\documentclass[aps,preprintnumbers,nofootinbib]{revtex4}

\usepackage{graphicx}
\usepackage{bm}
\newcommand{\bold}[1]{\mbox{\text{\normalsize\boldmath $#1$}}}
\def\CPT{$ C \! P \!T$ } 
\def\CP{$ C \! P$ } 
\def\CPn{$ C \! P$}

\newcommand{\Omegachi}{\Omega_{\chi}}
\newcommand{\mgaugino}{M_{1/2}}
\newcommand{\bsg}{B\to X_s \gamma}
\newcommand{\sign}{\:\!\text{sign}\:\!}
\newcommand{\mgut}{M_{\text{GUT}}}
\newcommand{\tb}{\tan\beta}
\newcommand{\gev}{\text{GeV}}

\newcommand{\amu}{a_{\mu}}
\newcommand{\dmu}{d_{\mu}}
\newcommand{\amuexp}{a_{\mu}^{\text{exp}}}
\newcommand{\amususy}{a_{\mu}^{\text{SUSY}}}
\newcommand{\amusm}{a_{\mu}^{\text{SM}}}
\newcommand{\dmusm}{d_{\mu}^{\text{SM}}}

\newcommand{\amunp}{a_{\mu}^{\text{NP}}}
\newcommand{\dmunp}{d_{\mu}^{\text{NP}}}
\newcommand{\phicp}{\phi_{\text{CP}}}
\newcommand{\text}[1]{ {\rm #1} }
\newcommand{\etal}{{\em et al.}}
\newcommand{\eg}{{\em e.g.}}

\newcommand{\km}{{\rm km}}
\newcommand{\cm}{{\rm cm}}
\newcommand{\yr}{{\rm yr}}
\newcommand{\s}{{\rm s}}
\newcommand{\ethr}{E_{\rm th}}
\newcommand{\eopt}{E_{\rm opt}}
\newcommand{\mlosp}{M_{\text{LOSP}}}
\newcommand{\ecm}{e~\text{cm}}
\renewcommand{\Re}{{\cal R}e}
\renewcommand{\Im}{{\cal I}m}
\newcommand{\gsim}{ \mathop{}_{\textstyle \sim}^{\textstyle >} }
\newcommand{\lsim}{ \mathop{}_{\textstyle \sim}^{\textstyle <} }
\newcommand{\ra}{\rightarrow}
\newcommand{\textfrac}[2]{{\textstyle\frac{#1}{#2}}}

\newcommand{\bs}{$B_s\rightarrow\mu^+\mu^-$}
\newcommand{\cbs}{{\cal B}(B_s\rightarrow\mu^+\mu^-)}

\begin{document}
\bibliographystyle{revtex}

\preprint{Bergen ISSN 0803-2696/2001-06, 
          CERN--TH/2001-322, Snowmass P3-42, hep-ph/0112312}

\title{Indirect Investigations of Supersymmetry}



\author{Gerald Eigen}
\email[]{Gerald.Eigen@fi.uib.no}
\affiliation{Department of Physics, 
             University of Bergen,
             Allegaten 55, 5007 Bergen, Norway}
\author{Rick Gaitskell}
\email[]{gaitskell@physics.brown.edu}
\affiliation{Department of Physics, 
             Brown University, 
             Providence, RI 02912, USA}
\author{Graham D. Kribs}
\email[]{kribs@pheno.physics.wisc.edu}
\affiliation{Department of Physics,
             University of Wisconsin,
             Madison, WI 53706-1390, USA}
\author{Konstantin T.~Matchev}
\email[]{Konstantin.Matchev@cern.ch}
\affiliation{Theory Division, CERN,
             CH--1211, Geneva 23, Switzerland}


\date{December 21, 2001}

\begin{abstract}
This is the summary report of the ``Indirect Investigations of SUSY''
subgroup of the P3 Physics Group at Snowmass 2001.
\end{abstract}

\maketitle


\section{Introduction}
\label{sec:intro}

In this report we consider indirect probes of supersymmetry (SUSY).
Following our charge, we will review the current experimental
status and discuss possible levels of improvements in various
measurements sensitive to supersymmetry through either virtual
or astrophysics effects. We mention the upcoming experiments which are
likely to achieve such precision, and outline which theoretical 
models and ideas can be tested by those experiments.

Since it is impossible to give a detailed review of every
single topic here, we have limited our discussion to a few representative
topics which were studied either at Snowmass or since then.
A large fraction of this report is based on the individual written
contributions~\cite{Feng:2001mq,Feng:2001hm,Ellis:2001us,Feng:2001ut,ge1,ge3}
to our subgroup as well as talks presented at Snowmass.

In Section~\ref{sec:gm2} we discuss the implications of the
recent $g_\mu-2$ measurement (and its possible improvements)
for supersymmetry. In Section~\ref{sec:cp} we review searches
for \CP violation and Section~\ref{sec:LFV} deals with lepton-flavor 
violation (LFV). Section~\ref{sec:B} is devoted to the future of B-physics.
In Section~\ref{sec:DMdirect} (\ref{sec:DMindirect})
we discuss direct (indirect) searches for supersymmetric
dark matter. We summarize and present our conclusions in 
Section~\ref{sec:discussion}.

\section{Anomalous Magnetic Moment of the Muon}
\label{sec:gm2}

Undoubtedly, among the most exciting news of the year was the 
announcement of the new measurement of the muon anomaly
at Brookhaven~\cite{Brown:2001mg}. The muon anomalous magnetic moment
$\amu$ was reported\footnote{Notice the recent questioning of 
the sign of the theoretical prediction for the light-by-light
scattering contribution~\cite{Knecht:2001qf,Knecht:2001qg}. 
With a consensus currently building towards the opposite 
sign~\cite{kinoshita,Blokland:2001pb,Bijnens:2001cq}, 
the deviation is less than $2\sigma$.
The purely hadronic contribution has also been under active
discussion~\cite{Melnikov:2001uw,DeTroconiz:2001wt,Cvetic:2001pg}.}
to differ from the Standard Model (SM) prediction by $2.6 \sigma$
\begin{equation}
\amuexp - \amusm = (43 \pm 16) \times 10^{-10}\ ,
\label{currentMDM}
\end{equation}
which is about three times larger than the Standard
Model's electroweak contribution~\cite{Czarnecki:2001pv}.
Deviations of roughly this order are expected in many models motivated
by attempts to understand electroweak symmetry breaking. A
supersymmetric interpretation is particularly attractive, since
supersymmetry naturally provides electroweak scale contributions that
are easily enhanced (by large $\tan\beta$) to produce deviations
of the required magnitude.  In addition, $\amu$ is both 
flavor- and \CP-conserving.  Thus, while the impact of supersymmetry 
on other low energy observables (see sections~\ref{sec:cp} 
and~\ref{sec:LFV}) can be highly suppressed by scalar degeneracy 
or small \CPn -violating phases, supersymmetric 
contributions to $\amu$ are generic.

Supersymmetric contributions to $\amu$ have been explored for many 
years~\cite{Fayet:1980yy,Grifols:1982vx,Ellis:1982by,Barbieri:1982aj,Kosower:1983yw,%
Yuan:1984ww,Moroi:1996yh,Carena:1997qa,Gabrielli:1997jp,Mahanthappa:1999ta}.  
Following the recent $\amu$ result, 
the implications for supersymmetry have been considered in numerous
studies~\cite{Feng:2001tr,Everett:2001tq,Baltz:2001ts,Chattopadhyay:2001vx,Komine:2001fz,%
Hisano:2001qz,Ibrahim:2001ym,Ellis:2001yu,Arnowitt:2001be,Choi:2001pz,%
Kim:2001se,Martin:2001st,Komine:2001hy,Baek:2001nz,Carvalho:2001ex,Baer:2001kn,%
Baek:2001kh,Cerdeno:2001aj,Chacko:2001xd,Blazek:2001zm,Cho:2001nf,Adhikari:2001ra,%
Byrne:2001yu,Komine:2001rm,Chattopadhyay:2001mj,Endo:2001ym}. 
The most significant consequence is that
at least two superpartners cannot decouple, if supersymmetry is to
explain the deviation, and one of these must be charged and so observable at
colliders.  Non-vanishing $\amususy$ thus imply upper bounds on
the mass $\mlosp$ of the lightest observable superpartner.  

\begin{figure}[tbp]
\includegraphics[width=.48\textwidth]{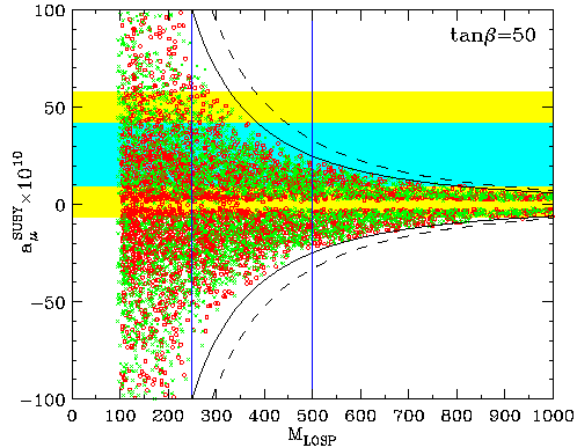}%
\caption{Allowed values of $\mlosp$, the mass of the lightest
observable supersymmetric particle (LOSP), and $\amususy$ from a scan of
parameter space with $M_1=M_2/2$, $A_{\mu} = 0$, and $\tb = 50$. Green
crosses (red circles) have smuons (charginos/neutralinos) as the LOSP.
A stable lightest supersymmetric particle (LSP) is assumed.  
Relaxing the relation $M_1=M_2/2$ leads to
the solid envelope curve, and further allowing arbitrary left-right
smuon mixing (large $A_{\mu}$) leads to the dashed curve.  The
envelope contours scale linearly with $\tb$.  The 1$\sigma$ 
(dark shaded, blue) and 2$\sigma$ (light shaded, yellow) 
allowed $\amususy$ ranges are shown, 
and the discovery reaches of linear colliders with 
$\sqrt{s} = 500~\gev$ and 1 TeV are
given by the vertical blue lines. (From Ref.~\cite{Feng:2001hm}.)}
\label{fig:mlosp}
\end{figure}

Fig.~\ref{fig:mlosp} shows the results from a series of high statistics
scans in the relevant supersymmetric parameter space, consistent
with slepton-flavor conservation: $M_1$, $M_2$, $\mu$, $\tb$, 
$m_{\tilde{\mu}_L}$, $m_{\tilde{\mu}_R}$, and $A_{\mu}$.
(For more details, see Ref.~\cite{Feng:2001tr}.)
The points are obtained by assuming gaugino mass unification $M_1= M_2/2$,
fixing $A_{\mu}=0$ and $\tb=50$, and scanning over the remaining parameters
up to 2.5 TeV. Collider bounds from supersymmetry searches are 
enforced and a neutral LSP is assumed. Relaxing the gaugino unification
assumption leads to possibilities bounded by the solid curve. 
Finally, allowing any $A_{\mu}$ in the interval $\rm [-100~TeV, 100~TeV]$
extends the envelope curve to the dashed contour of Fig.~\ref{fig:mlosp}. 
The envelope contours scale linearly with $\tb$ to excellent approximation.

{}From Fig.~\ref{fig:mlosp} we see that the measured deviation in
$\amu$ is in the range accessible to supersymmetric theories and is
easily explained by supersymmetric effects.  The case where the LSP 
decays visibly in collider detectors yields even lower bounds, 
and is examined in Ref.~\cite{Feng:2001tr}.

Such model-independent upper bounds have many implications. They
improve the prospects for observation of weakly-interacting
superpartners at the Tevatron and LHC. They also impact linear
colliders: for reference, the discovery reach of linear colliders with
$\sqrt{s} = 500~\gev$ and 1 TeV are given in Fig.~\ref{fig:mlosp}.  In
this highly model-independent framework, an observable supersymmetry
signal is very probable at a 1.2 TeV linear collider; 
the expected improvements in $\amu$
measurements may significantly strengthen such conclusions. 
(The ultimate target of the BNL experiment is
an experimental error of $4\times 10^{-10}$.) 
Finally, these bounds provide fresh impetus for searches for 
lepton-flavor violation, which is also mediated by sleptons and
charginos/neutralinos \cite{Graesser:2001ec,Chacko:2001xd}.

Turning now to specific models, we first consider the framework of minimal
supergravity, which is completely specified by four continuous
parameters and one binary choice: $m_0$, $\mgaugino$, $A_0$, $\tb$,
and $\sign(\mu)$.  The first three are the universal scalar, gaugino,
and trilinear coupling masses at the grand unified theory scale $\mgut
\simeq 2\times 10^{16}~\gev$.  

In minimal supergravity (mSUGRA), $\sign(\amususy) = \sign(\mu M_{1,2})$,
so the $\amu$ result prefers a particular sign of $\mu$ relative
to the gaugino masses. As is well-known, however,
the sign of $\mu$ also enters in the supersymmetric contributions to
$\bsg$.  Current constraints on $\bsg$ require $\mu M_3 > 0$ if $\tb$
is large (here $M_3$ is the gluino mass parameter). 
Gaugino mass unification implies $M_{1,2} M_3 > 0$, therefore, 
a large discrepancy in $\amu$ is only possible for $\amususy > 0$, in
accord with the new measurement. Minimal supergravity, and gaugino
unified models, in general, are generally consistent with the BNL
measurement~\cite{Feng:2001tr,Chattopadhyay:2001vx,Komine:2001fz,%
Ellis:2001yu,Arnowitt:2001be, Martin:2001st,Baer:2001kn,Komine:2001rm,%
Djouadi:2001yk}.

In contrast, the minimal model of anomaly-mediated supersymmetry breaking
seems to be disfavored.  One of the most striking predictions of
anomaly mediation is that the gaugino masses are proportional to the
corresponding beta function coefficients, and so $M_{1,2} M_3 < 0$.  
Anomaly-mediation, therefore, most naturally predicts 
$\amususy < 0$~\cite{Feng:1999hg,Chattopadhyay:2000ws},
in contrast to the observed deviation. The dependence of this
argument on the characteristic gaugino mass relations of anomaly
mediation suggests that similar conclusions will remain valid beyond
the minimal model.

In summary, the recently reported deviation in $\amu$ is easily
accommodated in supersymmetric models.  Its value provides {\em
model-independent} upper bounds on masses of observable superpartners
and already discriminates between well-motivated models.

\section{\CP Violation}
\label{sec:cp}

\CP violation is among the least understood phenomena in the 
Standard Model.
At present, \CP violation is observed in only a small number
of processes, such as in Kaon and B-meson mixing and decays.  
This can be accommodated through
the single phase that appears in the CKM matrix.  However, 
the CKM phase is not the only parameter that can lead 
to \CP violation in the SM.  The QCD $\theta$ term, 
\begin{equation}
\theta \frac{g_3^2}{32 \pi^2} G \tilde{G}\ ,
\end{equation}
where $G$ is the gluon field strength and $\tilde G$ is its dual,
also leads to \CP violation,
and indeed there are very strong constraints on this from limits 
on the electric dipole moment of the neutron (and mercury).
Furthermore, \CP violation is an essential ingredient of
almost all attempts to explain the matter-antimatter asymmetry of the
universe~\cite{Sakharov:1967dj}, yet the amount
of \CP violation present in the CKM matrix is insufficient to explain
the observed 
asymmetry~\cite{Farrar:1994hn,Gavela:1994ts,Gavela:1994dt,Huet:1995jb}.  
Hence, searches for \CP violation beyond the CKM matrix are an
important probe into physics beyond the Standard Model.

Electric dipole moments (EDMs) violate both parity (P) and time
reversal (T) invariance.  If \CPT is assumed to be an unbroken
symmetry, a permanent EDM is, then, a signature of \CP violation.  
A non-vanishing permanent EDM has not been measured for
any of the known elementary particles.  In the Standard Model, EDMs
are generated only at the multi-loop level and are predicted to be many
orders of magnitude below the sensitivity of foreseeable
experiments~\cite{Hoogeveen:1990cb,Khriplovich:1986jr}.  
A non-vanishing EDM, therefore, would be
unambiguous evidence for \CP violation beyond the CKM matrix, and
searches for permanent EDMs of fundamental particles are powerful
probes of extensions of the Standard Model.  In fact, current EDM
bounds are already some of the most stringent constraints on new
physics, and they are highly complementary to many other low energy
constraints, since they require \CP violation, but not flavor
violation. In this Section we review the experimental prospects 
for the EDMs of various systems, and the implications for
supersymmetry.

\subsection{Experimental limits on EDMs}

First, we summarize the current experimental bounds on EDMs
and briefly mention future prospects.  For the electron EDM, 
the current bound is \cite{Commins:1994gv}
\begin{equation}
d_e < 4 \times 10^{-27} \; e \; \mbox{cm} \; ,
\end{equation}
although a new result was recently informally announced,
$d_e < 1.5 \times 10^{-27} \; e$ cm~\cite{kaon2001}.  
It is unlikely 
significant further improvements can be expected due to
stray magnetic fields.  A new approach for the electron is a 
YbF molecule method \cite{EdHinds} that could allow getting down to 
the $10^{-30}$ level within a decade~\cite{Pendlebury:2000an}.  
The best current bound with this method is about an order 
of magnitude weaker than the bound quoted above.

For the neutron EDM, the current bound is \cite{Harris:1999jx}
\begin{equation}
d_n < 6.3 \times 10^{-26} \; e \; \mbox{cm} \; ,
\end{equation}
and the expectation is to strengthen the limit down to $10^{-26}$
by 2004~\cite{Pendlebury:2000an}.  
By 2012 the goal is to reach the $10^{-28} e$ cm level 
using a cryogenic apparatus.

For $^{199}$Hg atom EDM (Schiff moment), the current bound 
is \cite{Romalis:2000mg}
\begin{equation}
d_{Hg} < 2.1 \times 10^{-28} \; e \; \mbox{cm} \; ,
\end{equation}
from a 2001 analysis.  There may be hope to improve this by 
another factor of two, although the systematic errors will need
to be carefully investigated.

\subsection{Supersymmetric effects}

In the most general flavor non-preserving MSSM, there are over $40$ 
new complex phases \cite{Dimopoulos:1995ju}.
New complex phases arise, for example, in the Higgs mixing mass 
$\mu$, as well as in the soft SUSY-breaking terms in the Lagrangian: 
the trilinear scalar mixing masses, the bilinear Higgs mixing parameter, 
and the gaugino mass parameters.  Not all of these phases are physical,
however, and many may be removed by field redefinitions. 

In a wide range of supersymmetric models, these \CP phases are expected
to be ${\cal O}(1)$ \cite{Ibrahim:2001yv}.  Phases of this size 
lead to an EDM of the
electron, neutron, and mercury atom that are significantly larger
that experimental bounds.  There are several possible ways to avoid
these experimental constraints, with varying degrees of simplicity
and naturalness.  One solution is to simply take the phases to be 
small, of order $10^{-3}$ is sufficient \cite{Ellis:1982tk,Dugan:1984qf}.  
Another possibility is to allow arbitrary phases, but push (at least 
some of) the sparticle masses to the multi-TeV region which suppresses 
the supersymmetric 
contribution~\cite{Cohen:1996vb,Bagger:1999ty,Bagger:1999sy,Feng:2000bp}.  
Embedding supersymmetry in a left-right symmetric framework 
can also suppress the phases \cite{Babu:1999xf}.
Finally, it is possible that large phases are permitted due to
cancellations that conspire to render the SUSY contributions to EDMs 
below the experimental limits \cite{Ibrahim:1998je,Brhlik:1998zn}.

The dominant contributions to the lepton EDMs arise from the one-loop
chargino and one-loop neutralino graphs.  For the neutron EDM,
important contributions also arise from one-loop gluino graphs
and two-loop stop-top and sbottom-bottom graphs.  The operators
that contribute are the electric dipole operator
\begin{equation}
- \frac{i}{2} d_f \overline{\psi} \sigma_{\mu\nu} \gamma_5 \psi F^{\mu\nu} \; ,
\end{equation}
the chromoelectric dipole operator
\begin{equation}
- \frac{i}{2} \tilde{d}^C \overline{q} \sigma_{\mu\nu} \gamma_5 t^a q
G^{\mu\nu,a} \; ,
\end{equation}
and the purely gluonic dim-6 operator
\begin{equation}
- \frac{i}{6} \tilde{d}^G f_{\alpha\beta\gamma} G_{\alpha\mu\rho} 
G^\rho_{\beta\nu} G_{\gamma\lambda\sigma} \epsilon^{\mu\nu\lambda\sigma} \; .
\end{equation}
In extracting the effects of the chromoelectric and the purely gluonic
operators, one can use naive dimensional analysis to 
relate \cite{Manohar:1983md}
\begin{equation}
d_q^C = \frac{e}{4\pi} \tilde{d}^C_q \eta^C \; , \;
d_q^G = \frac{e M}{4\pi} \tilde{d}^G \eta^G \; , 
\end{equation}
where $\eta^C \simeq \eta^G \sim 3.4$ and $M = 1.19$ GeV is the 
chiral symmetry breaking scale.  The neutron EDM $d_n$ is estimated
using the $SU(6)$ quark model \cite{Manohar:1983md}
$d_n = (\textfrac{4}{3} d_d - \textfrac{1}{3} d_u)$.
(For an update using QCD sum rules, see \cite{Pospelov:2000bw}.)

The EDM of $^{199}$Hg arises from the T-odd nucleon-nucleon interaction
in supersymmetry \cite{Falk:1999tm}, induced mainly from the color 
operators with light quarks.  This interaction gives rise to an EDM 
of the mercury atom by inducing the Schiff moment of the mercury nucleus.  
The QCD uncertainties related to this calculation are actually 
smaller than for the case of the neutron EDM.  This interaction
can be calculated in terms of the MSSM phases, with the result that
there are somewhat better limits using mercury EDM than the
electron EDM \cite{Falk:1999tm}.  

\begin{figure*}[t]
\includegraphics[width=.8\textwidth]{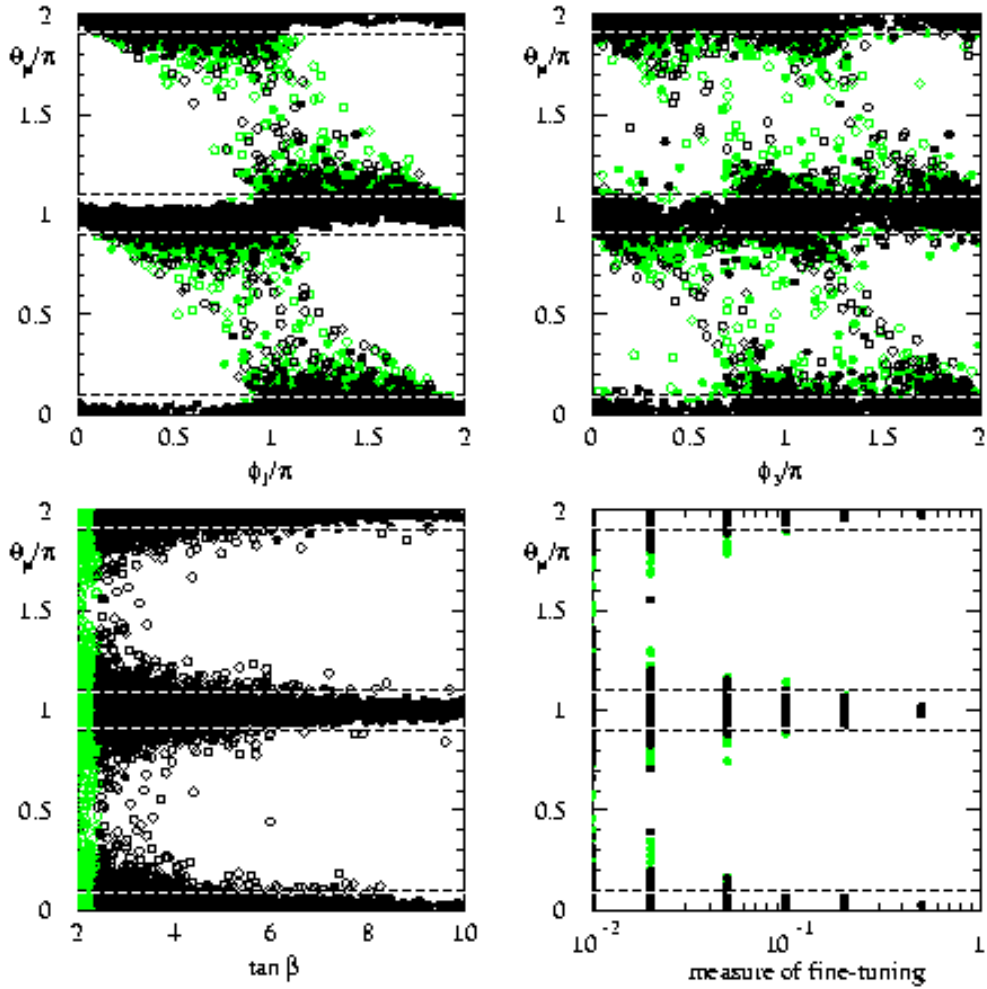}%
\caption{Parameter sets in the 15-parameter MSSM satisfying the
experimental limits on the electron, neutron, and mercury EDMs.
The constraints on $\theta_\mu \equiv {\rm arg}(\mu)$ are the strongest,
in comparison with $\phi_{1,3} \equiv {\rm arg}(M_{1,3})$.
Open circles suffer from larger fine-tuning ($< 0.01$), defined by
the maximum allowed variation of the input parameters such that the
point survives the cuts.  Lightly shaded (open or filled) circles have 
$m_h < 113$ GeV.  (From Ref.~\cite{Barger:2001nu}.)}
\label{fig:barger}
\end{figure*}

In fact, using the combination of the electron, neutron, and 
mercury EDM constraints severely limits the size of phases in the MSSM.
Although this cannot be done in general (due to the large number
of parameters), some information can be gleaned from a more
restricted, 15-parameter MSSM.  If we do not require gaugino
mass unification at the scale where the gauge couplings intersect,
then there are two independent phases in the gaugino masses since one can 
be made real by a $U(1)_R$ rotation.  The 15-parameter MSSM considered by
Ref.~\cite{Barger:2001nu} consists of the phases in $M_1$, $M_3$, 
$A_d$, $A_u$, $A_e$, $A_t$, and $\mu$, as well as the real
quantities $\tan \beta$, the gaugino masses, a common scalar
trilinear coupling, $|\mu|$, and the sfermions masses $m_{\tilde{e}_R},
m_{\tilde{\mu}_R}$.  These parameters were sampled using a 
Monte Carlo scan of nearly $10^9$ sets of parameters, and the remaining
solutions satisfying all collider bounds \emph{plus} the 
three EDM constraints were found.  The results are illustrated in 
Fig.~\ref{fig:barger}, in which the bounds on the phases of
$\mu$, $M_1$, and $M_3$ are shown.  Notice that the constraint
on ${\rm arg}(\mu)$ is strengthened significantly as $\tan\beta$ is 
increased.  Furthermore, while there are lightly populated diagonal bands 
in [${\rm arg}(\mu)$,${\rm arg}(M_{1,2})$] space representing 
cancellations between diagrams, these regions suffer from larger 
fine-tuning and/or are (borderline) excluded by the Higgs mass limit.
Clearly the parameter space is rather strongly constrained when
all experimental constraints are applied and some lower bound on
fine-tuning is imposed.

Finally, it is interesting to note that the operator giving rise to 
the electron EDM is similar to the operator for $g-2$, and this 
similarity can be utilized to find relations between these phenomena.  
In particular, 
assuming the BNL discrepancy is explained by supersymmetry, the phase of 
the electric dipole operator of the electron can be shown to be 
less than $2 \times 10^{-3}$ \cite{Graesser:2001ec}.

\subsection{EDM of the Muon}

The field of precision muon physics will be transformed in the next
few years. The EDM of the muon $\dmu$ is, therefore, of special interest. 
A new BNL experiment~\cite{Semertzidis:1999kv} has
been proposed to measure the muon EDM at the level of
\begin{equation}
\dmu \sim 10^{-24}~\ecm \ ,
\label{proposedEDM}
\end{equation}
more than five orders of magnitude below the current
bound~\cite{Bailey:1979mn}
\begin{equation}
\dmu = (3.7 \pm 3.4) \times 10^{-19}~\ecm \ ,
\label{currentEDM}
\end{equation}
and even higher precision might be attainable at a future 
neutrino factory complex \cite{Aysto:2001zs}.

The interest in the muon EDM is further heightened by the recent
measurement (\ref{currentMDM}) of the muon magnetic-dipole moment
(MDM)~\cite{Brown:2001mg}. 
The EDM and MDM arise from similar operators, 
and this tentative evidence for a non-Standard 
Model contribution to $\amu$ also motivates the search for 
the muon EDM~\cite{Feng:2001sq}.  In fact, the deviation 
of Eq.~(\ref{currentMDM}) may be partially, or even entirely 
attributed to a muon EDM \cite{Feng:2001sq}! 
This is because in modern experiments the muon MDM is 
deduced by measuring (the magnitude of) the muon 
spin precession frequency in a perpendicular and 
uniform magnetic field. However, the spin precession 
frequency receives contributions from both the MDM and the
EDM. For a muon traveling with velocity $\bold{\beta}$
perpendicular to both a magnetic field $\bold{B}$ and an electric
field $\bold{E}$, the anomalous spin precession vector is
\begin{equation}
\bold{\omega}_a = -a_{\mu} \frac{e}{m_{\mu}} \bold{B}
- d_{\mu} \frac{2c}{\hbar} \bold{\beta} \times \bold{B}
- d_{\mu} \frac{2}{\hbar} \bold{E} 
- \frac{e}{m_{\mu}c} \left(\frac{1}{\gamma^2-1} - a_{\mu}\right) 
\bold{\beta} \times \bold{E} \ . 
\label{omega}
\end{equation}
In recent experiments, the last term is removed by
running at the `magic' $\gamma \approx 29.3$, and the third term is
negligible.  For highly relativistic muons with $|\bold{\beta}|
\approx 1$, then, the anomalous precession frequency is found from
\begin{equation}
{|\bold{\omega}_a| \over |\bold{B}|}
\approx \left[ \left( \frac{e}{m_{\mu}} \right)^2
\left(\amusm + \amunp \right)^2 + 
\left(\frac{2c}{\hbar}\right)^2 {\dmunp}^2 \right]^{1/2} \ ,
\label{both}
\end{equation}
where NP (SM) denotes new physics (Standard Model) 
contributions, and assuming $\dmunp \gg \dmusm$.

\begin{figure}[t]
\includegraphics[width=.48\textwidth]{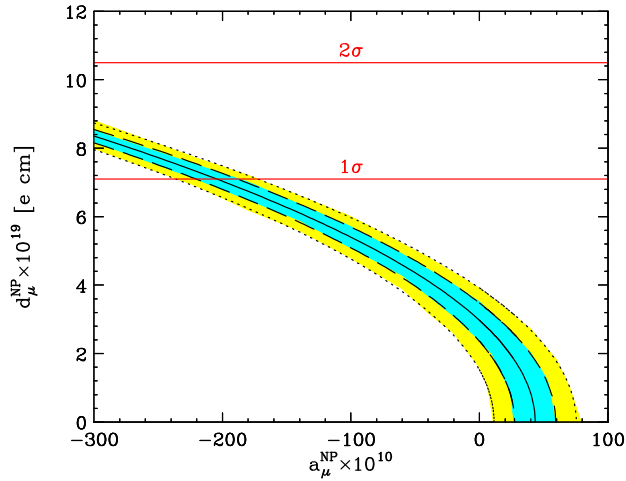}%
\caption[]{Regions in the $(\amunp,\dmunp)$ plane that are consistent
with the observed $|\bold{\omega}_a|$ at the 1$\sigma$ and 2$\sigma$
levels.  The current 1$\sigma$ and 2$\sigma$ bounds on
$\dmunp$~\protect\cite{Bailey:1979mn} are also shown.
(From Ref.~\cite{Feng:2001mq}.)}
\label{fig:amu_dmu}
\end{figure}

We see that the effect (\ref{currentMDM}) can also be
due to a combination of new physics MDM and EDM contributions.
Fig.~\ref{fig:amu_dmu} shows the regions in
the $(\amunp,\dmunp)$ plane that are consistent with the observed
deviation in $|\bold{\omega}_a|$. The current 1$\sigma$ and 2$\sigma$
upper bounds on $\dmunp$~\cite{Bailey:1979mn} are also given.  
It is evident from the figure
that a large fraction of the region allowed by both the current
$\amu$ measurement (\ref{currentMDM}) and the $\dmu$ bound
(\ref{currentEDM}) is already within the sensitivity of phase I of the
newly proposed experiment (with sensitivity $\sim 10^{-22}\ \ecm$).

The proposed dedicated muon EDM experiment will use a different setup,
by applying a constant radial electric field. In that case
a similar EDM $\leftrightarrow$ MDM ambiguity 
is present~\cite{Feng:2001mq}, and can be resolved by
up-down asymmetry measurements.

\begin{figure}[t]
\includegraphics[height=2.3in]{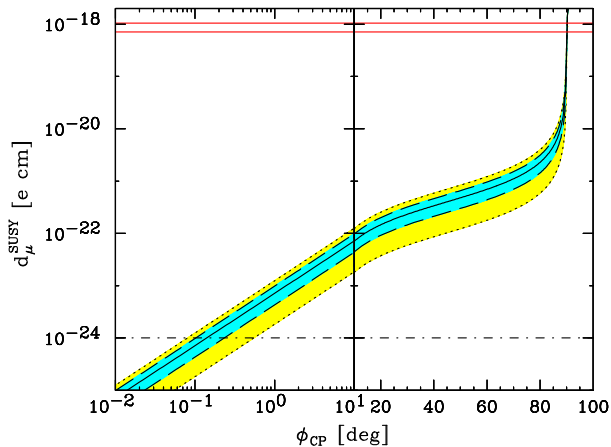}%
\caption{Regions of the $(\phicp, \dmunp)$ plane allowed by the
measured central value of $|\bold{\omega}_a|$ (solid) and its
1$\sigma$ and 2$\sigma$ preferred values (shaded).  The horizontal
dot-dashed line marks the proposed experimental sensitivity to $\dmunp$. 
The red horizontal solid lines denote the current 1$\sigma$ and 2$\sigma$
bounds on $\dmunp$~\protect\cite{Bailey:1979mn}.
(From Ref.~\cite{Feng:2001mq}.)}
\label{fig:dmu_phi}
\end{figure}

It is useful to write the new physics contributions to the
EDM and MDM operators as
\begin{equation}
\dmunp = \frac{e}{2m_\mu}\ \Im A \ , \qquad\qquad
\amunp = \Re A \ ,         \label{ImReA}
\end{equation}
with $A \equiv |A|e^{i\phicp}$.  This defines an experimentally 
measurable quantity $\phicp$ which quantifies the amount of \CP
violation in the new physics, independently of its energy scale.
Upon eliminating $|A|$, one finds
\begin{equation}
\dmunp = 4.0 \times 10^{-22}~\ecm\ \frac{\amunp}{43 \times 10^{-10}} 
\ \tan\phicp \ .
\label{phicp}
\end{equation}
The measured discrepancy in $|\bold{\omega}_a|$ then constrains
$\phicp$ and $\dmunp$. The preferred regions of the 
$(\phicp,\dmunp)$ plane are shown in Fig.~\ref{fig:dmu_phi}.  
For `natural' values of $\phicp \sim 1$,
$\dmunp$ is of order $10^{-22}~\ecm$.  With the proposed $\dmunp$
sensitivity of (\ref{proposedEDM}), all of the 2$\sigma$ allowed
region with $\phicp > 10^{-2} \ {\rm r}$ yields an observable
signal.

At the same time, while this model-independent analysis indicates that
natural values of $\phicp$ prefer $\dmunp$ well within reach of the
proposed muon EDM experiment, very large values of $\dmunp$ also
require highly fine-tuned $\phicp$.  For example, we see from
Fig.~\ref{fig:dmu_phi} that values of $\dmunp\gsim 10^{-20}\ \ecm$ are
possible only if $|\pi/2 - \phicp| \sim 10^{-3}$.  
Furthermore, in specific supersymmetric models it is 
difficult to achieve values of $\dmu$ large enough to 
affect the conventional interpretation of (\ref{currentMDM}).  For
example, in supersymmetry, assuming flavor conservation and taking
extreme values of sparticle masses ($\sim 100~\gev$) and $\tb$
($\tb \sim 50$) to maximize the effect, the largest possible value of
$\amu$ is $a_{\mu}^{\text{max}} \sim 10^{-7}$~\cite{Feng:2001tr}.
Very roughly, one therefore expects a maximal $d_{\mu}$ only of order 
$(e \hbar / 2 m_{\mu} c) a_{\mu}^{\text{max}} \sim 10^{-20}~\ecm$ in
supersymmetry. Similar conclusions hold for specific models as well
\cite{Romanino:2001zf}.

Simplest models relate the EDMs of the electron and muon by 
`naive scaling':
\begin{equation}
\label{naive}
d_\mu \approx {m_\mu\over m_e}\, d_e \ .
\end{equation}
Given the current published bound on the electron EDM,
`naive scaling' limits the muon EDM as
\begin{equation}
d_\mu \alt 9.1\times 10^{-25}~\ecm \ ,
\label{muedmlimit}
\end{equation}
at the 90\% CL, barely below the sensitivity of (\ref{proposedEDM}).
Naive scaling must be violated if a non-vanishing $\dmu$ is to be
observable at the proposed experiment, and this may happen in one 
of three ways:
\begin{itemize}
\item Departure from scalar degeneracy, \emph{i.e.\ }
generation-dependent slepton masses~\cite{Feng:2001sq}.
\item Departure from proportionality, \emph{i.e.\ }
the $A$ terms do not scale with the corresponding fermion mass
\cite{Ibrahim:2001jz}.
\item Flavor violation, \emph{i.e.\ }
non-vanishing flavor off-diagonal elements for the sfermion masses 
and the $A$-terms~\cite{Feng:2001sq,Graesser:2001ec}.
\end{itemize}

Such multitude of possibilities provides sufficient motivation 
and relatively good prospects for a dedicated muon EDM experiment.

\section{Lepton Flavor Violation}
\label{sec:LFV}

One of the most powerful probes of low energy supersymmetry are
the precise measurements and limits on flavor-violating processes.
In the squark sector, for example, the smallness of 
$K_0 \leftrightarrow \overline{K}_0$ mixing either requires 
tiny off-diagonal squark (mass)$^2$ elements, 
or pushes supersymmetry to embarrassingly high scales
$\sim {\cal O}(100\ {\rm TeV})$
\cite{Hagelin:1994tc,Gabbiani:1996hi,Bagger:1997gg}.  
Similarly strong
constraints are also present for certain elements in the slepton 
mass matrix.  Here, we discuss a few of the strongest constraints 
from limits on rare $\mu \ra e$ processes.

\subsection{Experimental status}

There are also several experimental probes that tightly constrain 
lepton-flavor violation.  The strongest constraints arise from the
rare $\mu \ra e$ processes:  $\mu \ra e\gamma$, $\mu \ra 3e$, and 
$\mu \ra e$ conversion.  The current bounds on these processes are
\begin{eqnarray*}
BR(\mu \ra e\gamma) &<& 1.2 \times 10^{-11} \; \mbox{\cite{MEGA}} \\
BR(\mu \ra 3e) &<& 1.0 \times 10^{-12} \; \mbox{\cite{SINDRUM}} \\
BR(\mu \ra e \; {\rm conversion}) &\lsim& 2 \times 10^{-12} \; 
\mbox{\cite{SINDRUMII}} \; ,
\end{eqnarray*}
with the precise bound on $\mu \ra e$ conversion dependent on
the particular host scattering nucleus.

Improvements to these bounds are on the horizon.  At PSI, there
is a proposal to improve the limit on $\mu \ra e\gamma$
down to $10^{-14}$ \cite{PSI-proposal}, which is likely to 
get at least to the level of a few $\times 10^{-13}$, since PSI has a
$\times 20$ advantage over LAMPF in duty cycle and the planned detector 
is probably at least as capable as that of MEGA (the previous experiment).  
This experiment could possibly be pursued further at BNL, since a very 
intense muon beam will be built for another experiment.

There is also a detailed proposal by MECO \cite{MECO-proposal} 
to improve the limit on 
$\mu \ra e$ conversion down to order $10^{-16}$ at the BNL AGS.   
No other facility can compete at the moment, but toward the end 
of the decade the Japanese Joint Project might get involved.
There is apparently a plan to build a dedicated cooled muon beam 
there at an early stage, which they claim would make possible an 
experiment at the $10^{-18}$ level.  Note that one can talk about 
such amazing sensitivity for this process because it has an 
excellent signature that does not require a coincidence and so 
is robust against high rates.

\subsection{Supersymmetric contributions}

Supersymmetric contributions to these processes dominantly arise 
when flavor violating contributions to the slepton mass matrix 
are present.  The sizes of these flavor-violating elements of
the slepton mass matrix are arbitrary in the MSSM.  Particular
models of supersymmetry breaking and mediation can in some cases 
predict the size of slepton-flavor violation.  Some examples include
supersymmetric unified theories (e.g., see \cite{Barbieri:1995tw}),
and also exponentially suppressed contributions from sequestering 
\cite{Randall:1998uk,Kaplan:1999ac,Chacko:1999mi}.

\begin{figure}[t]
\includegraphics[height=2.3in]{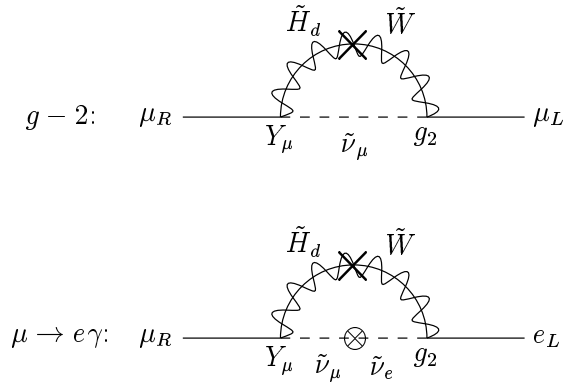}%
\caption{One example set of chargino-sneutrino contributions 
to muon $g-2$ and $\mu \ra e \gamma$ in the interaction 
eigenstate basis with incoming right-handed muons.  
The photon (not shown) is emitted from the chargino.
The chirality flip is shown by the $\times$
on the fermion line, while the the lepton flavor violating mass insertion
is shown by the $\otimes$.  (From Ref.~\cite{Chacko:2001xd}.)}
\label{fig:char-R-fig}
\end{figure}

The contributions to $\mu \ra e\gamma$ are particularly interesting,
due to the strong resemblance between the operators giving rise
to muon $g-2$ and those giving $\mu \ra e\gamma$
\cite{Graesser:2001ec,Chacko:2001xd}.  This correspondence is 
best revealed diagrammatically.  In particular, there is a precise 
correspondence between the diagrams that contribute to 
$\mu \ra e\gamma$ (and $\tau \ra \mu\gamma$) 
with those that contribute to the muon anomalous magnetic moment
\cite{Chacko:2001xd}.  This is illustrated for one class of diagrams
involving charginos and sneutrinos in Fig.~\ref{fig:char-R-fig}.
Other than emitting an electron instead of a muon, the contribution
has the same form except for the essential addition of a sneutrino 
flavor-mixing mass insertion and a propagator for the electron sneutrino.
Using just this process, is it easy to see that the amplitudes for
$g-2$ and $\mu \ra e\gamma$ are related by
\begin{equation}
a_{\mu \ra e\gamma} = \frac{m_{e\mu}^2}{m_{\tilde{\nu}_e}^2} a_\mu
\end{equation}
in the mass-insertion approximation, assuming for this example 
$m_{\tilde{\nu}_e} > m_{\tilde{\nu}_\mu},m_{\tilde{\chi}^\pm}$.  
Here, $m_{e\mu}^2$ is
the off-diagonal element in the sneutrino (mass)$^2$ matrix.
One can systematically go through all classes of one-loop diagrams 
and all possible sparticle-mass hierarchies to find \cite{Chacko:2001xd}
\begin{equation}
BR(\mu \ra e\gamma) \simeq 10^{-4} \,
\left( \frac{a_\mu}{4.3 \times 10^{-9}} \right)^2 \,
\frac{m_{e\mu}^2}{\tilde{m}^2}
\end{equation}
where $\tilde{m} = {\rm Max}[m_{\tilde{\chi}},m_{\tilde{\ell}_e}]$ 
is the largest mass in the loop.
So, if the deviation in the muon $g-2$ measured 
at the BNL experiment is interpreted as supersymmetry, we
obtain strong model-independent bounds on the sfermion mass mixing,
\begin{eqnarray}
m_{e\mu}^2/\tilde{m}^2    &\le& 2 \times 10^{-4} \\
m_{\mu\tau}^2/\tilde{m}^2 &\le& 0.1 \; .
\end{eqnarray}
(The bounds on the ``left-left'' and ``right-right'' 
slepton masses are essentially the same, using the approach of
\cite{Chacko:2001xd}.)
The bound on the second-third generation mass mixing is much
weaker due to the weaker limit on the decay $BR(\tau \ra \mu\gamma) 
< 1.1 \times 10^{-6}$ \cite{CLEO}.  These results are largely
insensitive to any supersymmetric parameters.
Indeed, for some mass ranges the bounds can be as much as an
order of magnitude better than quoted above.  The only assumption
is that there are no accidental cancellations resulting from
summing over diagrams, and even then the expectation is that the
bounds are only mildly relaxed.

Finally, there are exciting prospects for directly observing 
slepton-flavor physics at a linear collider \cite{Dine:2001cf}.
Even with a limited number of slepton states, the observation
(or non-observation) of slepton-flavor violation is expected
to provide important clues to the underlying supersymmetry breaking
structure.  The reason that observable sflavor violation at a LC
is possible is easy to understand.  In many proposals to solve the
SUSY-flavor problem, a high degree of degeneracy among sleptons
(and squarks) is predicted.  As a result, there is the potential for 
substantial mixing of flavor eigenstates.  This can lead to 
substantial and observable sflavor violation.  To be readily
observable, it is necessary that the mass splittings between
the states not be too much smaller than the decay widths, and that
the mixing angles not be terribly small.  In the case when the
off-diagonal mass terms are comparable or larger than the widths,
dramatic collider signatures are possible (see, e.g., 
Ref.~\cite{Krasnikov:1996qq,Arkani-Hamed:1996au,Arkani-Hamed:1997km}).

\section{B Physics}
\label{sec:B}

Rare decays and \CP violating asymmetries provide another interesting 
hunting ground for SUSY-mediated processes. In the $B$ system many rare 
decays involve $ b \rightarrow s (d)$ transitions, which are 
flavor-changing neutral-current (FCNC) processes that are forbidden in
SM at tree level but occur at loop level. The one-loop processes involve 
gluonic, electromagnetic or weak penguin diagrams as well as box diagrams. 
Though suppressed in SM they are relatively large because of the CKM 
structure and the top-quark dominating the loop. However,
SUSY processes may become competitive and
interfere with those in SM. Depending on the
sign of the interference term enhanced or depleted branching fractions are
obtained. Due to the presence of new weak phases SUSY processes
may affect \CP asymmetries as well. While \CP asymmetries of rare 
decays to flavor eigenstates are typically small in SM ($\leq 1\%$)
enhancements up to $20\%$ are possible in SUSY models. For \CP 
asymmetries of $B$ decays to \CP eigenstates, which are quite sizable in SM,
SUSY processes either may enhance or deplete the effect. 
Other interesting penguin modes are $B^0_s$ and $B^0_d$ decays into two 
charged leptons, which are highly suppressed in SM and, therefore, bear a high 
sensitivity for New Physics. To establish a
coherent picture of $B$ decays and uncover SUSY contributions it is important 
to perform several high-precision measurements of rare-decay branching 
fractions and \CP asymmetries.
 
CLEO was the first experiment to observe and study FCNC $B$ decays 
\cite{cleoa}. However, the CLEO data sample is limited to $\rm 9.1\ fb^{-1}$. 
At the asymmetric $B$ factories, which started operation in 1999, BABAR and 
BELLE have recorded already data samples of $60\ \rm \ fb^{-1}$ and 
$\rm 44\ fb^{-1}$, respectively. By summer 2002, BABAR expects 
$100 \ \rm \ fb^{-1}$. This will be increased to $500\ \rm \ fb^{-1}$
by summer 2005.
If luminosity upgrades are successful as planned each experiment should reach 
$1 \ \rm \ ab^{-1}$ by 2010. Experiments at the Tevatron (CDF, D0) will 
augment $B$ samples. However, starting 2006 
high-precision measurements are expected from BTEV at the Tevatron and 
LHCb, ATLAS and CMS at the LHC. 
Furthermore, there are ongoing discussions about a super $B$ factory 
operating with peak luminosities of $10^{36}\ \rm cm^{-2} s^{-1}$,
which would deliver $10 \ \rm \ ab^{-1}$ per year \cite{burchat, ge3}. 
Thus, high-statistics $B$ samples should be available in the future. 
In the following we summarize expectations regarding radiative penguin 
decays \cite{ge3}, \CP violation in $B$ decays \cite{burchat} and
$B$ decays into two charged leptons. Whenever possible, the extrapolations 
are based on present measurements.

\subsection{Radiative Penguin Decays}

Radiative penguin decays involve electroweak penguin loops or box diagrams. 
The largest decay is $B \rightarrow X_s \gamma$, which is
dominated by the magnetic-dipole operator $O_7$. The SM decay rate
contains the squares of the CKM matrix elements $\mid V_{ts} \mid$ and the
Wilson coefficient $C_7$. The latter accounts for all perturbative QCD 
contributions. Due to operator mixing an effective Wilson coefficient 
results. The non-perturbative contributions are absorbed into
the hadronic matrix element of the magnetic dipole operator. Because of
large model uncertainties one avoids the calculation of the hadronic matrix 
element by using the approximation that the ratio of decay rates of 
$b \rightarrow s \gamma$ and $b \rightarrow c e \bar \nu$ at the parton 
level is equal to that at the meson level (quark-hadron duality). 
SUSY processes yield additional contributions $C_7^{new}$ and $C_8^{new}$, 
where the latter arises from SUSY operators that are equivalent to the 
chromomagnetic dipole operator $O_8$.  

The branching fraction in next-to-leading order (NLO) in SM
is predicted to be 
$ {\cal B}(B \rightarrow X_s \gamma) = (3.28 \pm 0.33) \times 10^{-4}$
\cite{misiak1}. Recently, however, Gambino and Misiak argued for a different 
choice of the charm-quark mass, which increases the branching 
fraction to ${\cal B}(B \rightarrow X_s \gamma) =
(3.73 \pm 0.3) \times 10^{-4}$ \cite{misiak2}. The present 
theoretical uncertainty of $\sim 10\%$ is dominated by the mass ratio of 
the $c$-quark and $b$-quark and the choice of the scale parameter $\mu_b$. 
In an updated analysis using the full sample of $9.1 \rm \ fb^{-1}$ CLEO 
has measured $ {\cal B}(B \rightarrow X_s \gamma) = 
(3.21 \pm 0.43_{(stat)} \pm 0.27_{(sys)}$$_{-0.10\ (th)}^{+0.18}) 
\times 10^{-4}$ \cite{cleo1},
where the errors represent statistical, systematic, and theoretical 
uncertainties, respectively. Because of the large errors
this is consistent with the SM NLO prediction. 
Note that the signal region is dominated by continuum ($75\%$) and
$B \bar B$ ($12\%$) backgrounds which have to be subtracted. The present 
relative statistical error is $13.4\%$. Assuming that measurements improve 
with luminosity as $1/\sqrt{{\cal L}}$ the relative statistical error
$\sigma_{{\cal B}}/{\cal B}$ will be reduced to 
$4\% \ (1.3\%)$ for luminosities of $\rm 100 \ fb^{-1} \ (\rm 1 \ ab^{-1})$.
In a super $B$
factory one would expect $\sigma_{{\cal B}}/{\cal B} =0.4\%$ for 
$\rm 10 \ ab^{-1}$. The present relative systematic error is $8.4\%$.
It is expected that with increased statistics the systematic error can be 
reduced substantially by using appropriate data selections and by improving
measurements of the tracking efficiency, photon energy, photon efficiency
and $B$ counting. For $\rm 10\ ab^{-1}$ the hope is to reach 
a systematic error of $1-2\%$. 

\begin{figure}
\includegraphics[width=10cm]{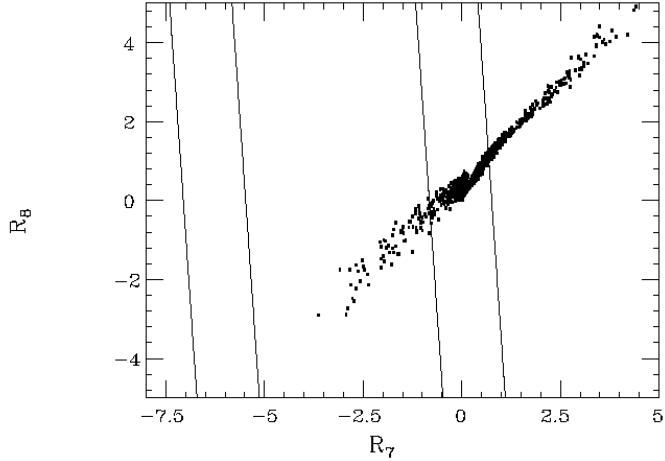}%
\caption{Scatter plot of $R_8$ versus $R_7$ for solutions obtained in the 
SUGRA model. The region allowed by the CLEO measurement lies inside the two 
sets of solid diagonal bands. 
\label{fig:susy}} 
\end{figure}

The CLEO ${\cal B} ( B \rightarrow X_s \gamma)$ measurement already
provides a significant constraint on the SUSY parameter space. 
For example, the new physics contributions to $B \rightarrow X_s \gamma$, 
$C_7^{new}$ and $C_8^{new}$, have been calculated using the minimal 
supergravity model (SUGRA) \cite{hewett1}. Many solutions have been
generated by varying the input parameters within the ranges 
$ 0 < m_0 < 500$~GeV, $50 < m_{1/2} < 250$~GeV, $ -3 < A_0/m_0 < 3$ and 
$ 2 < \tan \beta < 50$, where a common scalar mass $m_0$ for squarks and 
sleptons, a common gaugino mass $m_{1/2}$ and a common trilinear scalar 
coupling $A_0$ is assumed in SUGRA. As usual the ratio
of vacuum expectation values of the neutral components of the two Higgs 
doublets is parameterized by $\tan \beta$. The top-quark mass was kept fixed 
at $ m_t = 175$~GeV. Only solutions were retained that were 
not in violation with  SLC/LEP
constraints and Tevatron direct sparticle production limits. For these
the ratios $R_7= C_7^{new}(M_W)/C^{SM}_7(M_W)$ and 
$R_8= C_8^{new}(M_W)/C^{SM}_8(M_W)$ were determined. The results are depicted 
in Figure~\ref{fig:susy} \cite{hewett2}. The solid bands show the regions 
allowed by the CLEO measurement. It is interesting to note that many 
solutions are already in conflict with the data. However, due to the
theoretical uncertainties it will be difficult to uncover SUSY 
contributions at high luminosities, if the central value remains closely 
to the present result. 

The exclusive decay rate for $B \rightarrow K^* \gamma$ involves the 
hadronic matrix element of the magnetic dipole operator, which in general
is expressed in terms of three $q^2$-dependent form factors $T_i(q^2)$. 
For on-shell photons $T_3$ vanishes and $T_2$ is related to $T_1$. For the
determination of the form factors various techniques are used, introducing
additional theoretical uncertainties.
Recently, two NLO calculations were carried out, predicting SM branching 
fractions of 
${\cal B} (B  \rightarrow K^* \gamma) = (7.1^{+2.5}_{-2.3}) \times 10^{-5}$
\cite{bosch} and 
${\cal B} (B  \rightarrow K^* \gamma) = (7.9^{+3.5}_{-3.0}) \times 10^{-5}$
\cite{beneke}. 
The most precise branching-fraction measurements of the exclusive decays 
$B^0 \rightarrow K^{*0} \gamma$ and $B^+ \rightarrow K^{*+} \gamma$
have been achieved by BABAR. Utilizing kinematic constraints in the $B$ rest
frame provides a substantial reduction of $q \bar q$-continuum 
background here. In a sample of $\rm 20.7 \ fb^{-1}$
BABAR measured ${\cal B} (B^0  \rightarrow K^{*0} \gamma) = (4.39 \pm 
0.41 \pm 0.27) \times 10^{-5}$ in the $K^+ \pi^-$ final state \cite{babar1}.
Due to the large theoretical errors of $35-40\%$ the BABAR measurement
is still consistent with the NLO SM predictions. Note that the combined
statistical and systematic error is already more than a factor of three 
smaller than the theoretical uncertainty. The precision expected for
increased luminosities will be comparable to that in the inclusive mode. 
Thus, it will be difficult to use the exclusive modes for SUSY discoveries, 
unless the theoretical errors are considerably reduced or SUSY effects
are gigantic. In hadron colliders $B \rightarrow K^{*0} \gamma$ is 
also measurable. CDF expects to achieve a $7.6\%$ statistical error per 
$2 \ \rm fb^{-1}$, while BTEV \cite{tev}
and LHCb \cite{lhc} estimate a statistical error of 
$\sigma_{\cal B}/{\cal B} \sim 0.6\%$
per year of LHC running$ \ (\sim 2\ \rm fb^{-1})$.

Other interesting radiative penguin decays are the
$B \rightarrow X_s \ell^+ \ell^-$ modes, where $\ell^\pm$ is either 
an $e^\pm$ or a $\mu^\pm$. In SM, these decays are suppressed by about two
orders of magnitude with respect to corresponding $B \rightarrow X_s \gamma$ 
modes. The suppression by $\alpha$ is compensated partially by additional 
contributions from the $Z^0$-penguin diagram and a box diagram that involves
the semileptonic operators, $O_{9V}$ and $O_{10A}$. Each of them can receive 
additional SUSY contributions.
The branching fractions of the inclusive modes in SM
in NLO are predicted to be ${\cal B}(B \rightarrow X_s e^+ e^-)
= (6.3^{+1.0}_{-0.9})\times 10^{-6}$ and ${\cal B}(B \rightarrow 
X_s \mu^+ \mu^-) = (5.7\pm0.8)\times 10^{-6}$ \cite{ali1, misiak3, buras1}.
The theoretical uncertainties are about $16\%$. These modes have not been 
observed so far. The lowest branching-fraction upper limits $@90\%$ from
CLEO are about an order of magnitude above the SM predictions \cite{cleo4}. 
Using SM predictions and efficiencies determined by CLEO an 
observation of the $X_s e^+ e^-$ and $X_s \mu^+ \mu^-$ modes is expected 
in a sample of $\rm 100 \ fb^{-1}$ with statistical errors around 
$\sim 17\%$ and $\sim 19\%$, respectively. 
This will be improved to $\sim 5.3\% \ (\sim 1.7\%)$ and 
$\sim 6\% \ (\sim 1.9\%)$
in samples of $\rm 1 \ ab^{-1} \ (10 \ ab^{-1})$, respectively. 
Unless the SUSY
contributions lead to significant enhancements the theory errors need to
be reduced at the same time precise measurements are obtained in order
to increase the sensitivity for observing New Physics. 

Branching fractions of the exclusive modes are further suppressed.
Using SM predictions from two recent models and their uncertainties
yield the following ranges of branching fractions:
${\cal B}(B \rightarrow K \ell^+ \ell^-) = (4.7-7.5) \times 10^{-7}$,
${\cal B}(B \rightarrow K^* e^+ e^-) = (1.4-3.0) \times 10^{-6}$, and
${\cal B}(B \rightarrow K^* \mu^+ \mu^-) = (0.9-2.4) \times 10^{-6}$
\cite{qm, lcsm}. 
SUSY processes could enhance these branching fractions considerably.
As an example Figure~\ref{fig:bkll} depicts the dilepton-mass-squared spectrum
for $B \rightarrow K^* \mu^+ \mu^-$ calculated in SM, SUGRA models 
and minimal-insertion-approach SUSY models (MIA) \cite{lcsm}. 
Shown are both the pure penguin contribution and the sum of the penguin process
and the long-distance effects, displaying constructive interference
below the charmonium resonances and destructive interference above.  
The different models are characterized in terms of ratios of Wilson 
coefficients $R_i =1 +C_i^{new}/C_i^{SM}$ for $i=7,9,10$.
The SM prediction is the lowest but bears large uncertainties.  

Except for an unconfirmed signal in $B \rightarrow K^+ \mu^+ \mu^-$ seen
by BELLE, none of the exclusive $B \rightarrow K (K^*) \ell^+ \ell^-$ modes 
have been observed yet. The branching fraction of 
${\cal B}(B  \rightarrow K \mu^+ \mu^-) =(0.99 ^{+0.40+0.13}
_{-0.32-0.14}) \times 10^{-6} $ measured by BELLE \cite{belle3} is
barely consistent with the BABAR limit of 
${\cal B}(B \rightarrow K \ell^+ \ell^-) < 0.6 \times 10^{-6} \ @90\% \ 
CL$ \cite{babar1}. The BABAR limits of 
${\cal B}(B \rightarrow K^{*0} e^+ e^-) < 5.0 \times 10^{-6}$, and
${\cal B}(B \rightarrow K^{*0} \mu^+ \mu^-) < 3.6 \times 10^{-6}$ \cite{babar1}
lie less than a factor of two above the SM predictions.
In a sample of $\rm 100 \ fb^{-1}$ we expect first observation of these modes.
The statistical errors expected at high luminosities for 
$B \rightarrow K^{*0} \ell^+ \ell^-$ are about
a factor of two higher than those for the corresponding inclusive modes.
Experiments at the Tevatron and LHC will be competitive in the
$K^{*0} \mu^+ \mu^-$ and $K^+ \mu^+ \mu^-$ final states \cite{tev}.

The lepton forward-backward asymmetry ${\cal A}_{fb}(s)$ as a function of 
$s = m_{\ell \ell}^2$ is an observable that is very sensitive to SUSY
contributions. It reveals characteristic shapes in the SM both for
inclusive and exclusive final states. To avoid complications from the
charmonium resonances one restricts the range $s$ to masses below the
$J/\psi$, which accounts for $\sim 40\%$ of the entire
spectrum. Figure~\ref{fig:bkll} shows ${\cal A}_{fb}(s)$ 
for the $B \rightarrow K^{*0} \mu^+ \mu^-$ mode \cite{lcsm}.
In SM, the position $s_0$ of ${\cal A}_{fb}(s_0)=0$ is predicted to lie at 
$s_0 = 2.88 ^{+0.44}_{-0.28}\ \rm GeV^2$. Both, the shape and $s_0$
are expected to differ significantly in New Physics models. The shape is
very sensitive to the sign of $R_7$ and varies from model to model.
Thus, a precise measurement of  ${\cal A}_{fb}(q^2)$ may permit an
extraction of the coefficients $R_i$. 
However, to achieve sufficient precision a super $B$ factory is needed.
For Measuring 18 data points below $s = 9 \ \rm GeV^2$ with 
100~events each in the $B \rightarrow X_s \ell^+ \ell^-$
($B \rightarrow K^{*0} \ell^+ \ell^-$) modes at a super $B$ factory
($10 \ \rm ab^{-1}/y$) requires a run period of 0.3-0.4 (0.8-1.3) years. 
For comparison, LHCb expects to achieve the same precision in 
the $B \rightarrow K^{*0} \mu^+ \mu^-$ mode in about one year.

{
\begin{figure}
\includegraphics[width=7 cm]{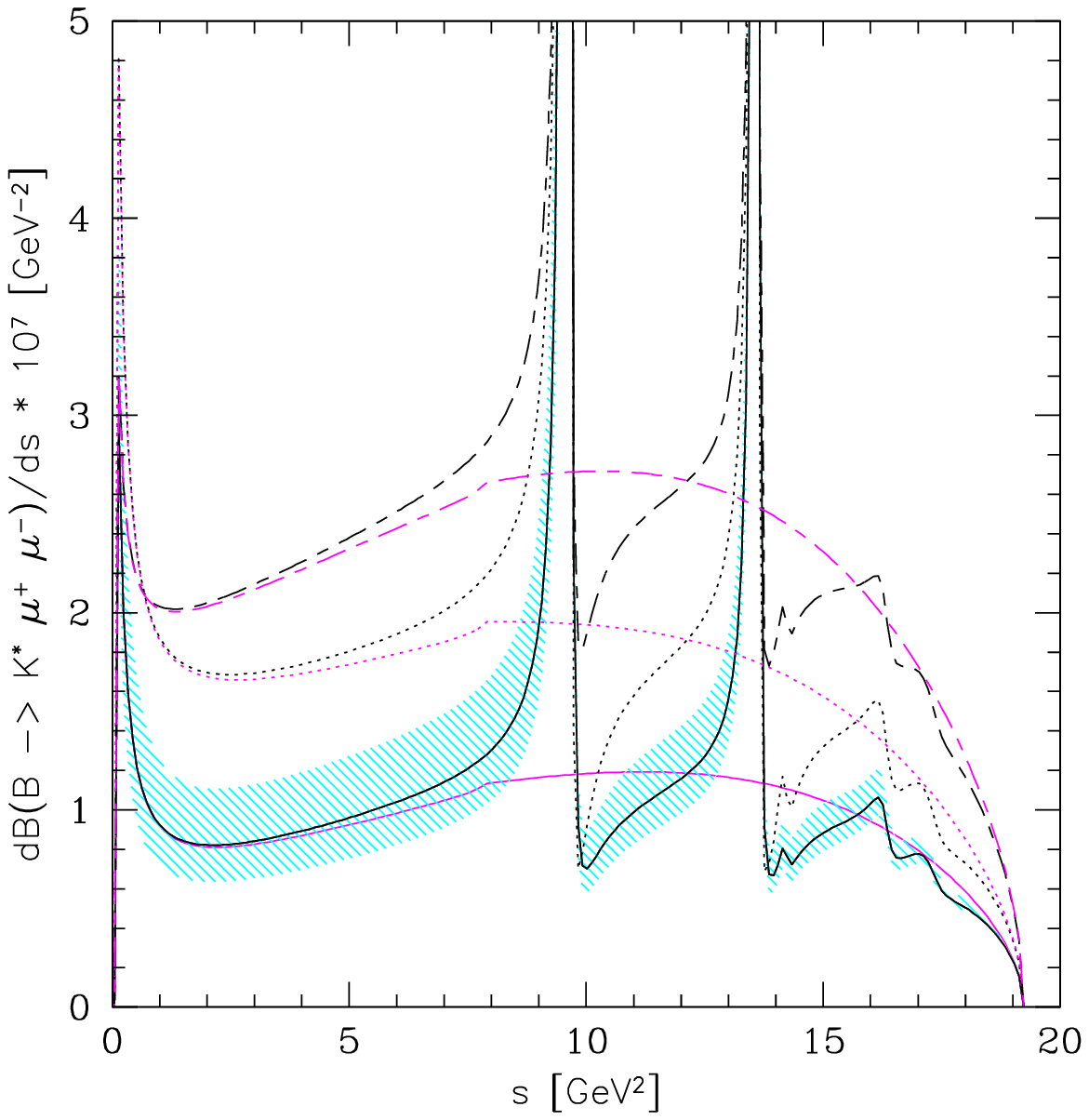}%
\includegraphics[width=7 cm]{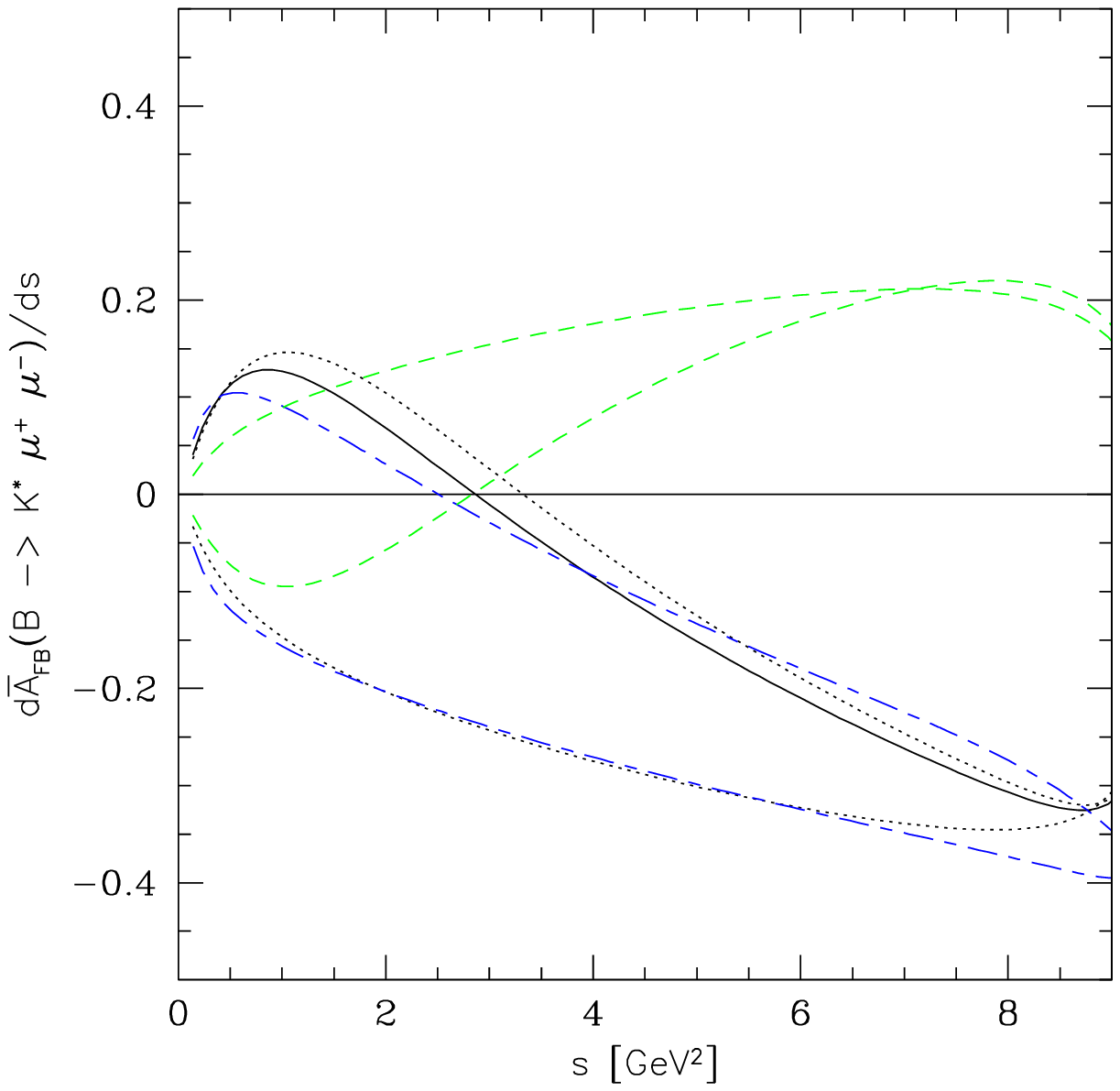}%
\caption{The dilepton invariant mass-squared spectrum (left) and the normalized
forward-backward asymmetry (right) as a function of $s = m^2_{\mu\mu}$ in
$B \rightarrow K^* \mu^+ \mu^-$ \cite{lcsm}. The solid lines denote
the SM prediction. The shaded region depicts form-factor related 
uncertainties. The dotted lines correspond to a SUGRA model ($R_7 =-1.2,
R_9=1.03, R_{10}=1$) and the dash-dotted lines to a MIA
model ($R_7 =-0.83,R_9=0.92, R_{10}=1.61$). In the $m^2_{\mu\mu}$
spectrum both the penguin contribution and the distribution including
long-distance effects are shown. In the ${\cal A}_{fb}$ plot
the upper and lower sets
of curves show the difference between $C^{(0)eff}_7 < 0$ and 
$C^{(0)eff}_7 > 0$, while the dashed curves give results for
another MIA model ($R_7 =\mp 0.83,R_9=0.79, R_{10}=-0.38$).
}
\label{fig:bkll}
\end{figure}
}

The modes bearing the smallest theoretical uncertainties are the
$B \rightarrow X_s \nu \bar \nu$ decays, since long distance effects are
absent and QCD corrections are small. Here, only weak penguin and box diagrams
contribute. However, experimentally both inclusive and exclusive modes 
are difficult to observe due to the two neutrinos. To reduce backgrounds 
from $q \bar q$ continuum and other
$B$ decays at least a partial reconstruction of the other $B$ is necessary.
Since branching fractions are slightly lower than those in the corresponding
$B \rightarrow X_s \mu^+ \mu^-$ channels \cite{gln, gln2, ali3, qm}, 
several years of running at a 
super $B$ factory (at $\rm 10^{36}\ cm^{-2} s^{-1}$) are needed to 
study these modes. For example, in a sample of $\rm 10 \ ab^{-1}$
the relative statistical error expected in the inclusive $B \rightarrow X_s 
\nu \bar \nu$ branching fraction is $\sigma_{\cal B}/{\cal B} 
\sim 5\%$. For the $K^* \nu \bar \nu$ and $K \nu \bar \nu$ final states the 
relative statistical errors are $\sigma_{\cal B}/{\cal B} \sim
10\%$ and  $\sigma_{\cal B}/{\cal B} \sim 13\%$, respectively \cite{ge3}.

\subsection{\CP Violation}

\CP violation can arise from different effects. We focus here on (i) \CP
violation resulting from
the interference between decays with and without mixing, which occurs when 
decay rates of a $B^0$ and $\bar B^0$ into a \CP eigenstate or into 
\CPn -conjugate states differ, and on (ii) \CP violation in decay, resulting
when the magnitudes of the amplitudes for a decay and for the \CPn -conjugate 
decay are different. In SM, the only weak phase besides the  
mixing phase is the phase of the CKM matrix. In SUSY-mediated
processes new phases arise as well as new contributions to $B^0 \bar B^0$ 
mixing. Experimentally, \CP violation is present when an asymmetry between
a process and its \CP conjugate is observed. At the $\Upsilon(4S)$ the two 
$B$ mesons are produced in opposite \CP eigenstates. For such systems 
time-integrated \CP asymmetries of $B^0$ and $\bar B^0$ mesons decaying into 
a \CP eigenstate or into \CPn -conjugate states vanish. To observe an effect, 
one needs to measure the \CP asymmetry as a function of the time difference 
$\Delta t$ between the two $B$ decays. Measurements of such time-dependent 
\CP asymmetries provide a determination of the angles $\alpha$, $\beta$ and 
$\gamma$ in the Unitarity Triangle \cite{BF}. 
For measurements of $\gamma$ also other methods exist. 
The focus here is on measurements accessible in neutral and charged
$B$ decays. 

The time-dependent \CP asymmetry for a $B$ 
decay into a \CP eigenstate, $f_{CP}$, is defined by

\begin{equation}
 {\cal A}_{CP} (\Delta t) = {\Gamma(B^0 \rightarrow f_{CP})(\Delta t)
-  \Gamma(\bar B^0 \rightarrow f_{CP})(\Delta t) \over
\Gamma(B^0 \rightarrow f_{CP})(\Delta t)
+  \Gamma(\bar B^0 \rightarrow f_{CP})(\Delta t)} =
C_f \cos( \Delta m_{B_d} \Delta t) + S_f \sin( \Delta m_{B_d} \Delta t),
\label{eq:cpas}
\end{equation}

\noindent
where

\begin{equation}
C_f = {(1 - \mid \lambda \mid^2) \over (1 + \mid \lambda \mid^2) },
\ \ 
S_f = 2 { {\cal I}m \lambda \over   (1 + \mid \lambda \mid^2) },
\ \ {\rm and} \ \ 
\lambda = \eta_f {q \over p} {\bar A \over A}.
\end{equation}

\noindent
The factor $\eta_f$ indicates the \CP eigenvalue of 
$f_{CP}$, yielding $\eta_f = \pm 1$ for \CPn $(f_{CP}) =\pm 1$, $ q / p$ 
represents the $B^0 \bar B^0$ mixing contribution and 
$\bar A/A$ is the amplitude ratio for
$\bar B^0  \rightarrow f_{CP}$ and $B^0 \rightarrow f_{CP}$, respectively.
\CP is violated if $\mid \lambda \mid \not= 1$ or 
${\cal I} m \lambda \not= 0$. Experimentally, the measurement of 
${\cal A}_{CP}(\Delta t)$ involves three steps:
(i) the reconstruction of \CP eigenstates states such as
$B \rightarrow J/\psi K^0_S$ or $B \rightarrow \pi^+ \pi^-$, 
(ii) the tagging of the $b$ flavor at production point 
using for example the charge of a lepton or a kaon observed in the other
$B$ meson, and (iii) the measurement of the time difference $\Delta t$
of the two $B$-decay vertices. To account for time-resolution effects,
the measured \CP asymmetry is parameterized by the periodic function in 
equation~\ref{eq:cpas} convoluted with a time-resolution function.  
To account for errors in the tagging procedure, the measured 
amplitudes $C_f$ and $S_f$
contain in addition a dilution factor $D=1-2w$, where $w$ represents 
the fraction of mistagged events. 
  
For $B \rightarrow J/\psi K^0_S$ and related modes one expects
$\mid \lambda \mid = 1$ in SM. Thus, the first term in equation 
\ref{eq:cpas} vanishes and \CP asymmetries measure $D \cdot S_f$ with
$S_f = \sin 2 \beta$.
 Using a sample of $\rm 29.7 \ fb^{-1}$ BABAR was the first to observe
\CP violation in the $B^0 \bar B^0$ system. The \CP sample
contained 803 events of which 640 events remained after tagging and vertexing.
By analyzing \CP asymmetries of $B^0 (\bar B^0)$ decays into
$J/\psi K^0_S$, $\psi(2S) K^0_S$, $\chi_c K^0_S$, and $J/\psi K^0_L$, 
\CP eigenstates as well as $J/\psi K^{*0}$ \CP conjugate states, 
BABAR measured
$\sin 2 \beta = 0.59 \pm 0.14_{(stat)} \pm 0.05_{(sys)}$ \cite{bbr}. 
BELLE confirmed the presence of \CP violation measuring 
$\sin 2 \beta = 0.99 \pm 0.14_{(stat)} \pm 0.06_{(sys)}$ \cite{bel}. 
The present world average of $\sin 2 \beta = 0.79 \pm 0.1$ is consistent 
with a value of $\beta$ obtained
from measurements of the three sides of the Unitarity Triangle. However, 
presently the errors are rather large, mainly because of large theoretical 
uncertainties in the extraction of CKM parameters from measurements of 
semileptonic branching fractions, $\epsilon_k$ (parameterizing \CP violation 
in $K^0 \bar K^0$ mixing), and $\Delta m_{B_d}$. 
The measurement of $\alpha$ is complicated by contributions from penguin
processes. Thus, both time-dependent terms in equation \ref{eq:cpas}
are present and ${\cal I} m \lambda = \sin 2 \alpha_{eff}$, where
$\alpha$ and $\alpha_{eff}$ differ by a penguin phase $\delta_P$.
Furthermore, $\alpha$-sensitive modes
are $b \rightarrow u$ transitions that are suppressed with respect to 
$b \rightarrow c$ transitions by 
$\mid V_{ub}/ V_{cb} \mid^2 = {\cal O} (10^{-2})$, thus requiring 
larger $B$ samples than for $b \rightarrow c$ processes. 
So far $\alpha$ has not been measured directly. 
In a sample of  $\rm 30.4 \ fb^{-1}$ 
BABAR has studied the time dependence of 65 $\pi^+ \pi^-$ events, yielding
$S_{\pi\pi} = 0.03^{+0.53}_{-0.56\ (stat)}\pm 0.11_{(sys)}$ and
$C_{\pi\pi} = -0.25^{+0.45}_{-0.47\ (stat)}\pm 0.14_{(sys)}$ \cite{bbr2}.

The measurements of the three angles in the Unitarity Triangle allow various 
tests. First, one checks for consistency between $\beta$ measured directly
in \CP asymmetries and $\beta$ obtained from measurements of the sides
of the Unitarity Triangle. 
Second, one compares $\sin 2 \beta$ measurements obtained in different 
quark processes, such as $b \rightarrow c \bar c s$, 
$b \rightarrow c \bar c d$ and  $b \rightarrow s \bar s s$. In SM, all 
measurements have to give the same result. SUSY contributions, however,  
may affect each quark process differently. Thus, it is necessary to measure 
$\sin 2 \beta$ also in $b \rightarrow c \bar c d$ processes such as
$B \rightarrow D^{(*)+} D^{(*)-}$ or $B \rightarrow J/\psi \pi^0$ and
in $b \rightarrow s \bar s s$ processes such as 
$B \rightarrow \phi K^0_S$ equally well as in $B \rightarrow J/\psi K^0_S$. 
Third, with the additional measurements of $\alpha$
the Unitarity Triangle is two-fold overconstrained. Fourth, \CP asymmetries 
measured in $B \rightarrow J/\psi K^0_S$ and $B \rightarrow \pi^+ \pi^-$, 
combined with measurement of the
$B^0 \bar B^0$ oscillation frequency $\Delta m_{B_d}$ and that 
of $\mid V_{ub} / V_{cb} \mid$ allow a model-independent analysis to
look for New Physics contributions. These can be parameterized in terms of
a scale parameter $r_d=\Delta m_{B_d}^{exp}/\Delta m_{B_d}^{SM}$, 
that accounts for new contributions 
to $B^0 \bar B^0$ mixing, and a new weak phase $\theta_d$ that represents
new sources of \CP violation \cite{SW, nir, bbrpb}.
Finally, with the measurement of $\gamma$ one can test for closure
of the Unitarity Triangle: $ \alpha + \beta + \gamma = \pi$ \cite{bbrpb}. 

In order to carry out these tests and to uncover SUSY contributions 
in \CP asymmetries high-precision measurements are important. 
The present statistical uncertainty of $\sin 2 \beta$ in BABAR is 0.14 for 
$\rm 29.7\ fb^{-1}$. The precision is expected to improve
as $1/\sqrt{{\cal L}}$. Assuming that the reconstruction efficiency and
tagging performance remain unchanged
extrapolations to $\rm 100 \ fb^{-1}$ and 
$\rm 1 \ ab^{-1}$ yield statistical errors of 
$\sigma_{\sin 2 \beta} = 0.076$ and $\sigma_{\sin 2 \beta} = 0.024$, 
respectively. Since BELLE should achieve similar precisions, 
the error in the world average of $\sin 2 \beta$ is reduced
by another factor of $\sqrt{2}$. Contributions
of the systematic error consists of tagging uncertainties, 
vertexing resolution, $B$ life-time precision and background 
treatment. With increased statistics, 
the individual systematic errors will be reduced, so that the total 
systematic uncertainty should remain of the order of the statistical error. 
For an annual luminosity of $\rm 10 \ ab^{-1}$ at a super $B$ factory  
the statistical error will be reduced to $\sigma_{\sin 2 \beta} = 0.0076$. 
This should be compared with expectations at the LHC. Based on Monte Carlo
simulations the statistical precision of $\sin 2 \beta$ 
measured in \CP asymmetries of $B \rightarrow J/\psi K^0_s$ is 
estimated to be 
$\sigma_{\sin 2 \beta}(\psi K_s) = 0.017$ for ATLAS,
$\sigma_{\sin 2 \beta}(\psi K_s) = 0.015$ for CMS and 
$\sigma_{\sin 2 \beta}(\psi K_s) = 0.021$ for LHCb
for one year of LHC running \cite{lhc}. 
In a sample of $2 fb^{-1}$ at the Tevatron \cite{tev}
precisions expected for $\sin 2 \beta$ are 
$\sigma_{\sin 2 \beta}(\psi K_s) = 0.05$ for CDF,
$\sigma_{\sin 2 \beta}(\psi K_s) = 0.04$ for D0, and
$\sigma_{\sin 2 \beta}(\psi K_s) = 0.025$ for BTEV. In the next 5-10 years, 
when the high luminosities will be achieved, the measurements of the sides 
of the Unitarity Triangle will also improve. Particularly, the theoretical
uncertainties associated with the side measurements are expected to improve
by a factor of 2-4 \cite{burchat}.

For a comparison of $\sin 2 \beta$ measurements in different quark
processes precisions expected for $ b \rightarrow s \bar s s$ and 
$b \rightarrow c \bar c d$ processes are not sufficient  
in present asymmetric $B$ factories. In a sample of 
$\rm 22 \ fb^{-1}$ BABAR observes 11 $B \rightarrow \phi K^0_S$ events
\cite{bbr3, cleob}.  
Using the results of the BABAR $\sin 2 \beta$ measurement and the observed
$\phi K^0_S$ yield gives statistical-error estimates of
$\sigma_{\sin 2 \beta}(\phi K_s) = 0.56$ for $\rm 100 \ fb^{-1}$,
$\sigma_{\sin 2 \beta} (\phi K_s) = 0.18$ for  $\rm 1 \ ab^{-1}$,
and $\sigma_{\sin 2 \beta} (\phi K_s) = 0.056$ for  $\rm 10 \ ab^{-1}$. 
At these levels of precision systematic errors are less important.
In the modes $B \rightarrow J/\psi \pi^0$ BABAR observes 13 events in a sample
of $\rm 23 \ fb^{-1}$ \cite{burchat}.
Following the same extrapolation procedure as
for $\phi K^0_S$ the precision expected for
$\sin 2 \beta $ measurements in $J/\psi \pi^0$
is  $\sigma_{\sin 2 \beta} (J/\psi \pi^0) = 0.52$ in  
$\rm 100 \ fb^{-1}$, $\sigma_{\sin 2 \beta} (J/\psi \pi^0) = 0.16$ in  
$\rm 1 \ ab^{-1}$ and
$\sigma_{\sin 2 \beta} (J/\psi \pi^0) = 0.052$ in  $\rm 10 \ ab^{-1}$. 

Measurements of $\sin 2 \alpha$ also require high statistics, since branching 
fractions for $ b \rightarrow u$ processes are small and are affected by 
competing penguin amplitudes. Extrapolating the present BABAR 
$B \rightarrow \pi^+ \pi^-$ results to high luminosities yields 
$\sigma_{C_{\pi \pi}} = 0.26,\ \sigma_{S_{\pi \pi}} = 0.32$ for 
$\rm 100\ fb^{-1}$,
$\sigma_{C_{\pi \pi}} = 0.09,\ \sigma_{S_{\pi \pi}} = 0.1$ for 
$\rm 1\ ab^{-1}$, and
$\sigma_{C_{\pi \pi}} = 0.026,\ \sigma_{S_{\pi \pi}} = 0.032$ for 
$\rm 10\ ab^{-1}$. To extract $\alpha$ from $S_{\pi\pi}$ one needs to measure
the penguin phase $\delta_P$. This is achieved by exploiting isospin
relations among the amplitudes of $B \rightarrow \pi \pi$ and 
$\bar B \rightarrow \pi \pi$ decays \cite{gronau}. 
In the absence of electroweak-penguin
amplitudes the isospin relations form two triangles (one for $B$ and one for
$\bar B$ decays) with a common amplitude base
$A(B^+ \rightarrow \pi^+ \pi^0) = A(B^- \rightarrow \pi^- \pi^0)$. The
angle between the two triangles is $2 \delta_{P}$. The presence of 
electroweak-penguin amplitudes introduces a small correction. The real
challenge in this analysis, however, is the measurement of 
${\cal B} (B^0  \rightarrow \pi^0 \pi^0)$ and 
${\cal B} (\bar B^0  \rightarrow \pi^0 \pi^0)$, since branching fractions
are rather small and the $\pi^0 \pi^0$ final state is affected by a 
large $q \bar q$-continuum background. The determination of $\delta_P$ as 
a function of ${\cal B} (B^0  \rightarrow \pi^0 \pi^0)$ is depicted in 
Figure~\ref{fig:dp} for three different luminosities. In a $\rm 10 \ ab^{-1}$
sample the uncertainty expected for $\delta_P$ is $\sigma_{\delta_P} = 5^o$,
which is about a factor of three larger than the uncertainty expected for 
$\alpha_{eff}$ from $S_{\pi \pi}$. For $\rm 2 \ fb^{-1}$ BTEV estimates
a statistical error of $\sqrt{\sigma_{C_{\pi\pi}}^2 + \sigma_{S_{\pi\pi}}^2}
= 0.024$ \cite{tev}. For comparison, LHCb quotes errors of 
$\sigma_{S_{\pi\pi}}= 0.07$ and  $\sigma_{C_{\pi\pi}}= 0.09$ per year of
LHC running. Using a Dalitz plot analysis in the $B \rightarrow \rho \pi$ 
channel is another promising method to extract $\alpha$ with high precision 
both at a super $B$ factory and in future experiments at hadron machines.   
For example, LHCb quotes an expected statistical error
of $\sigma_\alpha (\rho \pi) = 3-5^o$ per year of LHC running \cite{lhc}.

{
\begin{figure}
\includegraphics[width=7 cm]{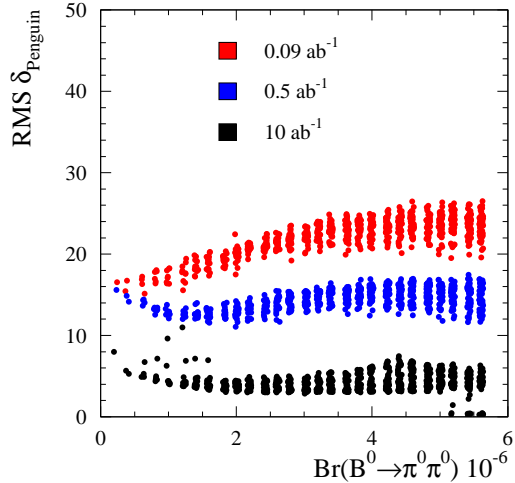}%
\caption{Uncertainty in extracting the angle 
$\delta_P = \alpha_{eff} - \alpha$ in degrees versus the 
$B^0 \rightarrow \pi^0 \pi^0$ branching fraction for samples of
$\rm 90 \ fb^{-1}$, $\rm 500 \ fb^{-1}$, and $\rm 10 \ ab^{-1}$.
}
\label{fig:dp}
\end{figure}
}

One promising method of measuring $\gamma$ at the $\Upsilon(4S)$
is based on studying 
$B \rightarrow D^{(*)} K^{(*)}$ decays \cite{soni}. 
Here the different CKM structure in $b \rightarrow c \bar u s$ and
$b \rightarrow u \bar c s$ processes is exploited, since the former decay has
no weak phase while the latter involves $\gamma$. For the extraction of
$\gamma$ at least two $D^0$ flavor eigenstates, ($K^- \pi^+$,
$K^- \pi^+ \pi^0$, or $K^- \pi^+\pi^+\pi^-$) need to be reconstructed, 
since besides $\gamma$ the $B^- \rightarrow \bar D^{(*)0} K^{(*)-}$ 
branching fraction and two strong phases need to be measured. Note, that this 
procedure leads to an eightfold ambiguity in the value of $\gamma$.
To gain sufficient statistics actually all possible $DK$, $D^*K$,   
$DK^*$ and $D^*K^*$ modes have to be combined. In a sample of
$\rm 600 \ fb^{-1}$
the statistical uncertainty in $\gamma$ is estimated to range around
$\sigma_\gamma \sim 5-10^o$ depending on the exact value of $\gamma$.
Thus, for a range of two-three standard deviations each solution covers
most of the $\gamma$ region. In a sample of $\rm 10 \ ab^{-1}$, however,
this is different, as the statistical uncertainty is expected to be reduced
to $\sigma_\gamma \sim 1-2.5^o$.
So, again high-statistics $B$ samples are required.
In $\rm 2 \ fb^{-1}$ BTEV expects to measure $\gamma$ with an error 
of about $7^o$, using $B^0_s \rightarrow D^-_s K^+$ decays \cite{tev}. 

Another method of measuring $\gamma$ in $B$ decays involves 
\CP asymmetries in $D^{(*)} \pi$ or  $D^{(*)} \rho$ modes \cite{sachs, 
aleksan, dunietz}. 
Here, the combination of CKM angles $2 \beta + \gamma$ is measured, which
results from an interference between the decay $b \rightarrow c \bar u d$  
and the $B^0 \bar B^0$-mixed process $\bar b \rightarrow \bar u c \bar d$,
where the latter depends on the angle $\gamma$. BABAR recently studied 
the $D^{*+} \pi^-$ mode \cite{burchat}, 
which has a large branching fraction and can 
be partially reconstructed with low backgrounds but yields a small 
\CP asymmetry. The main disadvantage of this technique is the necessity of 
measuring the ratio $r$ of the doubly CKM-suppressed decay 
$\bar b \rightarrow \bar u c \bar d$ to the allowed decay 
$b \rightarrow c \bar u d$. Extrapolating the BABAR study yields 
statistical-error estimates 
of $\sigma_{\sin (2 \beta + \gamma)}(D^* \pi) \sim 0.3$
for $\rm 100 \ fb^{-1}$, $\sigma_{\sin (2 \beta + \gamma)}(D^* \pi) \sim 0.1$
for $\rm 1 \ ab^{-1}$, and 
$\sigma_{\sin (2 \beta + \gamma)}(D^* \pi) \sim 0.03$ for $\rm 10 \ ab^{-1}$.
The probability that $r$ is not measured due to statistical fluctuations
is  $30\%, \ 10\%$ and $3\%$, respectively. For comparison, LHCb expects 
to achieve a statistical error of $\sigma_{sin (2 \beta + \gamma)} = 0.26$ 
per year of LHC running.

\CP asymmetries of rare decays into flavor eigenstates are also rather 
suited to search for SUSY-mediated processes. For example in $B \rightarrow
X_s \gamma$ decays \CP asymmetries are expected to be small in 
SM $(\leq 1\%)$ \cite{soares} 
but they may be enhanced up to $20\%$ in SUSY models \cite{kagan}. So far all
observed \CP asymmetries are consistent with zero. In the inclusive 
$B \rightarrow X_s \gamma$ mode CLEO measured 
${\cal A}_{CP}(B \rightarrow X_s \gamma) = 
(-0.079 \pm 0.108 \pm 0.022) \times (1.0
\pm 0.03)$ \cite{cleo3}, where the first error is statistical while 
the second and third errors represent additive and multiplicative systematic 
uncertainties, respectively. Though the \CP asymmetry has been corrected 
for contributions from $B \rightarrow X_d$, a separation of these events from
$B \rightarrow X_s \gamma$ events may become necessary on an event-by-event 
basis to rule out that the \CP asymmetry in 
$B \rightarrow X_s \gamma$ is canceled by that in $B \rightarrow X_d \gamma$.
In the exclusive $B^0 \rightarrow K^{*0} \gamma$ 
modes BABAR observed a \CP asymmetry of 
${\cal A}_{CP}(B \rightarrow K^{*0} \gamma) = -0.035 \pm 0.076 \pm 0.012$ 
using the three $K^*$ final states, $ K^+ \pi^-$, $K^+ \pi^0$ and 
$K_S^0 \pi^+$ \cite{babar1}. Extrapolating the measured statistical 
error to high luminosities yields $\sigma_{{\cal A}_{CP}} = 3.3\% \ (3.5\%)$
for $\rm 100 \ fb^{-1}$,  $\sigma_{{\cal A}_{CP}} = 1.0\% \ (1.1\%)$
for $\rm 1 \ ab^{-1}$ and  $\sigma_{{\cal A}_{CP}} = 0.33\% \ (0.35\%)$
for $\rm 10 \ ab^{-1}$ in $B \rightarrow X_s \gamma$ ($B \rightarrow K^* 
\gamma$) modes. LHCb expects to measure 
${\cal A}_{CP}(B \rightarrow K^* \gamma)$ with a precision of $1\%$ per
year of running.

\subsection{$B$ Decays into Two Charged Leptons}

Another interesting probe of supersymmetry is
the decay \bs . The SM prediction
is given by $\cbs = (3.7 \pm 1.2) \times 10^ {-9}$, with the
uncertainty ($\pm 25\%$) dominated by the decay constant $f_{B_s}$.
The current experimental bound on the branching ratio, (${\cal B}$),
has been set during  Run-I of the Tevatron, where CDF~\cite{CDFbmumu}
determined $\cbs < 2.6 \times 10^{-6}$ at 90\% C.L. 
In addition to the experimental challenge, the almost three 
orders of magnitude gap between the current experimental bound and
the SM prediction makes this mode an excellent laboratory for
new physics. In contrast to observables which enter the unitarity 
triangle~\cite{Bartl:2001wc}, in the MSSM, the branching 
ratio $\cbs$ grows like $\tan^6\beta$
\cite{Babu:1999hn,Huang:1998vb,Chankowski,Urban,Isidori}, 
with a possible several orders of
magnitude enhancement. More interestingly, it has been very recently 
shown~\cite{Dedes:2001fv} that in the mSUGRA
scenario there is a strong correlation with the muon
 anomalous magnetic moment $(g-2)_\mu$.
  An interpretation of the
  recently measured excess in $(g-2)_\mu$ in terms of mSUGRA 
  corrections implies a substantial supersymmetric enhancement of the
  branching ratio $\cbs$: if $(g-2)_\mu$ exceeds the Standard Model
  prediction by $(\delta a_\mu)_{SUSY}=4\times 10^{-9}$, 
  $\cbs$ is larger by a factor of { 10--100}
  than in the Standard Model and within reach of Run-II of the
  Tevatron. The single event sensitivity 
of CDF at Run-IIa is estimated to be $1.0
\times 10^{-8}$, for an integrated luminosity of 2 fb$^{-1}$
\cite{fnalbrep}. Thus if mSUGRA  corrections enhance $\cbs$ to
e.g.\ $5\times 10^{-7}$, one will see 50 events in Run-IIa. Run-IIb may
collect 10-20 fb$^{-1}$ of integrated luminosity, which implies 
250-500 events in this example.

{
\begin{figure}[tb]
\includegraphics[width=7 cm, angle=90]{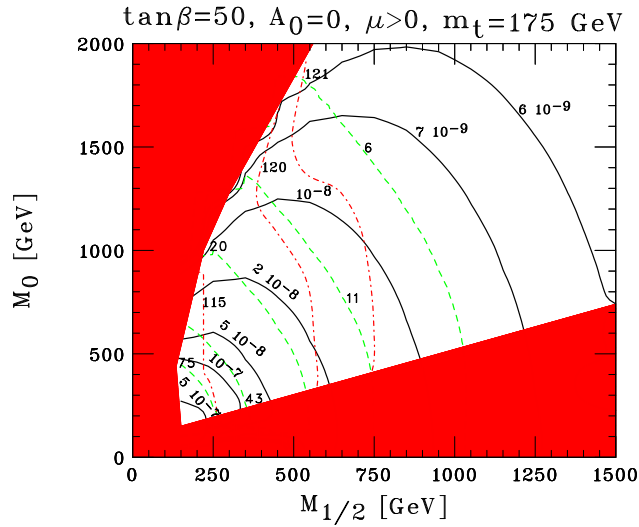}%
\caption{Contour plots  of the $\cbs$ (solid) and on $(\delta a_
  \mu)_{SUSY}$ (dashed)  in the ($M_{0},M_{1/2}$)-plane for
  mSUGRA parameter values  as shown. Contours of the light Higgs boson mass
  (dot-dash line) are also shown. From Ref.~\cite{Dedes:2001fv}. }
\label{fig1}
\end{figure}
}

Following Ref.~\cite{Dedes:2001fv}, 
in Fig.\ref{fig1} we show the contours of $\cbs$ (solid) and
$(\delta a_\mu)_{SUSY}$ (dashed) in this plane, for $\tan\beta=50$,
$A_0=0$, $\mu>0$ and $m_t=175 \, {\rm GeV}$.   The  shaded
region is excluded through the various theoretical and experimental
constraints.  A sensitivity of $\cbs\sim
2\times10^{-7}$ at CDF now corresponds to a sensitivity of
$M_{1/2}\sim 280 \, {\rm GeV}$ and $M_0\sim 400 { \,{\rm GeV}}$, 
respectively.
While CDF is not able to see squark masses directly up to 0.7 TeV
(corresponding to $M_{1/2}=M_0\simeq 300$ GeV), it will nevertheless
be able to prepare the ground for LHC by observing the \bs\ mode.
Even better, after 10$\,{\rm fb}^{-1}$ CDF will probe $M_{1/2} \sim
450$ GeV and $M_0 \sim 600\,{\rm GeV}$ (for $\tan \beta=50$) which in
mSUGRA corresponds to masses for the heaviest superpartners of 1 TeV.
We conclude the discussion of Fig.\ref{fig1} with the prediction of
the light Higgs boson mass $M_h$ (dot-dashed line) for $\tan\beta=50$
in the mSUGRA scenario~\cite{ASBS}.  Any measurement of $\cbs$ by
itself implies a useful \emph{upper}\ bound on $M_h$. The simultaneous
information of $\cbs$ and $\delta a_{ \mu}$ fixes $M_h$ in most
regions of the $(M_{1/2},M_0)$-plane.  A Higgs mass around 115.6 GeV
results in $ 10^{-8} \sim \cbs \sim 3\times 10^{-7}$ which would most
likely be measured before the Higgs boson is discovered.

\section{Direct Dark Matter Searches}
\label{sec:DMdirect}

Neutralinos in supersymmetry are well-motivated candidates
to provide much or all of the non-baryonic dark matter.
In many models, the neutralino is the LSP and is effectively stable.
In general, the lightest neutralino is a mixture of
the superpartners of Higgs and electroweak gauge bosons.
Remarkably, detailed calculations of the thermal relic density of
neutralinos have shown that neutralinos may indeed account
for most of the missing mass of the universe.

Typically, there can be several different cosmologically preferred
regions in parameter space. For example, in minimal supergravity,
there are four: a `bulk' region at relatively low $m_0$ and $m_{1/2}$,
a `focus-point' region~\cite{Feng:1999mn,Feng:1999zg,Feng:2000gh}
at relatively large $m_0$, a co-annihilation `tail'
extending out to relatively
large $m_{1/2}$~\cite{Ellis:1998kh,Ellis:1999mm,Gomez:1999dk},
and a possible `funnel' between the focus-point and
co-annihilation regions due to rapid annihilation via
direct-channel Higgs boson poles~\cite{Ellis:2001ms,Lahanas:2001yr}.
A set of benchmark supersymmetric model parameter choices was
proposed~\cite{Battaglia:2001zp} just prior to Snowmass 2001,
with points representing each of those regions.
In this and the next section, we shall use this set of benchmarks
to illustrate our discussion of the various signals
of supersymmetric dark matter~\cite{Ellis:2001hv}.

Neutralinos are very weakly interacting and
they pass through collider detectors without leaving a trace.
Therefore, it is practically impossible to observe
them in collider experiments directly. Existing bounds on
neutralinos must rely on model-dependent correlations between
their properties and those of other supersymmetric particles.
However, if neutralinos make up a significant portion of the
halo dark matter, many additional avenues for their detection
open up.  They may deposit energy as they scatter off nuclei in
terrestrial, usually sub-terrestrial, detectors.

Searches for this signal in low background detectors have
been underway for around 20 years.
The first decade was dominated by conventional HPGe (and Si) 
semiconductor detectors.
The design of these detectors was to
some extent ``off-the-shelf'' and progress was achieved, for the most part,
by improving the radioactive
backgrounds around the detectors.
In the mid-90's results from NaI scintillator detectors became competitive.
They were able to employ pulse shape discrimination to make statistical
distinctions between populations of
electron recoil events, and nuclear recoil events.
In principle, the intrinsic background of the detector and
environment were no longer the limiting factors, since with sufficient exposure
time and target mass the
limits could be driven down. However, the relatively poor quality of the NaI
discrimination meant that
systematic effects rapidly dominate, halting any further improvement
with exposure. The NaI detector
technology could also be described as off-the-shelf, however,
the low background and high light yield
housing systems were very definitely novel.
At the end of 90's we finally
saw results, from new cryogenic ($<$1 K) detector technology that 
had been developed
specifically for direct detection, take the
lead in terms of sensitivity.

Current sensitivities are now at a
WIMP-nucleon normalized cross-section of $4\times 10^{-6}$~pb, which for
a Ge target with a detection threshold of 10 keVr is an integrated event rate of
1.0 ~kg$^{-1}$~d$^{-1}$. In this discussion keVr will be used to
indicate actual recoil energy of an event, whereas keVee (electron equivalent), 
will be used to indicate the visible energy of a nuclear recoil 
event based on a energy scale
for electron recoil events.  For example, in NaI scintillator,
an iodine nuclear recoil of 22 keVr, generates the same light output as a 2
keVee electron event. To first approximation the differential recoil spectrum for
WIMP nucleon events has a characteristic,  but far from unique, exponential decay
shape. This text is not the right place to discuss the details of  what
determines the spectrum, or the details of the cross-section normalization.  See
instead~\cite{Lewin:1996, Jungman:1996df}.

We should touch on the DAMA experiment, which has reported a $4\sigma$
observation of annual modulation in the count rate of events in their lowest
 energy bins (2-6 keVee, which corresponds to 22-66 keVr for nuclear
recoil events scattering on iodine) over a time spanning
4 years \cite{Bernabei:1998, Bernabei:2001}.
The modulation has a phase (maximum in June) and
a period (1 year) which would be consistent with the
fluctuations in the observation arising from a contribution from
a WIMP recoil spectrum which will be modulated by
changes in the Earth's velocity in
the galaxy. (Baseline model assumes an isothermal, non-co-rotating WIMP
population.)  It is enormously important that this signal be studied and the
source  clearly determined. The DAMA experiment will check its own results  with
two new years of data already available for analysis
\cite{Belli:2001},  and further data from running for the seventh year just
beginning.  It could be argued that a substantial overhaul in the data taking
strategy  should occur before further data taking, since the significance of the
modulation effect is really  systematics, rather than statistics limited, at this
stage. The importance  of a ``beam-off'' component of the data taking cannot be
emphasized  enough to establish the long term
stability  of data in the lowest energy bins
when looking for few \% fluctuation.
While one cannot turn the WIMP wind off to order,
a revised acquisition strategy that
included direct calibration of the stability of the acceptance of the
2-3, 3-4, 4-5 keVee bins (in which
all of the WIMP signal is expected) would
help demonstrate that the modulation signal is not an artifact.
Multiple scattering events should be stored, rather than
vetoed since they could provide a non-WIMP calibration in real time.
Additionally, on-line light pulses,
or daily gamma calibrations, could be used to demonstrate the
stability of the count
rate in the bins if even higher statistics are required.
The DAMA Collaboration is now constructing LIBRA,
which will be a larger 250~kg array which by virtue of employing
lower background NaI should be able to make a more sensitive check
of the annual modulation signal. Using a different experimental setup
will also help address the systematics question, although its operation
at the same location, by the same collaboration doesn't allow all
possible systematics to be checked. To this end the Boulby DM
Collaboration may be able to carry through on the intention of running
around 50 kg of NaI to cross-check the results directly with
the same target material. Results from this Boulby program are
unlikely to be available before 2004, however.

The currently reported results from the CDMS~I \cite{Abusaidi:2000}
and EDELWEISS \cite{Benoit:2001} experiments can be
viewed as being inconsistent with the size of the annual modulation
signal seen in DAMA. This is most readily seen by converting the
published modulation amplitude of DAMA, into a cross-section on
nucleon ($1.4\times10^{-5}$~pb), assuming scalar WIMP nucleon
couplings, and then for comparison with CDMS~I, 
to an estimated number of WIMP events (40)
that would be observed in the 10.6~kg-live days exposure of Ge.
This experiment saw 13 single scatter nuclear recoil
events in Ge during this period which is clearly inconsistent with 40.
In fact, the observation of an additional 4 multiple scattering
nuclear recoil events during the same run, can be attributed 
unambiguously to neutrons.
This permits a modest reduction of the (90\% CL) upper limit 
in CDMS~I single
scatter nuclear recoil data to 8 WIMP events.  
On this basis, CDMS~I and DAMA are
inconsistent at 99.98\%. The Edelweiss experiment has a
similar WIMP sensitivity using detectors that are like the ones
used in CDMS~I.
The data reported to date is for a smaller exposure
of only 3 kg-days, however, it appears free of neutrons above a
25 keV analysis threshold due to the much deeper site location.
Further data from longer exposure is eagerly awaited.

A number of avenues exist for reducing this
incompatibility between the current results of the nuclear recoil
discriminating Ge detectors and DAMA NaI annual modulation.
Firstly, the DAMA collaboration reports an allowed region that
combines not just the positive signal from the annual modulation amplitude,
but also negative results/upper limits from other NaI derived data.
This has the effect of reducing the headline expectation for the WIMP
cross section  by approximately a factor 3. This makes the disagreement between
the experiments not statistically significant, but does beg the question whether
the various DAMA experimental results themselves are mutually compatible.
Alternative routes for explaining the apparently contradictory results have
looked at the nature of WIMPs themselves.
Comparison of Na, I and Ge recoil rates requires the
assumption of a model. As has been said before the
primary analysis of the experimental data is developed
assuming scalar, or spin independent interactions.
The interaction rates of the spin independent process often dominates
because of the enhancement by the coherent scattering across multiple
nucleons. However, in the absence of spin independent couplings, the
relative spin dependent
couplings will be stronger for an NaI target, in which
both elements are monoisotopic with odd proton spin, compared to Ge,
which has only 1 odd neutron spin isotope, with an 8\% abundance in the natural
element. Existing indirect  WIMP detection experiments, are able to (partially)
test the spin dependent hypothesis, ruling out some, but not all possible
solutions (see next section).  Another entertaining model is to suggest that the
WIMPs themselves  inelastically scatter from nuclei (i.e. WIMPs require internal
transition to scatter,  between low energy non-degenerate states.) Tuning of the
excitation energy can  ensure that a heavier nuclei (such as I) may be able to
scatter  WIMPs at a significantly enhanced rate relative to Ge
~\cite{Smith:2001hy}.

If we now look forward at some of the predicted goals of a few
experiments over the next decade, it is
immediately apparent that the forecast rate of progress appears to be rapidly
accelerating when compared to historical progress. The question is
whether this is simply a ``triumph of hope over expectation'',
or represents a genuinely improved rate of
progress that stems from applying detector technologies (2-phase Xe,
semiconductor and scintillator
cryogenic detectors, naked HPGe) that were ``birthed'' with this specific
application in mind. We believe that the optimism is justified.
At present, there are a number of experiments under construction (or that will be
shortly) that will test for WIMP interactions in the range 1.0
~kg$^{-1}$~d$^{-1}$ - 0.01 ~kg$^{-1}$~d$^{-1}$.
The CDMS~II, EDELWEISS~II and
CRESST~II which all employ low energy threshold ($<10$ keV), nuclear recoil
discriminating,  cryogenic detectors will be deploying $\approx$ 10 kg of target.

There is also significant interest in liquid Xe for WIMP detection.
In principle, Xe is well placed, having a photon yield similar to that of NaI, while
potentially being capable of creating a radioactively cleaner target
(especially if 85Kr is removed during isotopic
enrichment stage).  Nuclear recoil versus gamma discrimination is also possible, 
either by pulse shape analysis, which gives a relatively weak separation,
and using a
simulatneous measurement of the photon and electron-ion charge yield in Xe
which is a much
stronger method of discrimination. The Rome group \cite{Bernabei:1998} achieved early operation
of Xe deep underground. A new program has been started by  a Japanese group in Kamioka
\cite{Suzuki:2000}.  There is a major effort on Xe by the Boulby DM Collaboration
\cite{Spooner:2000} which
is constructing a series of liquid Xe experiments at Boulby mine in the UK.
ZEPLIN I, now running at
Boulby, is based on pulse shape discrimination.
ZEPLIN II makes use of
simultaneous collection of scintillation and charge to achieve factors of 10-100 improved
sensitivity, and ZEPLIN III incorporates a high E-yield in the liquid to enhance the recoil
signal.
A new US group, XENON, has also recently been proposed, centered at Columbia
that would also use high field operation of two phase liquid-gas Xe.

A prototype 1 m$^{3}$ low pressure gas (CS2) detector, DRIFT, has also just
started low background operation at Boulby \cite{Martoff:2000}.

The GENIUS \cite{Klapdor:2000, Baudis:2000} and MAJORANA experiments have proposed to improve limits in
WIMP searches by significant reduction in radioactive background levels,
and exploitation of
active Ge self-shielding. Nuclear recoil discrimination is not available for low
energy events in conventional HP Ge detectors.
It is worth noting that if these experiments can
reduce low energy backgrounds by a factor 3000 from current levels
(as projected by GENIUS for the complete 14 m liquid nitrogen shield)
the limiting background becomes events from electron scattering of pp solar neutrino flux.

In order to move beyond a sensitivity level of $2\times 10^{-9}$~pb targets of 100-1000~kg (equivalent Ge)
with very good nuclear recoil discrimination (rejecting $>$99.99\% of electron events) will be necessary. A
limiting sensitivity for realistic detector arrays, based on existing cryogenic detectors (such as CryoArray
\cite{Gaitskell:2001}), or liquid Xe (such as ZEPLIN-MAX \cite{Spooner:2001} will be at an event level of 1
event per~100~kg per~1~year which is equivalent to a WIMP-nucleon cross-section of
$10^{-10}$~pb.

\begin{figure}[t]
\includegraphics[height=2.3in]{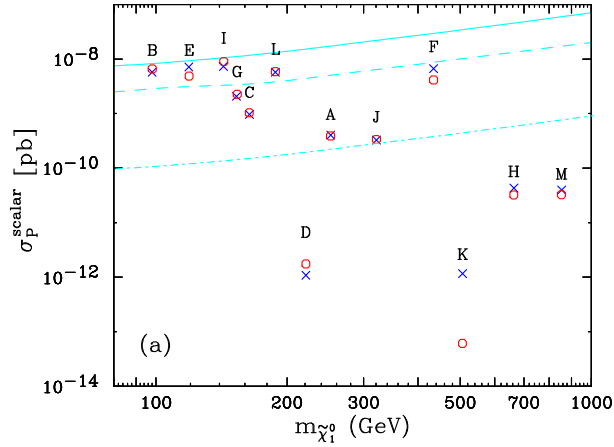}%
\caption{\it Elastic cross sections for spin-independent
neutralino-proton scattering.  The predictions of {\tt SSARD} (blue
crosses) and {\tt Neutdriver} (red circles) are compared.
Projected sensitivities for
CDMS II~\cite{Schnee:1998gf} or CRESST~\cite{Bravin:1999fc} (solid);
 GENIUS~\cite{GENIUS} (dashed);
 CryoArray~\cite{Gaitskell:2001} or ZEPLIN-MAX~\cite{Spooner:2001} (dot-dashed)
are also shown. (See Ref.~\cite{Ellis:2001us}.)
Further theoretical and experimental direct detection data can be plotted using
an interactive web plotter ~\cite{GM:2000web}.  A cross section of
$4\times 10^{-6}$~pb at a WIMP mass of 100 GeV (in line with
current experimental sensitivities)  would give an
integrated event rate of 1.0 (1.2) ~kg$^{-1}$~d$^{-1}$ for a Ge (Xe) 
target with
a detection threshold of 10 keVr.  A cross section of
$10^{-10}$~pb would give
an integrated event rate of 1.0 (1.2) ~(100~kg)$^{-1}$~year$^{-1}$
for Ge (Xe).}
\label{fig:direct}
\end{figure}

Fig.~\ref{fig:direct} shows the spin-independent
cross-section for neutralino-proton scattering for the 
benchmark points of~\cite{Battaglia:2001zp}, 
using two different codes: {\tt Neutdriver}~\cite{Jungman:1996df}
and {\tt SSARD}~\cite{ssard}. (Experiments sensitive to spin-dependent
scattering have inferior reach~\cite{Ellis:2001hv}.) One finds reasonable
agreement, with the largest differences arising for points D and K,
where the cross-section is abnormally small due to
cancellations~\cite{Ellis:2000ds,Ellis:2001qm}. For any given $\tan\beta$, the
cancellations occur only for a specific limited range in the
neutralino mass.  Unfortunately, points D and K fall exactly into this
category.

Fig.~\ref{fig:direct} also shows the projected sensitivities for a number of
experiments listed in the figure caption.
It should be emphasized that the CDMS~II experimental reach, based on a modest target mass of 5~kg of Ge, will test a significant subset of SUSY models, and comes close to testing 4 of the benchmarks I, B, E and L.
These could be reached by the larger (100~kg)/lower background GENIUS detector.
Benchmarks G, F, C, A and J can be reached by 1~tonne discriminating detectors, such as CryoArray, or
ZEPLIN-MAX, discussed above. Some of the benchmarks within reach of the direct detection experiements will not
be reached by accelerators in this decade, illustrating the complementary nature of direct detection results.

\section{Indirect Dark Matter Searches}
\label{sec:DMindirect}

Indirect dark matter signals arise from enhanced pair annihilation
rates of dark matter particles trapped in the gravitational wells at
the centers of astrophysical bodies. The different signals can be classified 
according to the nature of the emitted particles. 

While most of the neutralino annihilation products
are quickly absorbed, neutrinos may propagate for long distances and
be detected near the Earth's surface through their charged-current
conversion to muons.  High-energy muons produced by neutrinos from the
centers of the Sun and Earth are, therefore, prominent signals for
indirect dark matter detection~\cite{Silk:1985ax,Freese:1985qw,Krauss:1985aa,%
Bergstrom:1998xh,Bottino:1998vw,Corsetti:1999ma,Feng:2001zu,Barger:2001ur}.

\begin{figure}[t]
\includegraphics[height=2.3in]{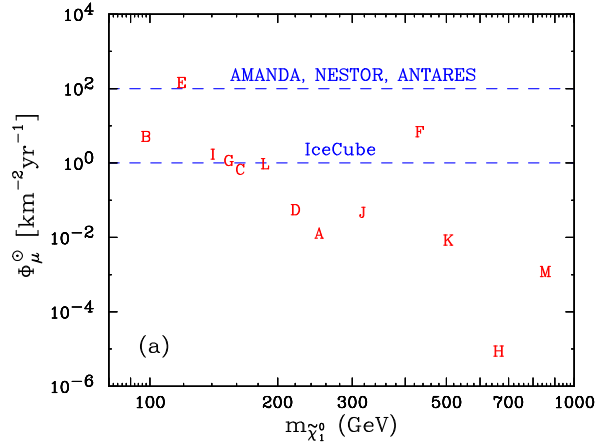}%
\caption{\it Muon fluxes from neutrinos originating from relic
annihilations inside the Sun. Approximate sensitivities of near future
neutrino telescopes ($\Phi_{\mu} = 10^2~\km^{-2}~\yr^{-1}$ for AMANDA
II~\cite{AMANDA}, NESTOR~\cite{NESTOR}, and ANTARES~\cite{ANTARES},
and $\Phi_{\mu} = 1~\km^{-2}~\yr^{-1}$ for IceCube~\cite{IceCube}) are
also indicated. (From Ref.~\cite{Ellis:2001us}.) }
\label{fig:muons}
\end{figure}

Muon fluxes for each of the benchmark points are given in
Fig.~\ref{fig:muons}, using {\tt Neutdriver} with a fixed constant
local density $\rho_0 = 0.3~\gev/\cm^3$ and neutralino velocity
dispersion $\bar{v} = 270~\km/\s$ (for further details, see
\cite{Ellis:2001hv}).  For the points considered, rates
from the Sun are far more promising than rates 
from the Earth~\cite{Ellis:2001hv,Feng:2001zu}.  For
the Sun, muon fluxes are for the most part anti-correlated with
neutralino mass, with two strong exceptions: the focus
point models (E and F) have anomalously large fluxes.  
In these cases, the dark matter's Higgsino content, 
though still small, is significant, leading to annihilations 
to gauge boson pairs, hard neutrinos, and enhanced detection rates.

The exact reach of neutrino telescopes
depends on the salient features of the particular detector, 
\eg, physical dimensions, muon energy threshold, etc., 
and the expected characteristics of the signal, \eg, 
angular dispersion, energy spectrum and source (Sun or Earth).  
Two sensitivities, which are roughly indicative of the
potential of upcoming neutrino telescope experiments, are given in
Fig.~\ref{fig:muons}. For focus point model E, where the neutralino is
both light and significantly different from pure Bino-like, detection
is possible already in the near future at AMANDA II~\cite{AMANDA}, 
NESTOR~\cite{NESTOR}, and ANTARES~\cite{ANTARES}.  Point F may be
within reach of IceCube~\cite{IceCube}, as the neutralino's significant Higgsino
component compensates for its large mass.  For point B, and possibly
also points I, G, C, and L, the neutralino is nearly pure Bino, but is
sufficiently light that detection at IceCube may also be possible.

As with the centers of the Sun and Earth, the center of the galaxy may
attract a significant overabundance of relic dark matter
particles~\cite{Urban:1992ej,Berezinsky:1992mx,Berezinsky:1994wv}.  
Relic pair annihilation at the galactic center will then produce 
an excess of photons, which may be observed in gamma ray detectors.  
While monoenergetic signals from
$\chi \chi \rightarrow \gamma \gamma$ and $\chi \chi \rightarrow
\gamma Z$ would be spectacular~\cite{Bergstrom:1998fj}, they are
loop-suppressed and rarely observable. 

\begin{figure}[t]
\includegraphics[height=2.3in]{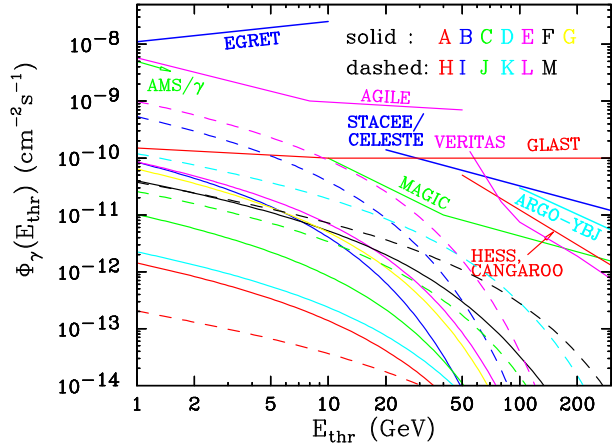}%
\caption{\it The integrated photon flux $\Phi_\gamma(\ethr)$ as a
function of photon energy threshold $\ethr$ for photons produced by
relic annihilations in the galactic center. A moderate halo parameter
$\bar{J} = 500$ is assumed~\cite{Bergstrom:1998fj}.  Point source flux
sensitivities for various gamma ray detectors are also shown. 
(From Ref.~\cite{Ellis:2001us}.)}
\label{fig:photon_spectra}
\end{figure}

Fig.~\ref{fig:photon_spectra} shows the integrated photon flux 
$\Phi_\gamma(\ethr)$ in the direction of the galactic center,
computed following the procedure of~\cite{Feng:2001zu}. 
Estimates for point source flux sensitivities of several gamma
ray detectors, both current and planned, are also shown.  
The space-based detectors EGRET, AMS/$\gamma$ and GLAST 
can detect soft photons, but are limited in
flux sensitivity by their small effective areas.  Ground-based
telescopes, such as MAGIC, HESS, CANGAROO and VERITAS, are much larger
and so sensitive to lower fluxes, but are limited by higher energy
thresholds.  These sensitivities are not strictly valid for
observations of the galactic center.  Nevertheless, they provide rough
guidelines for what sensitivities may be expected in coming years.
For a discussion of these estimates, their derivation, and references
to the original literature, see~\cite{Feng:2001zu}.

Fig.~\ref{fig:photon_spectra} suggests that space-based detectors offer
good prospects for detecting a photon signal, while ground-based
telescopes have a relatively limited reach.  GLAST appears to be
particularly promising, with points I and L giving observable signals.
One should keep in mind that all predicted fluxes scale linearly with
$\bar{J}$, and for a different halo profiles may be enhanced or suppressed
by up to two orders of magnitude. Such an enhancement may 
lead to detectable signals in GLAST for almost all
points, and at MAGIC for the majority of benchmark points.

Relic neutralino annihilations in the galactic halo may also be
detected through positron excesses in space-based and balloon
experiments~\cite{Tylka:1989xj,Turner:1990kg,Kamionkowski:1991ty,%
Moskalenko:1999sb}.  Ref.~\cite{Ellis:2001hv} also estimated the
observability of a positron excess, following the procedure
advocated in~\cite{Feng:2001zu}.  For each benchmark spectrum, 
one first finds the positron energy $\eopt$ at which the positron
signal to background ratio $S/B$ is maximized, and then requires
$S/B$ at $\eopt$ to be within reach of the experiment.  
The sensitivities of a variety of experiments have
been estimated in~\cite{Feng:2001zu}.
Among these experiments, the most promising is AMS~\cite{AMS}, the
anti-matter detector to be placed on the International Space Station.
AMS will detect unprecedented numbers of positrons in a wide energy
range.  We estimate that a 1\% excess in an fairly narrow energy bin,
as is characteristic of the neutralino signal, will be statistically
significant. Unfortunately, all benchmark points yield positron signals 
below the AMS sensitivity~\cite{Ellis:2001hv}. Similar rather 
pessimistic conclusions were derived in~\cite{Kane:2001fz,Baltz:2001ir}.
Of course, one should 
be aware that as with the photon signal, positron rates are sensitive 
to the halo model assumed; for clumpy halos~\cite{Silk:1992bh}, 
the rate may be again enhanced by orders of 
magnitude~\cite{Moskalenko:1999sb}.

Other possible indirect dark matter signals are
antiprotons~\cite{Chardonnet:1996ca,Bottino:1998tw,Bottino:1998vw} 
and antideuterium~\cite{Donato:1999gy},
but those were not pursued during Snowmass 2001.

\section{Discussion and Conclusions}
\label{sec:discussion}

\begin{figure}[t]
\includegraphics[height=2.3in]{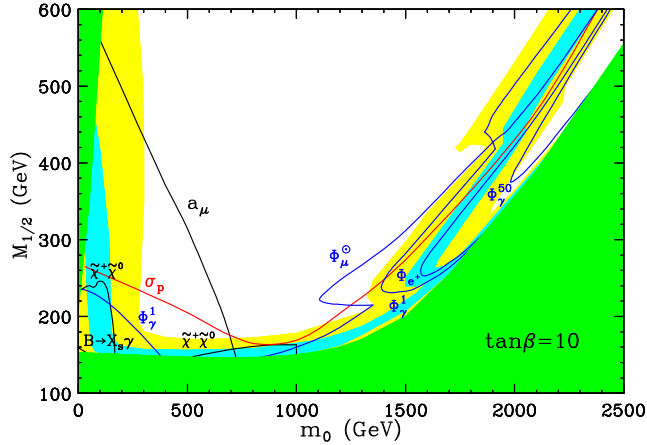}%
\caption[]{Estimated reaches of various high-energy collider and
low-energy precision searches, direct dark matter searches, 
and indirect dark matter searches before the LHC begins
operation, for $\tb=10$. The projected sensitivities used are given in
Ref.~\cite{Feng:2001ut}. The darker shaded (green) regions are excluded
by the requirement that the LSP be neutral (left) and by the LEP
chargino mass limit (bottom and right).  The
regions with potentially interesting values of the LSP relic
abundance: $0.025\le \Omegachi h^2 \le 1$ (light-shaded, yellow) 
and $0.1 \le \Omegachi h^2 \le 0.3$ (medium-shaded, light blue),
have also been delineated. 
The regions probed extend the curves toward the forbidden regions. 
(From Ref.~\cite{Feng:2001ut}.)}
\label{fig:reach}
\end{figure}

Fig.~\ref{fig:reach} shows a compilation of many 
pre-LHC experiments in astrophysics, as well as particle
physics at both the energy frontier and lower energies.
The signals considered, the projected sensitivities, 
and the experiments likely to achieve them, are discussed 
in \cite{Feng:2001ut}.
On the particle physics side, the signals considered were
supersymmetry searches at LEP~\cite{SUSYWG} and the 
Tevatron~\cite{Matchev:1999nb,Baer:1999bq,Barger:1998hp,%
Matchev:1999yn,Lykken:1999kp}, 
the improved measurement of the $B\rightarrow X_s\gamma$ 
branching ratio at B-factories, as well as the projected 
final sensitivity of the Brookhaven $g_\mu-2$ experiment.
On the astrophysics side, the figure shows the projected reach
of the upcoming direct dark matter detection experiments, 
as well as the multitude of other experiments to detect indirect
neutrino, photon or positron signals from neutralino annihilations,
as discussed earlier.

Several striking features emerge from Fig.~\ref{fig:reach}.  
First, we see that, within the minimal
supergravity framework, nearly all of the cosmologically preferred
models will be probed by at least one experiment\footnote{While 
this is strictly true for low $\tb$, at higher $\tb$ some of 
the preferred region may escape all probes, but this requires 
heavy superpartners and a significant fine-tuning of the electroweak
scale~\cite{Feng:2001zu}. Furthermore, the large $\tan\beta$ 
region is also effectively probed via the \bs 
signal~\cite{Dedes:2001fv} which was not considered here.}. 
In the most natural regions, all models in which neutralinos
form a significant fraction of dark matter will yield some signal
before the LHC begins operation.

Also noteworthy is the complementarity of traditional particle physics
searches and indirect dark matter searches.  Collider searches
require, of course, light superpartners.  High precision probes at low
energy also require light superpartners, as the virtual effects of
superpartners quickly decouple as they become heavy.  Thus, the LEP
and Tevatron reaches are confined to the lower left-hand corner, as
are, to a lesser extent, the searches for deviations in $B \to X_s
\gamma$ and $a_{\mu}$.  These bounds, and all others of this type, are
easily satisfied in the focus point models with large $m_0$, and
indeed this is one of the virtues of these models.  However, in the
focus point models, {\em all} of the indirect searches are maximally
sensitive, as the dark matter contains a significant Higgsino
component.  Direct dark matter probes share features with both
traditional and indirect searches, and have sensitivity in both
regions.  It is only by combining all of these experiments, that the
preferred region may be completely explored. 

\begin{figure}[t]
\includegraphics[height=2.3in]{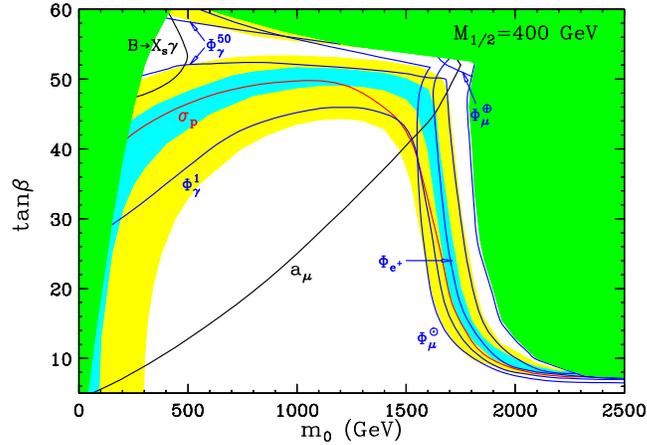}%
\caption[]{As in Fig.~\ref{fig:reach}, but in the $(m_0, \tb)$ plane
for fixed $\mgaugino=400~\gev$, $A_0 = 0$, and $\mu > 0$.  The regions
probed are toward the green regions, except for $\Phi_{\gamma}^{50}$,
where it is between the two contours.  The top excluded region is
forbidden by limits on the \CPn -odd Higgs scalar mass.
(From Ref.~\cite{Feng:2001ut}.)}
\label{fig:reach400}
\end{figure}

Finally, these results have implications for future colliders.  In the
cosmologically preferred regions of parameter space with $0.1 <
\Omegachi h^2 < 0.3$, all models with charginos or sleptons lighter
than 300 GeV will produce observable signals in at least one
experiment.  This is evident for $\tb=10$ in Fig.~\ref{fig:reach}.  
In Fig.~\ref{fig:reach400}, we vary $\tb$, fixing $\mgaugino$ to 400 GeV,
which roughly corresponds to 300 GeV charginos.  We see that the
preferred region is probed for any choice of $\tb$. (For extremely low
$\tb$ and $m_0$, there appears to be a region that is not probed.
However, this is excluded by current Higgs mass limits for $A_0 = 0$.
These limits might be evaded if $A_0$ is also tuned to some extreme
value, but in this case, top squark searches in Run II of the Tevatron
\cite{Demina:1999ty} will provide an additional constraint.)

These results imply that if any superpartners are to be within reach
of a 500 GeV lepton collider, some hint of supersymmetry must be seen
before the LHC begins collecting data.  This conclusion is independent
of naturalness considerations.  While our quantitative analysis is
confined to minimal supergravity, we expect this result to be valid
more generally.  For moderate values of $\tb$, if the dark matter is
made up of neutralinos, they must either be light, Bino-like, or a
gaugino-Higgsino mixture.  If they are light, charginos will be
discovered.  If they are Bino-like, light sfermions are required to
mediate their annihilation, and there will be anomalies in low energy
precision measurements. And if they are a gaugino-Higgsino mixture, at
least one indirect dark matter search will see a signal.  For large
$\tb$, low energy probes become much more effective and again there is
sensitivity to all superpartner spectra with light
superpartners. Thus it appears, on qualitative grounds, that all
models in which the scalar masses are not widely separated, and the
charginos are not extravagantly heavy, will be accessible prior to LHC
operation.

The most sensitive tests for SUSY contributions in $B$ decays 
before the start of the LHC will come from $\sin 2 \beta$ measurements in 
charmonium modes and \CPn -asymmetry measurements of $B \rightarrow X_s 
\gamma$ and $B \rightarrow K^* \gamma$ by BABAR and BELLE. 
For the latter studies the large SM theoretical uncertainties present in 
the branching-fraction measurements that make an extraction of SUSY 
contributions difficult are absent. In the LHC era improved measurements
on $\sin 2 \beta$ and the \CP asymmetry in $K^{*0} \gamma$ will be 
carried out. In addition, measurements of 
branching-fractions and lepton forward-backward asymmetries of 
$B \rightarrow s \ell^+ \ell^-$ modes will provide very sensitive tests
for uncovering SUSY effects. While hadron-collider experiments will focus on 
$K^{(*)} \mu^+ \mu^-$ modes, asymmetric $B$-factory experiments will study
$B \rightarrow X_s \ell^+ \ell^-$ and $B \rightarrow K^{(*)} \ell^+ \ell^-$ 
decays. At a super $B$ factory the precision can be significantly 
improved in these modes. While the first studies of $\alpha$ will be performed 
by BABAR and BELLE, precise measurements will come from
from BTEV and LHCb. In particular, the hadron-collider experiments produce
high statistics $B_s$ samples, which allow for precise measurements 
of $\gamma$. Additional precise measurements of $\gamma$ at the $\Upsilon(4S)$ 
can only come from a super $B$ factory. By this time a considerable 
reduction of the theoretical uncertainties in the extraction 
of $|V_{ub}/V_{cb}|$, $|V_{td}|$ and $|V_{ts}|$ from data is also expected, 
allowing to perform a sensitive model-independent analysis for extracting a 
new weak phase and new contributions to $B^0 \bar B^0$ mixing. 
In order to observe the modes $ B \rightarrow X_s \nu \bar \nu$, 
$ B \rightarrow K^+ \nu \bar \nu$, or
$ B \rightarrow K^{*0} \nu \bar \nu$, which bear the least theoretical 
uncertainties of radiative penguin decays, and study their properties a 
super $B$ factory is a prerequisite. So, for at least the next ten years 
a vivacious $B$ physics program ensures many high-precision measurements 
of \CP asymmetries and radiative penguin decay properties that are suitable
for extracting SUSY contributions.

\begin{acknowledgments}
We thank E.~A.~Hinds, L.~Littenberg, D.~Miller and J.~M.~Pendlebury
for sharing their wisdom.
\end{acknowledgments}

%
%

%
%

\bibliography{P3_42}

\begin{thebibliography}{222}
\expandafter\ifx\csname natexlab\endcsname\relax\def\natexlab#1{#1}\fi
\expandafter\ifx\csname bibnamefont\endcsname\relax
  \def\bibnamefont#1{#1}\fi
\expandafter\ifx\csname bibfnamefont\endcsname\relax
  \def\bibfnamefont#1{#1}\fi
\expandafter\ifx\csname citenamefont\endcsname\relax
  \def\citenamefont#1{#1}\fi
\expandafter\ifx\csname url\endcsname\relax
  \def\url#1{\texttt{#1}}\fi
\expandafter\ifx\csname urlprefix\endcsname\relax\def\urlprefix{URL }\fi
\providecommand{\bibinfo}[2]{#2}
\providecommand{\eprint}[2][]{\url{#2}}

\bibitem[{\citenamefont{Feng et~al.}(2001{\natexlab{a}})\citenamefont{Feng,
  Matchev, and Shadmi}}]{Feng:2001mq}
\bibinfo{author}{\bibfnamefont{J.~L.} \bibnamefont{Feng}},
  \bibinfo{author}{\bibfnamefont{K.~T.} \bibnamefont{Matchev}},
  \bibnamefont{and} \bibinfo{author}{\bibfnamefont{Y.}~\bibnamefont{Shadmi}}
  (\bibinfo{year}{2001}{\natexlab{a}}),
  \eprint[http://arXiv.org/abs]{hep-ph/0110157}.

\bibitem[{\citenamefont{Feng and Matchev}(2001{\natexlab{a}})}]{Feng:2001hm}
\bibinfo{author}{\bibfnamefont{J.~L.} \bibnamefont{Feng}} \bibnamefont{and}
  \bibinfo{author}{\bibfnamefont{K.~T.} \bibnamefont{Matchev}}
  (\bibinfo{year}{2001}{\natexlab{a}}),
  \eprint[http://arXiv.org/abs]{hep-ph/0111004}.

\bibitem[{\citenamefont{Ellis et~al.}(2001{\natexlab{a}})\citenamefont{Ellis,
  Feng, Ferstl, Matchev, and Olive}}]{Ellis:2001us}
\bibinfo{author}{\bibfnamefont{J.~R.} \bibnamefont{Ellis}},
  \bibinfo{author}{\bibfnamefont{J.~L.} \bibnamefont{Feng}},
  \bibinfo{author}{\bibfnamefont{A.}~\bibnamefont{Ferstl}},
  \bibinfo{author}{\bibfnamefont{K.~T.} \bibnamefont{Matchev}},
  \bibnamefont{and} \bibinfo{author}{\bibfnamefont{K.~A.} \bibnamefont{Olive}}
  (\bibinfo{year}{2001}{\natexlab{a}}),
  \eprint[http://arXiv.org/abs]{hep-ph/0111294}.

\bibitem[{\citenamefont{Feng et~al.}(2001{\natexlab{b}})\citenamefont{Feng,
  Matchev, and Wilczek}}]{Feng:2001ut}
\bibinfo{author}{\bibfnamefont{J.~L.} \bibnamefont{Feng}},
  \bibinfo{author}{\bibfnamefont{K.~T.} \bibnamefont{Matchev}},
  \bibnamefont{and} \bibinfo{author}{\bibfnamefont{F.}~\bibnamefont{Wilczek}}
  (\bibinfo{year}{2001}{\natexlab{b}}),
  \eprint[http://arXiv.org/abs]{hep-ph/0111295}.

\bibitem[{\citenamefont{Eigen}({\natexlab{a}})}]{ge1}
\bibinfo{author}{\bibfnamefont{G.}~\bibnamefont{Eigen}}, \eprint{talk in P3-2
  session, preprint P3-41, Proceedings of Snowmass Wokshop}.

\bibitem[{\citenamefont{Eigen}({\natexlab{b}})}]{ge3}
\bibinfo{author}{\bibfnamefont{G.}~\bibnamefont{Eigen}}, \eprint{talk in E2
  session, preprint E2-41, Proceedings of Snowmass Wokshop}.

\bibitem[{\citenamefont{Brown et~al.}(2001)}]{Brown:2001mg}
\bibinfo{author}{\bibfnamefont{H.~N.} \bibnamefont{Brown}} \bibnamefont{et~al.}
  (\bibinfo{collaboration}{Muon g-2}), \bibinfo{journal}{Phys. Rev. Lett.}
  \textbf{\bibinfo{volume}{86}}, \bibinfo{pages}{2227} (\bibinfo{year}{2001}),
  \eprint[http://arXiv.org/abs]{hep-ex/0102017}.

\bibitem[{\citenamefont{Knecht and Nyffeler}(2001)}]{Knecht:2001qf}
\bibinfo{author}{\bibfnamefont{M.}~\bibnamefont{Knecht}} \bibnamefont{and}
  \bibinfo{author}{\bibfnamefont{A.}~\bibnamefont{Nyffeler}}
  (\bibinfo{year}{2001}), \eprint[http://arXiv.org/abs]{hep-ph/0111058}.

\bibitem[{\citenamefont{Knecht et~al.}(2001)\citenamefont{Knecht, Nyffeler,
  Perrottet, and De~Rafael}}]{Knecht:2001qg}
\bibinfo{author}{\bibfnamefont{M.}~\bibnamefont{Knecht}},
  \bibinfo{author}{\bibfnamefont{A.}~\bibnamefont{Nyffeler}},
  \bibinfo{author}{\bibfnamefont{M.}~\bibnamefont{Perrottet}},
  \bibnamefont{and} \bibinfo{author}{\bibfnamefont{E.}~\bibnamefont{De~Rafael}}
  (\bibinfo{year}{2001}), \eprint[http://arXiv.org/abs]{hep-ph/0111059}.

\bibitem[{\citenamefont{Masashi and Kinoshita}(2001)}]{kinoshita}
\bibinfo{author}{\bibfnamefont{H.}~\bibnamefont{Masashi}} \bibnamefont{and}
  \bibinfo{author}{\bibfnamefont{T.}~\bibnamefont{Kinoshita}}
  (\bibinfo{year}{2001}), \eprint[http://arXiv.org/abs]{hep-ph/0112102}.

\bibitem[{\citenamefont{Blokland et~al.}(2001)\citenamefont{Blokland,
  Czarnecki, and Melnikov}}]{Blokland:2001pb}
\bibinfo{author}{\bibfnamefont{I.}~\bibnamefont{Blokland}},
  \bibinfo{author}{\bibfnamefont{A.}~\bibnamefont{Czarnecki}},
  \bibnamefont{and} \bibinfo{author}{\bibfnamefont{K.}~\bibnamefont{Melnikov}}
  (\bibinfo{year}{2001}), \eprint[http://arXiv.org/abs]{hep-ph/0112117}.

\bibitem[{\citenamefont{Bijnens et~al.}(2001)\citenamefont{Bijnens, Pallante,
  and Prades}}]{Bijnens:2001cq}
\bibinfo{author}{\bibfnamefont{J.}~\bibnamefont{Bijnens}},
  \bibinfo{author}{\bibfnamefont{E.}~\bibnamefont{Pallante}}, \bibnamefont{and}
  \bibinfo{author}{\bibfnamefont{J.}~\bibnamefont{Prades}}
  (\bibinfo{year}{2001}), \eprint[http://arXiv.org/abs]{hep-ph/0112255}.

\bibitem[{\citenamefont{Melnikov}(2001)}]{Melnikov:2001uw}
\bibinfo{author}{\bibfnamefont{K.}~\bibnamefont{Melnikov}},
  \bibinfo{journal}{Int. J. Mod. Phys.} \textbf{\bibinfo{volume}{A16}},
  \bibinfo{pages}{4591} (\bibinfo{year}{2001}),
  \eprint[http://arXiv.org/abs]{hep-ph/0105267}.

\bibitem[{\citenamefont{De~Troconiz and Yndurain}(2001)}]{DeTroconiz:2001wt}
\bibinfo{author}{\bibfnamefont{J.~F.} \bibnamefont{De~Troconiz}}
  \bibnamefont{and} \bibinfo{author}{\bibfnamefont{F.~J.}
  \bibnamefont{Yndurain}} (\bibinfo{year}{2001}),
  \eprint[http://arXiv.org/abs]{hep-ph/0106025}.

\bibitem[{\citenamefont{Cvetic et~al.}(2001)\citenamefont{Cvetic, Lee, and
  Schmidt}}]{Cvetic:2001pg}
\bibinfo{author}{\bibfnamefont{G.}~\bibnamefont{Cvetic}},
  \bibinfo{author}{\bibfnamefont{T.}~\bibnamefont{Lee}}, \bibnamefont{and}
  \bibinfo{author}{\bibfnamefont{I.}~\bibnamefont{Schmidt}},
  \bibinfo{journal}{Phys. Lett.} \textbf{\bibinfo{volume}{B520}},
  \bibinfo{pages}{222} (\bibinfo{year}{2001}),
  \eprint[http://arXiv.org/abs]{hep-ph/0107069}.

\bibitem[{\citenamefont{Czarnecki and Marciano}(2001)}]{Czarnecki:2001pv}
\bibinfo{author}{\bibfnamefont{A.}~\bibnamefont{Czarnecki}} \bibnamefont{and}
  \bibinfo{author}{\bibfnamefont{W.~J.} \bibnamefont{Marciano}},
  \bibinfo{journal}{Phys. Rev.} \textbf{\bibinfo{volume}{D64}},
  \bibinfo{pages}{013014} (\bibinfo{year}{2001}),
  \eprint[http://arXiv.org/abs]{hep-ph/0102122}.

\bibitem[{\citenamefont{Fayet}(1980)}]{Fayet:1980yy}
\bibinfo{author}{\bibfnamefont{P.}~\bibnamefont{Fayet}} (\bibinfo{year}{1980}),
  \bibinfo{note}{contributed to Europhysics Study Conf. on Unification of
  Fundamental Interactons, Erice, Italy, Mar 17-24, 1980}.

\bibitem[{\citenamefont{Grifols and Mendez}(1982)}]{Grifols:1982vx}
\bibinfo{author}{\bibfnamefont{J.~A.} \bibnamefont{Grifols}} \bibnamefont{and}
  \bibinfo{author}{\bibfnamefont{A.}~\bibnamefont{Mendez}},
  \bibinfo{journal}{Phys. Rev.} \textbf{\bibinfo{volume}{D26}},
  \bibinfo{pages}{1809} (\bibinfo{year}{1982}).

\bibitem[{\citenamefont{Ellis et~al.}(1982{\natexlab{a}})\citenamefont{Ellis,
  Hagelin, and Nanopoulos}}]{Ellis:1982by}
\bibinfo{author}{\bibfnamefont{J.~R.} \bibnamefont{Ellis}},
  \bibinfo{author}{\bibfnamefont{J.~S.} \bibnamefont{Hagelin}},
  \bibnamefont{and} \bibinfo{author}{\bibfnamefont{D.~V.}
  \bibnamefont{Nanopoulos}}, \bibinfo{journal}{Phys. Lett.}
  \textbf{\bibinfo{volume}{B116}}, \bibinfo{pages}{283}
  (\bibinfo{year}{1982}{\natexlab{a}}).

\bibitem[{\citenamefont{Barbieri and Maiani}(1982)}]{Barbieri:1982aj}
\bibinfo{author}{\bibfnamefont{R.}~\bibnamefont{Barbieri}} \bibnamefont{and}
  \bibinfo{author}{\bibfnamefont{L.}~\bibnamefont{Maiani}},
  \bibinfo{journal}{Phys. Lett.} \textbf{\bibinfo{volume}{B117}},
  \bibinfo{pages}{203} (\bibinfo{year}{1982}).

\bibitem[{\citenamefont{Kosower et~al.}(1983)\citenamefont{Kosower, Krauss, and
  Sakai}}]{Kosower:1983yw}
\bibinfo{author}{\bibfnamefont{D.~A.} \bibnamefont{Kosower}},
  \bibinfo{author}{\bibfnamefont{L.~M.} \bibnamefont{Krauss}},
  \bibnamefont{and} \bibinfo{author}{\bibfnamefont{N.}~\bibnamefont{Sakai}},
  \bibinfo{journal}{Phys. Lett.} \textbf{\bibinfo{volume}{B133}},
  \bibinfo{pages}{305} (\bibinfo{year}{1983}).

\bibitem[{\citenamefont{Yuan et~al.}(1984)\citenamefont{Yuan, Arnowitt,
  Chamseddine, and Nath}}]{Yuan:1984ww}
\bibinfo{author}{\bibfnamefont{T.~C.} \bibnamefont{Yuan}},
  \bibinfo{author}{\bibfnamefont{R.}~\bibnamefont{Arnowitt}},
  \bibinfo{author}{\bibfnamefont{A.~H.} \bibnamefont{Chamseddine}},
  \bibnamefont{and} \bibinfo{author}{\bibfnamefont{P.}~\bibnamefont{Nath}},
  \bibinfo{journal}{Zeit. Phys.} \textbf{\bibinfo{volume}{C26}},
  \bibinfo{pages}{407} (\bibinfo{year}{1984}).

\bibitem[{\citenamefont{Moroi}(1996)}]{Moroi:1996yh}
\bibinfo{author}{\bibfnamefont{T.}~\bibnamefont{Moroi}},
  \bibinfo{journal}{Phys. Rev.} \textbf{\bibinfo{volume}{D53}},
  \bibinfo{pages}{6565} (\bibinfo{year}{1996}),
  \eprint[http://arXiv.org/abs]{hep-ph/9512396}.

\bibitem[{\citenamefont{Carena et~al.}(1997)\citenamefont{Carena, Giudice, and
  Wagner}}]{Carena:1997qa}
\bibinfo{author}{\bibfnamefont{M.}~\bibnamefont{Carena}},
  \bibinfo{author}{\bibfnamefont{G.~F.} \bibnamefont{Giudice}},
  \bibnamefont{and} \bibinfo{author}{\bibfnamefont{C.~E.~M.}
  \bibnamefont{Wagner}}, \bibinfo{journal}{Phys. Lett.}
  \textbf{\bibinfo{volume}{B390}}, \bibinfo{pages}{234} (\bibinfo{year}{1997}),
  \eprint[http://arXiv.org/abs]{hep-ph/9610233}.

\bibitem[{\citenamefont{Gabrielli and Sarid}(1997)}]{Gabrielli:1997jp}
\bibinfo{author}{\bibfnamefont{E.}~\bibnamefont{Gabrielli}} \bibnamefont{and}
  \bibinfo{author}{\bibfnamefont{U.}~\bibnamefont{Sarid}},
  \bibinfo{journal}{Phys. Rev. Lett.} \textbf{\bibinfo{volume}{79}},
  \bibinfo{pages}{4752} (\bibinfo{year}{1997}),
  \eprint[http://arXiv.org/abs]{hep-ph/9707546}.

\bibitem[{\citenamefont{Mahanthappa and Oh}(2000)}]{Mahanthappa:1999ta}
\bibinfo{author}{\bibfnamefont{K.~T.} \bibnamefont{Mahanthappa}}
  \bibnamefont{and} \bibinfo{author}{\bibfnamefont{S.}~\bibnamefont{Oh}},
  \bibinfo{journal}{Phys. Rev.} \textbf{\bibinfo{volume}{D62}},
  \bibinfo{pages}{015012} (\bibinfo{year}{2000}),
  \eprint[http://arXiv.org/abs]{hep-ph/9908531}.

\bibitem[{\citenamefont{Feng and Matchev}(2001{\natexlab{b}})}]{Feng:2001tr}
\bibinfo{author}{\bibfnamefont{J.~L.} \bibnamefont{Feng}} \bibnamefont{and}
  \bibinfo{author}{\bibfnamefont{K.~T.} \bibnamefont{Matchev}},
  \bibinfo{journal}{Phys. Rev. Lett.} \textbf{\bibinfo{volume}{86}},
  \bibinfo{pages}{3480} (\bibinfo{year}{2001}{\natexlab{b}}),
  \eprint[http://arXiv.org/abs]{hep-ph/0102146}.

\bibitem[{\citenamefont{Everett et~al.}(2001)\citenamefont{Everett, Kane,
  Rigolin, and Wang}}]{Everett:2001tq}
\bibinfo{author}{\bibfnamefont{L.~L.} \bibnamefont{Everett}},
  \bibinfo{author}{\bibfnamefont{G.~L.} \bibnamefont{Kane}},
  \bibinfo{author}{\bibfnamefont{S.}~\bibnamefont{Rigolin}}, \bibnamefont{and}
  \bibinfo{author}{\bibfnamefont{L.-T.} \bibnamefont{Wang}},
  \bibinfo{journal}{Phys. Rev. Lett.} \textbf{\bibinfo{volume}{86}},
  \bibinfo{pages}{3484} (\bibinfo{year}{2001}),
  \eprint[http://arXiv.org/abs]{hep-ph/0102145}.

\bibitem[{\citenamefont{Baltz and Gondolo}(2001)}]{Baltz:2001ts}
\bibinfo{author}{\bibfnamefont{E.~A.} \bibnamefont{Baltz}} \bibnamefont{and}
  \bibinfo{author}{\bibfnamefont{P.}~\bibnamefont{Gondolo}},
  \bibinfo{journal}{Phys. Rev. Lett.} \textbf{\bibinfo{volume}{86}},
  \bibinfo{pages}{5004} (\bibinfo{year}{2001}),
  \eprint[http://arXiv.org/abs]{hep-ph/0102147}.

\bibitem[{\citenamefont{Chattopadhyay and
  Nath}(2001{\natexlab{a}})}]{Chattopadhyay:2001vx}
\bibinfo{author}{\bibfnamefont{U.}~\bibnamefont{Chattopadhyay}}
  \bibnamefont{and} \bibinfo{author}{\bibfnamefont{P.}~\bibnamefont{Nath}},
  \bibinfo{journal}{Phys. Rev. Lett.} \textbf{\bibinfo{volume}{86}},
  \bibinfo{pages}{5854} (\bibinfo{year}{2001}{\natexlab{a}}),
  \eprint[http://arXiv.org/abs]{hep-ph/0102157}.

\bibitem[{\citenamefont{Komine et~al.}(2001{\natexlab{a}})\citenamefont{Komine,
  Moroi, and Yamaguchi}}]{Komine:2001fz}
\bibinfo{author}{\bibfnamefont{S.}~\bibnamefont{Komine}},
  \bibinfo{author}{\bibfnamefont{T.}~\bibnamefont{Moroi}}, \bibnamefont{and}
  \bibinfo{author}{\bibfnamefont{M.}~\bibnamefont{Yamaguchi}},
  \bibinfo{journal}{Phys. Lett.} \textbf{\bibinfo{volume}{B506}},
  \bibinfo{pages}{93} (\bibinfo{year}{2001}{\natexlab{a}}),
  \eprint[http://arXiv.org/abs]{hep-ph/0102204}.

\bibitem[{\citenamefont{Hisano and Tobe}(2001)}]{Hisano:2001qz}
\bibinfo{author}{\bibfnamefont{J.}~\bibnamefont{Hisano}} \bibnamefont{and}
  \bibinfo{author}{\bibfnamefont{K.}~\bibnamefont{Tobe}},
  \bibinfo{journal}{Phys. Lett.} \textbf{\bibinfo{volume}{B510}},
  \bibinfo{pages}{197} (\bibinfo{year}{2001}),
  \eprint[http://arXiv.org/abs]{hep-ph/0102315}.

\bibitem[{\citenamefont{Ibrahim et~al.}(2001)\citenamefont{Ibrahim,
  Chattopadhyay, and Nath}}]{Ibrahim:2001ym}
\bibinfo{author}{\bibfnamefont{T.}~\bibnamefont{Ibrahim}},
  \bibinfo{author}{\bibfnamefont{U.}~\bibnamefont{Chattopadhyay}},
  \bibnamefont{and} \bibinfo{author}{\bibfnamefont{P.}~\bibnamefont{Nath}},
  \bibinfo{journal}{Phys. Rev.} \textbf{\bibinfo{volume}{D64}},
  \bibinfo{pages}{016010} (\bibinfo{year}{2001}),
  \eprint[http://arXiv.org/abs]{hep-ph/0102324}.

\bibitem[{\citenamefont{Ellis et~al.}(2001{\natexlab{b}})\citenamefont{Ellis,
  Nanopoulos, and Olive}}]{Ellis:2001yu}
\bibinfo{author}{\bibfnamefont{J.~R.} \bibnamefont{Ellis}},
  \bibinfo{author}{\bibfnamefont{D.~V.} \bibnamefont{Nanopoulos}},
  \bibnamefont{and} \bibinfo{author}{\bibfnamefont{K.~A.} \bibnamefont{Olive}},
  \bibinfo{journal}{Phys. Lett.} \textbf{\bibinfo{volume}{B508}},
  \bibinfo{pages}{65} (\bibinfo{year}{2001}{\natexlab{b}}),
  \eprint[http://arXiv.org/abs]{hep-ph/0102331}.

\bibitem[{\citenamefont{Arnowitt et~al.}(2001)\citenamefont{Arnowitt, Dutta,
  Hu, and Santoso}}]{Arnowitt:2001be}
\bibinfo{author}{\bibfnamefont{R.}~\bibnamefont{Arnowitt}},
  \bibinfo{author}{\bibfnamefont{B.}~\bibnamefont{Dutta}},
  \bibinfo{author}{\bibfnamefont{B.}~\bibnamefont{Hu}}, \bibnamefont{and}
  \bibinfo{author}{\bibfnamefont{Y.}~\bibnamefont{Santoso}},
  \bibinfo{journal}{Phys. Lett.} \textbf{\bibinfo{volume}{B505}},
  \bibinfo{pages}{177} (\bibinfo{year}{2001}),
  \eprint[http://arXiv.org/abs]{hep-ph/0102344}.

\bibitem[{\citenamefont{Choi et~al.}(2001)\citenamefont{Choi, Hwang, Kang, Lee,
  and Song}}]{Choi:2001pz}
\bibinfo{author}{\bibfnamefont{K.}~\bibnamefont{Choi}},
  \bibinfo{author}{\bibfnamefont{K.}~\bibnamefont{Hwang}},
  \bibinfo{author}{\bibfnamefont{S.~K.} \bibnamefont{Kang}},
  \bibinfo{author}{\bibfnamefont{K.~Y.} \bibnamefont{Lee}}, \bibnamefont{and}
  \bibinfo{author}{\bibfnamefont{W.~Y.} \bibnamefont{Song}},
  \bibinfo{journal}{Phys. Rev.} \textbf{\bibinfo{volume}{D64}},
  \bibinfo{pages}{055001} (\bibinfo{year}{2001}),
  \eprint[http://arXiv.org/abs]{hep-ph/0103048}.

\bibitem[{\citenamefont{Kim et~al.}(2001)\citenamefont{Kim, Kyae, and
  Lee}}]{Kim:2001se}
\bibinfo{author}{\bibfnamefont{J.~E.} \bibnamefont{Kim}},
  \bibinfo{author}{\bibfnamefont{B.}~\bibnamefont{Kyae}}, \bibnamefont{and}
  \bibinfo{author}{\bibfnamefont{H.~M.} \bibnamefont{Lee}},
  \bibinfo{journal}{Phys. Lett.} \textbf{\bibinfo{volume}{B520}},
  \bibinfo{pages}{298} (\bibinfo{year}{2001}),
  \eprint[http://arXiv.org/abs]{hep-ph/0103054}.

\bibitem[{\citenamefont{Martin and Wells}(2001)}]{Martin:2001st}
\bibinfo{author}{\bibfnamefont{S.~P.} \bibnamefont{Martin}} \bibnamefont{and}
  \bibinfo{author}{\bibfnamefont{J.~D.} \bibnamefont{Wells}},
  \bibinfo{journal}{Phys. Rev.} \textbf{\bibinfo{volume}{D64}},
  \bibinfo{pages}{035003} (\bibinfo{year}{2001}),
  \eprint[http://arXiv.org/abs]{hep-ph/0103067}.

\bibitem[{\citenamefont{Komine et~al.}(2001{\natexlab{b}})\citenamefont{Komine,
  Moroi, and Yamaguchi}}]{Komine:2001hy}
\bibinfo{author}{\bibfnamefont{S.}~\bibnamefont{Komine}},
  \bibinfo{author}{\bibfnamefont{T.}~\bibnamefont{Moroi}}, \bibnamefont{and}
  \bibinfo{author}{\bibfnamefont{M.}~\bibnamefont{Yamaguchi}},
  \bibinfo{journal}{Phys. Lett.} \textbf{\bibinfo{volume}{B507}},
  \bibinfo{pages}{224} (\bibinfo{year}{2001}{\natexlab{b}}),
  \eprint[http://arXiv.org/abs]{hep-ph/0103182}.

\bibitem[{\citenamefont{Baek et~al.}(2001{\natexlab{a}})\citenamefont{Baek, Ko,
  and Lee}}]{Baek:2001nz}
\bibinfo{author}{\bibfnamefont{S.-w.} \bibnamefont{Baek}},
  \bibinfo{author}{\bibfnamefont{P.}~\bibnamefont{Ko}}, \bibnamefont{and}
  \bibinfo{author}{\bibfnamefont{H.~S.} \bibnamefont{Lee}}
  (\bibinfo{year}{2001}{\natexlab{a}}),
  \eprint[http://arXiv.org/abs]{hep-ph/0103218}.

\bibitem[{\citenamefont{Carvalho et~al.}(2001)\citenamefont{Carvalho, Ellis,
  Gomez, and Lola}}]{Carvalho:2001ex}
\bibinfo{author}{\bibfnamefont{D.~F.} \bibnamefont{Carvalho}},
  \bibinfo{author}{\bibfnamefont{J.~R.} \bibnamefont{Ellis}},
  \bibinfo{author}{\bibfnamefont{M.~E.} \bibnamefont{Gomez}}, \bibnamefont{and}
  \bibinfo{author}{\bibfnamefont{S.}~\bibnamefont{Lola}},
  \bibinfo{journal}{Phys. Lett.} \textbf{\bibinfo{volume}{B515}},
  \bibinfo{pages}{323} (\bibinfo{year}{2001}),
  \eprint[http://arXiv.org/abs]{hep-ph/0103256}.

\bibitem[{\citenamefont{Baer et~al.}(2001)\citenamefont{Baer, Balazs,
  Ferrandis, and Tata}}]{Baer:2001kn}
\bibinfo{author}{\bibfnamefont{H.}~\bibnamefont{Baer}},
  \bibinfo{author}{\bibfnamefont{C.}~\bibnamefont{Balazs}},
  \bibinfo{author}{\bibfnamefont{J.}~\bibnamefont{Ferrandis}},
  \bibnamefont{and} \bibinfo{author}{\bibfnamefont{X.}~\bibnamefont{Tata}},
  \bibinfo{journal}{Phys. Rev.} \textbf{\bibinfo{volume}{D64}},
  \bibinfo{pages}{035004} (\bibinfo{year}{2001}),
  \eprint[http://arXiv.org/abs]{hep-ph/0103280}.

\bibitem[{\citenamefont{Baek et~al.}(2001{\natexlab{b}})\citenamefont{Baek,
  Goto, Okada, and Okumura}}]{Baek:2001kh}
\bibinfo{author}{\bibfnamefont{S.-w.} \bibnamefont{Baek}},
  \bibinfo{author}{\bibfnamefont{T.}~\bibnamefont{Goto}},
  \bibinfo{author}{\bibfnamefont{Y.}~\bibnamefont{Okada}}, \bibnamefont{and}
  \bibinfo{author}{\bibfnamefont{K.-i.} \bibnamefont{Okumura}},
  \bibinfo{journal}{Phys. Rev.} \textbf{\bibinfo{volume}{D64}},
  \bibinfo{pages}{095001} (\bibinfo{year}{2001}{\natexlab{b}}),
  \eprint[http://arXiv.org/abs]{hep-ph/0104146}.

\bibitem[{\citenamefont{Cerdeno et~al.}(2001)\citenamefont{Cerdeno, Gabrielli,
  Khalil, Munoz, and Torrente-Lujan}}]{Cerdeno:2001aj}
\bibinfo{author}{\bibfnamefont{D.~G.} \bibnamefont{Cerdeno}},
  \bibinfo{author}{\bibfnamefont{E.}~\bibnamefont{Gabrielli}},
  \bibinfo{author}{\bibfnamefont{S.}~\bibnamefont{Khalil}},
  \bibinfo{author}{\bibfnamefont{C.}~\bibnamefont{Munoz}}, \bibnamefont{and}
  \bibinfo{author}{\bibfnamefont{E.}~\bibnamefont{Torrente-Lujan}},
  \bibinfo{journal}{Phys. Rev.} \textbf{\bibinfo{volume}{D64}},
  \bibinfo{pages}{093012} (\bibinfo{year}{2001}),
  \eprint[http://arXiv.org/abs]{hep-ph/0104242}.

\bibitem[{\citenamefont{Chacko and Kribs}(2001)}]{Chacko:2001xd}
\bibinfo{author}{\bibfnamefont{Z.}~\bibnamefont{Chacko}} \bibnamefont{and}
  \bibinfo{author}{\bibfnamefont{G.~D.} \bibnamefont{Kribs}},
  \bibinfo{journal}{Phys. Rev.} \textbf{\bibinfo{volume}{D64}},
  \bibinfo{pages}{075015} (\bibinfo{year}{2001}),
  \eprint[http://arXiv.org/abs]{hep-ph/0104317}.

\bibitem[{\citenamefont{Blazek and King}(2001)}]{Blazek:2001zm}
\bibinfo{author}{\bibfnamefont{T.}~\bibnamefont{Blazek}} \bibnamefont{and}
  \bibinfo{author}{\bibfnamefont{S.~F.} \bibnamefont{King}},
  \bibinfo{journal}{Phys. Lett.} \textbf{\bibinfo{volume}{B518}},
  \bibinfo{pages}{109} (\bibinfo{year}{2001}),
  \eprint[http://arXiv.org/abs]{hep-ph/0105005}.

\bibitem[{\citenamefont{Cho and Hagiwara}(2001)}]{Cho:2001nf}
\bibinfo{author}{\bibfnamefont{G.-C.} \bibnamefont{Cho}} \bibnamefont{and}
  \bibinfo{author}{\bibfnamefont{K.}~\bibnamefont{Hagiwara}},
  \bibinfo{journal}{Phys. Lett.} \textbf{\bibinfo{volume}{B514}},
  \bibinfo{pages}{123} (\bibinfo{year}{2001}),
  \eprint[http://arXiv.org/abs]{hep-ph/0105037}.

\bibitem[{\citenamefont{Adhikari and Rajasekaran}(2001)}]{Adhikari:2001ra}
\bibinfo{author}{\bibfnamefont{R.}~\bibnamefont{Adhikari}} \bibnamefont{and}
  \bibinfo{author}{\bibfnamefont{G.}~\bibnamefont{Rajasekaran}}
  (\bibinfo{year}{2001}), \eprint[http://arXiv.org/abs]{hep-ph/0107279}.

\bibitem[{\citenamefont{Byrne et~al.}(2001)\citenamefont{Byrne, Kolda, and
  Lennon}}]{Byrne:2001yu}
\bibinfo{author}{\bibfnamefont{M.}~\bibnamefont{Byrne}},
  \bibinfo{author}{\bibfnamefont{C.}~\bibnamefont{Kolda}}, \bibnamefont{and}
  \bibinfo{author}{\bibfnamefont{J.~E.} \bibnamefont{Lennon}}
  (\bibinfo{year}{2001}), \eprint[http://arXiv.org/abs]{hep-ph/0108122}.

\bibitem[{\citenamefont{Komine and Yamaguchi}(2001)}]{Komine:2001rm}
\bibinfo{author}{\bibfnamefont{S.}~\bibnamefont{Komine}} \bibnamefont{and}
  \bibinfo{author}{\bibfnamefont{M.}~\bibnamefont{Yamaguchi}}
  (\bibinfo{year}{2001}), \eprint[http://arXiv.org/abs]{hep-ph/0110032}.

\bibitem[{\citenamefont{Chattopadhyay and
  Nath}(2001{\natexlab{b}})}]{Chattopadhyay:2001mj}
\bibinfo{author}{\bibfnamefont{U.}~\bibnamefont{Chattopadhyay}}
  \bibnamefont{and} \bibinfo{author}{\bibfnamefont{P.}~\bibnamefont{Nath}}
  (\bibinfo{year}{2001}{\natexlab{b}}),
  \eprint[http://arXiv.org/abs]{hep-ph/0110341}.

\bibitem[{\citenamefont{Endo and Moroi}(2001)}]{Endo:2001ym}
\bibinfo{author}{\bibfnamefont{M.}~\bibnamefont{Endo}} \bibnamefont{and}
  \bibinfo{author}{\bibfnamefont{T.}~\bibnamefont{Moroi}}
  (\bibinfo{year}{2001}), \eprint[http://arXiv.org/abs]{hep-ph/0110383}.

\bibitem[{\citenamefont{Graesser and Thomas}(2001)}]{Graesser:2001ec}
\bibinfo{author}{\bibfnamefont{M.}~\bibnamefont{Graesser}} \bibnamefont{and}
  \bibinfo{author}{\bibfnamefont{S.}~\bibnamefont{Thomas}}
  (\bibinfo{year}{2001}), \eprint[http://arXiv.org/abs]{hep-ph/0104254}.

\bibitem[{\citenamefont{Djouadi et~al.}(2001)\citenamefont{Djouadi, Drees, and
  Kneur}}]{Djouadi:2001yk}
\bibinfo{author}{\bibfnamefont{A.}~\bibnamefont{Djouadi}},
  \bibinfo{author}{\bibfnamefont{M.}~\bibnamefont{Drees}}, \bibnamefont{and}
  \bibinfo{author}{\bibfnamefont{J.~L.} \bibnamefont{Kneur}},
  \bibinfo{journal}{JHEP} \textbf{\bibinfo{volume}{08}}, \bibinfo{pages}{055}
  (\bibinfo{year}{2001}), \eprint[http://arXiv.org/abs]{hep-ph/0107316}.

\bibitem[{\citenamefont{Feng and Moroi}(2000)}]{Feng:1999hg}
\bibinfo{author}{\bibfnamefont{J.~L.} \bibnamefont{Feng}} \bibnamefont{and}
  \bibinfo{author}{\bibfnamefont{T.}~\bibnamefont{Moroi}},
  \bibinfo{journal}{Phys. Rev.} \textbf{\bibinfo{volume}{D61}},
  \bibinfo{pages}{095004} (\bibinfo{year}{2000}),
  \eprint[http://arXiv.org/abs]{hep-ph/9907319}.

\bibitem[{\citenamefont{Chattopadhyay et~al.}(2000)\citenamefont{Chattopadhyay,
  Ghosh, and Roy}}]{Chattopadhyay:2000ws}
\bibinfo{author}{\bibfnamefont{U.}~\bibnamefont{Chattopadhyay}},
  \bibinfo{author}{\bibfnamefont{D.~K.} \bibnamefont{Ghosh}}, \bibnamefont{and}
  \bibinfo{author}{\bibfnamefont{S.}~\bibnamefont{Roy}},
  \bibinfo{journal}{Phys. Rev.} \textbf{\bibinfo{volume}{D62}},
  \bibinfo{pages}{115001} (\bibinfo{year}{2000}),
  \eprint[http://arXiv.org/abs]{hep-ph/0006049}.

\bibitem[{\citenamefont{Sakharov}(1967)}]{Sakharov:1967dj}
\bibinfo{author}{\bibfnamefont{A.~D.} \bibnamefont{Sakharov}},
  \bibinfo{journal}{Pisma Zh. Eksp. Teor. Fiz.} \textbf{\bibinfo{volume}{5}},
  \bibinfo{pages}{32} (\bibinfo{year}{1967}).

\bibitem[{\citenamefont{Farrar and Shaposhnikov}(1994)}]{Farrar:1994hn}
\bibinfo{author}{\bibfnamefont{G.~R.} \bibnamefont{Farrar}} \bibnamefont{and}
  \bibinfo{author}{\bibfnamefont{M.~E.} \bibnamefont{Shaposhnikov}},
  \bibinfo{journal}{Phys. Rev.} \textbf{\bibinfo{volume}{D50}},
  \bibinfo{pages}{774} (\bibinfo{year}{1994}),
  \eprint[http://arXiv.org/abs]{hep-ph/9305275}.

\bibitem[{\citenamefont{Gavela et~al.}(1994{\natexlab{a}})\citenamefont{Gavela,
  Hernandez, Orloff, and Pene}}]{Gavela:1994ts}
\bibinfo{author}{\bibfnamefont{M.~B.} \bibnamefont{Gavela}},
  \bibinfo{author}{\bibfnamefont{P.}~\bibnamefont{Hernandez}},
  \bibinfo{author}{\bibfnamefont{J.}~\bibnamefont{Orloff}}, \bibnamefont{and}
  \bibinfo{author}{\bibfnamefont{O.}~\bibnamefont{Pene}},
  \bibinfo{journal}{Mod. Phys. Lett.} \textbf{\bibinfo{volume}{A9}},
  \bibinfo{pages}{795} (\bibinfo{year}{1994}{\natexlab{a}}),
  \eprint[http://arXiv.org/abs]{hep-ph/9312215}.

\bibitem[{\citenamefont{Gavela et~al.}(1994{\natexlab{b}})\citenamefont{Gavela,
  Hernandez, Orloff, Pene, and Quimbay}}]{Gavela:1994dt}
\bibinfo{author}{\bibfnamefont{M.~B.} \bibnamefont{Gavela}},
  \bibinfo{author}{\bibfnamefont{P.}~\bibnamefont{Hernandez}},
  \bibinfo{author}{\bibfnamefont{J.}~\bibnamefont{Orloff}},
  \bibinfo{author}{\bibfnamefont{O.}~\bibnamefont{Pene}}, \bibnamefont{and}
  \bibinfo{author}{\bibfnamefont{C.}~\bibnamefont{Quimbay}},
  \bibinfo{journal}{Nucl. Phys.} \textbf{\bibinfo{volume}{B430}},
  \bibinfo{pages}{382} (\bibinfo{year}{1994}{\natexlab{b}}),
  \eprint[http://arXiv.org/abs]{hep-ph/9406289}.

\bibitem[{\citenamefont{Huet and Sather}(1995)}]{Huet:1995jb}
\bibinfo{author}{\bibfnamefont{P.}~\bibnamefont{Huet}} \bibnamefont{and}
  \bibinfo{author}{\bibfnamefont{E.}~\bibnamefont{Sather}},
  \bibinfo{journal}{Phys. Rev.} \textbf{\bibinfo{volume}{D51}},
  \bibinfo{pages}{379} (\bibinfo{year}{1995}),
  \eprint[http://arXiv.org/abs]{hep-ph/9404302}.

\bibitem[{\citenamefont{Hoogeveen}(1990)}]{Hoogeveen:1990cb}
\bibinfo{author}{\bibfnamefont{F.}~\bibnamefont{Hoogeveen}},
  \bibinfo{journal}{Nucl. Phys.} \textbf{\bibinfo{volume}{B341}},
  \bibinfo{pages}{322} (\bibinfo{year}{1990}).

\bibitem[{\citenamefont{Khriplovich}(1986)}]{Khriplovich:1986jr}
\bibinfo{author}{\bibfnamefont{I.~B.} \bibnamefont{Khriplovich}},
  \bibinfo{journal}{Phys. Lett.} \textbf{\bibinfo{volume}{B173}},
  \bibinfo{pages}{193} (\bibinfo{year}{1986}).

\bibitem[{\citenamefont{Commins et~al.}(1994)\citenamefont{Commins, Ross,
  DeMille, and Regan}}]{Commins:1994gv}
\bibinfo{author}{\bibfnamefont{E.~D.} \bibnamefont{Commins}},
  \bibinfo{author}{\bibfnamefont{S.~B.} \bibnamefont{Ross}},
  \bibinfo{author}{\bibfnamefont{D.}~\bibnamefont{DeMille}}, \bibnamefont{and}
  \bibinfo{author}{\bibfnamefont{B.~C.} \bibnamefont{Regan}},
  \bibinfo{journal}{Phys. Rev.} \textbf{\bibinfo{volume}{A50}},
  \bibinfo{pages}{2960} (\bibinfo{year}{1994}).

\bibitem[{\citenamefont{Hinds}(2001)}]{kaon2001}
\bibinfo{author}{\bibfnamefont{E.}~\bibnamefont{Hinds}} (\bibinfo{year}{2001}),
  \bibinfo{note}{talk given at `KAON 2001', Pisa, June 12-17, 2001}.

\bibitem[{\citenamefont{Hinds}()}]{EdHinds}
\bibinfo{author}{\bibfnamefont{E.}~\bibnamefont{Hinds}},
  \bibinfo{note}{informal communication}.

\bibitem[{\citenamefont{Pendlebury and Hinds}(2000)}]{Pendlebury:2000an}
\bibinfo{author}{\bibfnamefont{J.~M.} \bibnamefont{Pendlebury}}
  \bibnamefont{and} \bibinfo{author}{\bibfnamefont{E.~A.} \bibnamefont{Hinds}},
  \bibinfo{journal}{Nucl. Instrum. Meth.} \textbf{\bibinfo{volume}{A440}},
  \bibinfo{pages}{471} (\bibinfo{year}{2000}).

\bibitem[{\citenamefont{Harris et~al.}(1999)}]{Harris:1999jx}
\bibinfo{author}{\bibfnamefont{P.~G.} \bibnamefont{Harris}}
  \bibnamefont{et~al.}, \bibinfo{journal}{Phys. Rev. Lett.}
  \textbf{\bibinfo{volume}{82}}, \bibinfo{pages}{904} (\bibinfo{year}{1999}).

\bibitem[{\citenamefont{Romalis et~al.}(2001)\citenamefont{Romalis, Griffith,
  and Fortson}}]{Romalis:2000mg}
\bibinfo{author}{\bibfnamefont{M.~V.} \bibnamefont{Romalis}},
  \bibinfo{author}{\bibfnamefont{W.~C.} \bibnamefont{Griffith}},
  \bibnamefont{and} \bibinfo{author}{\bibfnamefont{E.~N.}
  \bibnamefont{Fortson}}, \bibinfo{journal}{Phys. Rev. Lett.}
  \textbf{\bibinfo{volume}{86}}, \bibinfo{pages}{2505} (\bibinfo{year}{2001}),
  \eprint[http://arXiv.org/abs]{hep-ex/0012001}.

\bibitem[{\citenamefont{Dimopoulos and Sutter}(1995)}]{Dimopoulos:1995ju}
\bibinfo{author}{\bibfnamefont{S.}~\bibnamefont{Dimopoulos}} \bibnamefont{and}
  \bibinfo{author}{\bibfnamefont{D.~W.} \bibnamefont{Sutter}},
  \bibinfo{journal}{Nucl. Phys.} \textbf{\bibinfo{volume}{B452}},
  \bibinfo{pages}{496} (\bibinfo{year}{1995}),
  \eprint[http://arXiv.org/abs]{hep-ph/9504415}.

\bibitem[{\citenamefont{Ibrahim and Nath}(2001{\natexlab{a}})}]{Ibrahim:2001yv}
\bibinfo{author}{\bibfnamefont{T.}~\bibnamefont{Ibrahim}} \bibnamefont{and}
  \bibinfo{author}{\bibfnamefont{P.}~\bibnamefont{Nath}}
  (\bibinfo{year}{2001}{\natexlab{a}}),
  \eprint[http://arXiv.org/abs]{hep-ph/0107325}.

\bibitem[{\citenamefont{Ellis et~al.}(1982{\natexlab{b}})\citenamefont{Ellis,
  Ferrara, and Nanopoulos}}]{Ellis:1982tk}
\bibinfo{author}{\bibfnamefont{J.~R.} \bibnamefont{Ellis}},
  \bibinfo{author}{\bibfnamefont{S.}~\bibnamefont{Ferrara}}, \bibnamefont{and}
  \bibinfo{author}{\bibfnamefont{D.~V.} \bibnamefont{Nanopoulos}},
  \bibinfo{journal}{Phys. Lett.} \textbf{\bibinfo{volume}{B114}},
  \bibinfo{pages}{231} (\bibinfo{year}{1982}{\natexlab{b}}).

\bibitem[{\citenamefont{Dugan et~al.}(1985)\citenamefont{Dugan, Grinstein, and
  Hall}}]{Dugan:1984qf}
\bibinfo{author}{\bibfnamefont{M.}~\bibnamefont{Dugan}},
  \bibinfo{author}{\bibfnamefont{B.}~\bibnamefont{Grinstein}},
  \bibnamefont{and} \bibinfo{author}{\bibfnamefont{L.~J.} \bibnamefont{Hall}},
  \bibinfo{journal}{Nucl. Phys.} \textbf{\bibinfo{volume}{B255}},
  \bibinfo{pages}{413} (\bibinfo{year}{1985}).

\bibitem[{\citenamefont{Cohen et~al.}(1996)\citenamefont{Cohen, Kaplan, and
  Nelson}}]{Cohen:1996vb}
\bibinfo{author}{\bibfnamefont{A.~G.} \bibnamefont{Cohen}},
  \bibinfo{author}{\bibfnamefont{D.~B.} \bibnamefont{Kaplan}},
  \bibnamefont{and} \bibinfo{author}{\bibfnamefont{A.~E.}
  \bibnamefont{Nelson}}, \bibinfo{journal}{Phys. Lett.}
  \textbf{\bibinfo{volume}{B388}}, \bibinfo{pages}{588} (\bibinfo{year}{1996}),
  \eprint[http://arXiv.org/abs]{hep-ph/9607394}.

\bibitem[{\citenamefont{Bagger et~al.}(1999)\citenamefont{Bagger, Feng, and
  Polonsky}}]{Bagger:1999ty}
\bibinfo{author}{\bibfnamefont{J.}~\bibnamefont{Bagger}},
  \bibinfo{author}{\bibfnamefont{J.~L.} \bibnamefont{Feng}}, \bibnamefont{and}
  \bibinfo{author}{\bibfnamefont{N.}~\bibnamefont{Polonsky}},
  \bibinfo{journal}{Nucl. Phys.} \textbf{\bibinfo{volume}{B563}},
  \bibinfo{pages}{3} (\bibinfo{year}{1999}),
  \eprint[http://arXiv.org/abs]{hep-ph/9905292}.

\bibitem[{\citenamefont{Bagger et~al.}(2000)\citenamefont{Bagger, Feng,
  Polonsky, and Zhang}}]{Bagger:1999sy}
\bibinfo{author}{\bibfnamefont{J.~A.} \bibnamefont{Bagger}},
  \bibinfo{author}{\bibfnamefont{J.~L.} \bibnamefont{Feng}},
  \bibinfo{author}{\bibfnamefont{N.}~\bibnamefont{Polonsky}}, \bibnamefont{and}
  \bibinfo{author}{\bibfnamefont{R.-J.} \bibnamefont{Zhang}},
  \bibinfo{journal}{Phys. Lett.} \textbf{\bibinfo{volume}{B473}},
  \bibinfo{pages}{264} (\bibinfo{year}{2000}),
  \eprint[http://arXiv.org/abs]{hep-ph/9911255}.

\bibitem[{\citenamefont{Feng and Matchev}(2001{\natexlab{c}})}]{Feng:2000bp}
\bibinfo{author}{\bibfnamefont{J.~L.} \bibnamefont{Feng}} \bibnamefont{and}
  \bibinfo{author}{\bibfnamefont{K.~T.} \bibnamefont{Matchev}},
  \bibinfo{journal}{Phys. Rev.} \textbf{\bibinfo{volume}{D63}},
  \bibinfo{pages}{095003} (\bibinfo{year}{2001}{\natexlab{c}}),
  \eprint[http://arXiv.org/abs]{hep-ph/0011356}.

\bibitem[{\citenamefont{Babu et~al.}(2000)\citenamefont{Babu, Dutta, and
  Mohapatra}}]{Babu:1999xf}
\bibinfo{author}{\bibfnamefont{K.~S.} \bibnamefont{Babu}},
  \bibinfo{author}{\bibfnamefont{B.}~\bibnamefont{Dutta}}, \bibnamefont{and}
  \bibinfo{author}{\bibfnamefont{R.~N.} \bibnamefont{Mohapatra}},
  \bibinfo{journal}{Phys. Rev.} \textbf{\bibinfo{volume}{D61}},
  \bibinfo{pages}{091701} (\bibinfo{year}{2000}),
  \eprint[http://arXiv.org/abs]{hep-ph/9905464}.

\bibitem[{\citenamefont{Ibrahim and Nath}(1998)}]{Ibrahim:1998je}
\bibinfo{author}{\bibfnamefont{T.}~\bibnamefont{Ibrahim}} \bibnamefont{and}
  \bibinfo{author}{\bibfnamefont{P.}~\bibnamefont{Nath}},
  \bibinfo{journal}{Phys. Rev.} \textbf{\bibinfo{volume}{D58}},
  \bibinfo{pages}{111301} (\bibinfo{year}{1998}),
  \eprint[http://arXiv.org/abs]{hep-ph/9807501}.

\bibitem[{\citenamefont{Brhlik et~al.}(1999)\citenamefont{Brhlik, Good, and
  Kane}}]{Brhlik:1998zn}
\bibinfo{author}{\bibfnamefont{M.}~\bibnamefont{Brhlik}},
  \bibinfo{author}{\bibfnamefont{G.~J.} \bibnamefont{Good}}, \bibnamefont{and}
  \bibinfo{author}{\bibfnamefont{G.~L.} \bibnamefont{Kane}},
  \bibinfo{journal}{Phys. Rev.} \textbf{\bibinfo{volume}{D59}},
  \bibinfo{pages}{115004} (\bibinfo{year}{1999}),
  \eprint[http://arXiv.org/abs]{hep-ph/9810457}.

\bibitem[{\citenamefont{Manohar and Georgi}(1984)}]{Manohar:1983md}
\bibinfo{author}{\bibfnamefont{A.}~\bibnamefont{Manohar}} \bibnamefont{and}
  \bibinfo{author}{\bibfnamefont{H.}~\bibnamefont{Georgi}},
  \bibinfo{journal}{Nucl. Phys.} \textbf{\bibinfo{volume}{B234}},
  \bibinfo{pages}{189} (\bibinfo{year}{1984}).

\bibitem[{\citenamefont{Pospelov and Ritz}(2001)}]{Pospelov:2000bw}
\bibinfo{author}{\bibfnamefont{M.}~\bibnamefont{Pospelov}} \bibnamefont{and}
  \bibinfo{author}{\bibfnamefont{A.}~\bibnamefont{Ritz}},
  \bibinfo{journal}{Phys. Rev.} \textbf{\bibinfo{volume}{D63}},
  \bibinfo{pages}{073015} (\bibinfo{year}{2001}),
  \eprint[http://arXiv.org/abs]{hep-ph/0010037}.

\bibitem[{\citenamefont{Falk et~al.}(1999)\citenamefont{Falk, Olive, Pospelov,
  and Roiban}}]{Falk:1999tm}
\bibinfo{author}{\bibfnamefont{T.}~\bibnamefont{Falk}},
  \bibinfo{author}{\bibfnamefont{K.~A.} \bibnamefont{Olive}},
  \bibinfo{author}{\bibfnamefont{M.}~\bibnamefont{Pospelov}}, \bibnamefont{and}
  \bibinfo{author}{\bibfnamefont{R.}~\bibnamefont{Roiban}},
  \bibinfo{journal}{Nucl. Phys.} \textbf{\bibinfo{volume}{B560}},
  \bibinfo{pages}{3} (\bibinfo{year}{1999}),
  \eprint[http://arXiv.org/abs]{hep-ph/9904393}.

\bibitem[{\citenamefont{Barger et~al.}(2001{\natexlab{a}})}]{Barger:2001nu}
\bibinfo{author}{\bibfnamefont{V.~D.} \bibnamefont{Barger}}
  \bibnamefont{et~al.}, \bibinfo{journal}{Phys. Rev.}
  \textbf{\bibinfo{volume}{D64}}, \bibinfo{pages}{056007}
  (\bibinfo{year}{2001}{\natexlab{a}}),
  \eprint[http://arXiv.org/abs]{hep-ph/0101106}.

\bibitem[{\citenamefont{Semertzidis et~al.}(1999)}]{Semertzidis:1999kv}
\bibinfo{author}{\bibfnamefont{Y.~K.} \bibnamefont{Semertzidis}}
  \bibnamefont{et~al.} (\bibinfo{year}{1999}),
  \eprint[http://arXiv.org/abs]{hep-ph/0012087}.

\bibitem[{\citenamefont{Bailey et~al.}(1979)}]{Bailey:1979mn}
\bibinfo{author}{\bibfnamefont{J.}~\bibnamefont{Bailey}} \bibnamefont{et~al.}
  (\bibinfo{collaboration}{CERN-Mainz-Daresbury}), \bibinfo{journal}{Nucl.
  Phys.} \textbf{\bibinfo{volume}{B150}}, \bibinfo{pages}{1}
  (\bibinfo{year}{1979}).

\bibitem[{\citenamefont{Aysto et~al.}(2001)}]{Aysto:2001zs}
\bibinfo{author}{\bibfnamefont{J.}~\bibnamefont{Aysto}} \bibnamefont{et~al.}
  (\bibinfo{year}{2001}), \eprint[http://arXiv.org/abs]{hep-ph/0109217}.

\bibitem[{\citenamefont{Feng et~al.}(2001{\natexlab{c}})\citenamefont{Feng,
  Matchev, and Shadmi}}]{Feng:2001sq}
\bibinfo{author}{\bibfnamefont{J.~L.} \bibnamefont{Feng}},
  \bibinfo{author}{\bibfnamefont{K.~T.} \bibnamefont{Matchev}},
  \bibnamefont{and} \bibinfo{author}{\bibfnamefont{Y.}~\bibnamefont{Shadmi}},
  \bibinfo{journal}{Nucl. Phys.} \textbf{\bibinfo{volume}{B613}},
  \bibinfo{pages}{366} (\bibinfo{year}{2001}{\natexlab{c}}),
  \eprint[http://arXiv.org/abs]{hep-ph/0107182}.

\bibitem[{\citenamefont{Romanino and Strumia}(2001)}]{Romanino:2001zf}
\bibinfo{author}{\bibfnamefont{A.}~\bibnamefont{Romanino}} \bibnamefont{and}
  \bibinfo{author}{\bibfnamefont{A.}~\bibnamefont{Strumia}}
  (\bibinfo{year}{2001}), \eprint[http://arXiv.org/abs]{hep-ph/0108275}.

\bibitem[{\citenamefont{Ibrahim and Nath}(2001{\natexlab{b}})}]{Ibrahim:2001jz}
\bibinfo{author}{\bibfnamefont{T.}~\bibnamefont{Ibrahim}} \bibnamefont{and}
  \bibinfo{author}{\bibfnamefont{P.}~\bibnamefont{Nath}},
  \bibinfo{journal}{Phys. Rev.} \textbf{\bibinfo{volume}{D64}},
  \bibinfo{pages}{093002} (\bibinfo{year}{2001}{\natexlab{b}}),
  \eprint[http://arXiv.org/abs]{hep-ph/0105025}.

\bibitem[{\citenamefont{Hagelin et~al.}(1994)\citenamefont{Hagelin, Kelley, and
  Tanaka}}]{Hagelin:1994tc}
\bibinfo{author}{\bibfnamefont{J.~S.} \bibnamefont{Hagelin}},
  \bibinfo{author}{\bibfnamefont{S.}~\bibnamefont{Kelley}}, \bibnamefont{and}
  \bibinfo{author}{\bibfnamefont{T.}~\bibnamefont{Tanaka}},
  \bibinfo{journal}{Nucl. Phys.} \textbf{\bibinfo{volume}{B415}},
  \bibinfo{pages}{293} (\bibinfo{year}{1994}).

\bibitem[{\citenamefont{Gabbiani et~al.}(1996)\citenamefont{Gabbiani,
  Gabrielli, Masiero, and Silvestrini}}]{Gabbiani:1996hi}
\bibinfo{author}{\bibfnamefont{F.}~\bibnamefont{Gabbiani}},
  \bibinfo{author}{\bibfnamefont{E.}~\bibnamefont{Gabrielli}},
  \bibinfo{author}{\bibfnamefont{A.}~\bibnamefont{Masiero}}, \bibnamefont{and}
  \bibinfo{author}{\bibfnamefont{L.}~\bibnamefont{Silvestrini}},
  \bibinfo{journal}{Nucl. Phys.} \textbf{\bibinfo{volume}{B477}},
  \bibinfo{pages}{321} (\bibinfo{year}{1996}),
  \eprint[http://arXiv.org/abs]{hep-ph/9604387}.

\bibitem[{\citenamefont{Bagger et~al.}(1997)\citenamefont{Bagger, Matchev, and
  Zhang}}]{Bagger:1997gg}
\bibinfo{author}{\bibfnamefont{J.~A.} \bibnamefont{Bagger}},
  \bibinfo{author}{\bibfnamefont{K.~T.} \bibnamefont{Matchev}},
  \bibnamefont{and} \bibinfo{author}{\bibfnamefont{R.-J.} \bibnamefont{Zhang}},
  \bibinfo{journal}{Phys. Lett.} \textbf{\bibinfo{volume}{B412}},
  \bibinfo{pages}{77} (\bibinfo{year}{1997}),
  \eprint[http://arXiv.org/abs]{hep-ph/9707225}.

\bibitem[{\citenamefont{Brooks et~al.}(1999)}]{MEGA}
\bibinfo{author}{\bibfnamefont{M.~L.} \bibnamefont{Brooks}}
  \bibnamefont{et~al.} (\bibinfo{collaboration}{MEGA}), \bibinfo{journal}{Phys.
  Rev. Lett.} \textbf{\bibinfo{volume}{83}}, \bibinfo{pages}{1521}
  (\bibinfo{year}{1999}), \eprint[http://arXiv.org/abs]{hep-ex/9905013}.

\bibitem[{\citenamefont{Bellgardt et~al.}(1988)}]{SINDRUM}
\bibinfo{author}{\bibfnamefont{U.}~\bibnamefont{Bellgardt}}
  \bibnamefont{et~al.} (\bibinfo{collaboration}{SINDRUM}),
  \bibinfo{journal}{Nucl. Phys.} \textbf{\bibinfo{volume}{B299}},
  \bibinfo{pages}{1} (\bibinfo{year}{1988}).

\bibitem[{\citenamefont{Kaulard et~al.}(1998)}]{SINDRUMII}
\bibinfo{author}{\bibfnamefont{J.}~\bibnamefont{Kaulard}} \bibnamefont{et~al.}
  (\bibinfo{collaboration}{SINDRUM II}), \bibinfo{journal}{Phys. Lett.}
  \textbf{\bibinfo{volume}{B422}}, \bibinfo{pages}{334} (\bibinfo{year}{1998}).

\bibitem[{\citenamefont{Mori and {\it et al.}}(1999)}]{PSI-proposal}
\bibinfo{author}{\bibfnamefont{T.}~\bibnamefont{Mori}} \bibnamefont{and}
  \bibinfo{author}{\bibnamefont{{\it et al.}}} (\bibinfo{year}{1999}),
  \bibinfo{note}{http://meg.psi.ch/doc/prop/index.html}.

\bibitem[{\citenamefont{Bachman and {\it et al.}}(1997)}]{MECO-proposal}
\bibinfo{author}{\bibfnamefont{M.}~\bibnamefont{Bachman}} \bibnamefont{and}
  \bibinfo{author}{\bibnamefont{{\it et al.}}} (\bibinfo{year}{1997}),
  \bibinfo{note}{http://meco.ps.uci.edu}.

\bibitem[{\citenamefont{Barbieri et~al.}(1995)\citenamefont{Barbieri, Hall, and
  Strumia}}]{Barbieri:1995tw}
\bibinfo{author}{\bibfnamefont{R.}~\bibnamefont{Barbieri}},
  \bibinfo{author}{\bibfnamefont{L.~J.} \bibnamefont{Hall}}, \bibnamefont{and}
  \bibinfo{author}{\bibfnamefont{A.}~\bibnamefont{Strumia}},
  \bibinfo{journal}{Nucl. Phys.} \textbf{\bibinfo{volume}{B445}},
  \bibinfo{pages}{219} (\bibinfo{year}{1995}),
  \eprint[http://arXiv.org/abs]{hep-ph/9501334}.

\bibitem[{\citenamefont{Randall and Sundrum}(1999)}]{Randall:1998uk}
\bibinfo{author}{\bibfnamefont{L.}~\bibnamefont{Randall}} \bibnamefont{and}
  \bibinfo{author}{\bibfnamefont{R.}~\bibnamefont{Sundrum}},
  \bibinfo{journal}{Nucl. Phys.} \textbf{\bibinfo{volume}{B557}},
  \bibinfo{pages}{79} (\bibinfo{year}{1999}),
  \eprint[http://arXiv.org/abs]{hep-th/9810155}.

\bibitem[{\citenamefont{Kaplan et~al.}(2000)\citenamefont{Kaplan, Kribs, and
  Schmaltz}}]{Kaplan:1999ac}
\bibinfo{author}{\bibfnamefont{D.~E.} \bibnamefont{Kaplan}},
  \bibinfo{author}{\bibfnamefont{G.~D.} \bibnamefont{Kribs}}, \bibnamefont{and}
  \bibinfo{author}{\bibfnamefont{M.}~\bibnamefont{Schmaltz}},
  \bibinfo{journal}{Phys. Rev.} \textbf{\bibinfo{volume}{D62}},
  \bibinfo{pages}{035010} (\bibinfo{year}{2000}),
  \eprint[http://arXiv.org/abs]{hep-ph/9911293}.

\bibitem[{\citenamefont{Chacko et~al.}(2000)\citenamefont{Chacko, Luty, Nelson,
  and Ponton}}]{Chacko:1999mi}
\bibinfo{author}{\bibfnamefont{Z.}~\bibnamefont{Chacko}},
  \bibinfo{author}{\bibfnamefont{M.~A.} \bibnamefont{Luty}},
  \bibinfo{author}{\bibfnamefont{A.~E.} \bibnamefont{Nelson}},
  \bibnamefont{and} \bibinfo{author}{\bibfnamefont{E.}~\bibnamefont{Ponton}},
  \bibinfo{journal}{JHEP} \textbf{\bibinfo{volume}{01}}, \bibinfo{pages}{003}
  (\bibinfo{year}{2000}), \eprint[http://arXiv.org/abs]{hep-ph/9911323}.

\bibitem[{\citenamefont{Ahmed et~al.}(2000)}]{CLEO}
\bibinfo{author}{\bibfnamefont{S.}~\bibnamefont{Ahmed}} \bibnamefont{et~al.}
  (\bibinfo{collaboration}{CLEO}), \bibinfo{journal}{Phys. Rev.}
  \textbf{\bibinfo{volume}{D61}}, \bibinfo{pages}{071101}
  (\bibinfo{year}{2000}), \eprint[http://arXiv.org/abs]{hep-ex/9910060}.

\bibitem[{\citenamefont{Dine et~al.}(2001)\citenamefont{Dine, Grossman, and
  Thomas}}]{Dine:2001cf}
\bibinfo{author}{\bibfnamefont{M.}~\bibnamefont{Dine}},
  \bibinfo{author}{\bibfnamefont{Y.}~\bibnamefont{Grossman}}, \bibnamefont{and}
  \bibinfo{author}{\bibfnamefont{S.}~\bibnamefont{Thomas}}
  (\bibinfo{year}{2001}), \eprint[http://arXiv.org/abs]{hep-ph/0111154}.

\bibitem[{\citenamefont{Krasnikov}(1996)}]{Krasnikov:1996qq}
\bibinfo{author}{\bibfnamefont{N.~V.} \bibnamefont{Krasnikov}},
  \bibinfo{journal}{Phys. Lett.} \textbf{\bibinfo{volume}{B388}},
  \bibinfo{pages}{783} (\bibinfo{year}{1996}),
  \eprint[http://arXiv.org/abs]{hep-ph/9511464}.

\bibitem[{\citenamefont{Arkani-Hamed et~al.}(1996)\citenamefont{Arkani-Hamed,
  Cheng, Feng, and Hall}}]{Arkani-Hamed:1996au}
\bibinfo{author}{\bibfnamefont{N.}~\bibnamefont{Arkani-Hamed}},
  \bibinfo{author}{\bibfnamefont{H.-C.} \bibnamefont{Cheng}},
  \bibinfo{author}{\bibfnamefont{J.~L.} \bibnamefont{Feng}}, \bibnamefont{and}
  \bibinfo{author}{\bibfnamefont{L.~J.} \bibnamefont{Hall}},
  \bibinfo{journal}{Phys. Rev. Lett.} \textbf{\bibinfo{volume}{77}},
  \bibinfo{pages}{1937} (\bibinfo{year}{1996}),
  \eprint[http://arXiv.org/abs]{hep-ph/9603431}.

\bibitem[{\citenamefont{Arkani-Hamed et~al.}(1997)\citenamefont{Arkani-Hamed,
  Feng, Hall, and Cheng}}]{Arkani-Hamed:1997km}
\bibinfo{author}{\bibfnamefont{N.}~\bibnamefont{Arkani-Hamed}},
  \bibinfo{author}{\bibfnamefont{J.~L.} \bibnamefont{Feng}},
  \bibinfo{author}{\bibfnamefont{L.~J.} \bibnamefont{Hall}}, \bibnamefont{and}
  \bibinfo{author}{\bibfnamefont{H.-C.} \bibnamefont{Cheng}},
  \bibinfo{journal}{Nucl. Phys.} \textbf{\bibinfo{volume}{B505}},
  \bibinfo{pages}{3} (\bibinfo{year}{1997}),
  \eprint[http://arXiv.org/abs]{hep-ph/9704205}.

\bibitem[{\citenamefont{Ammar and {\it et
  al.}~(CLEO~collaboration)}(1993)}]{cleoa}
\bibinfo{author}{\bibfnamefont{R.}~\bibnamefont{Ammar}} \bibnamefont{and}
  \bibinfo{author}{\bibnamefont{{\it et al.}~(CLEO~collaboration)}},
  \bibinfo{journal}{Phys.\ Rev. \ Lett.} \textbf{\bibinfo{volume}{70}},
  \bibinfo{pages}{138} (\bibinfo{year}{1993}).

\bibitem[{\citenamefont{Burchat and {\it et al.}}()}]{burchat}
\bibinfo{author}{\bibfnamefont{P.}~\bibnamefont{Burchat}} \bibnamefont{and}
  \bibinfo{author}{\bibnamefont{{\it et al.}}}, \eprint{Physics at $10^{36}$,
  Proceedings of Snowmass Wokshop}.

\bibitem[{\citenamefont{Chetyrkin et~al.}(1997)\citenamefont{Chetyrkin, Misiak,
  and Munz}}]{misiak1}
\bibinfo{author}{\bibfnamefont{K.}~\bibnamefont{Chetyrkin}},
  \bibinfo{author}{\bibfnamefont{M.}~\bibnamefont{Misiak}}, \bibnamefont{and}
  \bibinfo{author}{\bibfnamefont{M.}~\bibnamefont{Munz}},
  \bibinfo{journal}{Phys.\ Lett.} \textbf{\bibinfo{volume}{B400}},
  \bibinfo{pages}{206} (\bibinfo{year}{1997}).

\bibitem[{\citenamefont{Gambino and Misiak}(2001)}]{misiak2}
\bibinfo{author}{\bibfnamefont{P.}~\bibnamefont{Gambino}} \bibnamefont{and}
  \bibinfo{author}{\bibfnamefont{M.}~\bibnamefont{Misiak}},
  \bibinfo{journal}{Nucl.\ Phys.} \textbf{\bibinfo{volume}{B611}},
  \bibinfo{pages}{338} (\bibinfo{year}{2001}).

\bibitem[{\citenamefont{Chen and {\it et al.}~(CLEO~collaboration)}()}]{cleo1}
\bibinfo{author}{\bibfnamefont{S.}~\bibnamefont{Chen}} \bibnamefont{and}
  \bibinfo{author}{\bibnamefont{{\it et al.}~(CLEO~collaboration)}},
  \eprint{hep-ex/0108032}.

\bibitem[{\citenamefont{Hewett and Wells}(1996)}]{hewett1}
\bibinfo{author}{\bibfnamefont{J.}~\bibnamefont{Hewett}} \bibnamefont{and}
  \bibinfo{author}{\bibfnamefont{J.}~\bibnamefont{Wells}},
  \bibinfo{journal}{Phys.\ Rev.} \textbf{\bibinfo{volume}{D55}},
  \bibinfo{pages}{5549} (\bibinfo{year}{1996}).

\bibitem[{\citenamefont{Hewett}()}]{hewett2}
\bibinfo{author}{\bibfnamefont{J.}~\bibnamefont{Hewett}}, \eprint{Private \
  communication}.

\bibitem[{\citenamefont{Bosch and Buchalla}()}]{bosch}
\bibinfo{author}{\bibfnamefont{S.}~\bibnamefont{Bosch}} \bibnamefont{and}
  \bibinfo{author}{\bibfnamefont{G.}~\bibnamefont{Buchalla}},
  \eprint{hep-ph/0108081}.

\bibitem[{\citenamefont{Beneke et~al.}(2001)\citenamefont{Beneke, Feldmann, and
  Seidel}}]{beneke}
\bibinfo{author}{\bibfnamefont{M.}~\bibnamefont{Beneke}},
  \bibinfo{author}{\bibfnamefont{T.}~\bibnamefont{Feldmann}}, \bibnamefont{and}
  \bibinfo{author}{\bibfnamefont{D.}~\bibnamefont{Seidel}},
  \bibinfo{journal}{Nucl.\ Phys.} \textbf{\bibinfo{volume}{B612}},
  \bibinfo{pages}{25} (\bibinfo{year}{2001}).

\bibitem[{\citenamefont{Ryd and {\it et al.}~(BABAR~collaboration)}()}]{babar1}
\bibinfo{author}{\bibfnamefont{A.}~\bibnamefont{Ryd}} \bibnamefont{and}
  \bibinfo{author}{\bibnamefont{{\it et al.}~(BABAR~collaboration)}},
  \eprint{Proc.\ of \ Int. \ Symp. \ on \ Heavy \ Flavor \ Physics \ 9}.

\bibitem[{\citenamefont{Atwood and {\it et al.}}()}]{tev}
\bibinfo{author}{\bibfnamefont{D.}~\bibnamefont{Atwood}} \bibnamefont{and}
  \bibinfo{author}{\bibnamefont{{\it et al.}}}, \eprint{Fermilab Pub 01/197}.

\bibitem[{\citenamefont{Ball and {\it et al.}}()}]{lhc}
\bibinfo{author}{\bibfnamefont{P.}~\bibnamefont{Ball}} \bibnamefont{and}
  \bibinfo{author}{\bibnamefont{{\it et al.}}}, \eprint{hep-ph/0003238}.

\bibitem[{\citenamefont{Ali et~al.}(1991)\citenamefont{Ali, Mannel, and
  Morozumi}}]{ali1}
\bibinfo{author}{\bibfnamefont{A.}~\bibnamefont{Ali}},
  \bibinfo{author}{\bibfnamefont{T.}~\bibnamefont{Mannel}}, \bibnamefont{and}
  \bibinfo{author}{\bibfnamefont{T.}~\bibnamefont{Morozumi}},
  \bibinfo{journal}{Phys.\ Lett.} \textbf{\bibinfo{volume}{B273}},
  \bibinfo{pages}{505} (\bibinfo{year}{1991}).

\bibitem[{\citenamefont{Misiak}(1993)}]{misiak3}
\bibinfo{author}{\bibfnamefont{M.}~\bibnamefont{Misiak}},
  \bibinfo{journal}{Nucl.\ Phys.} \textbf{\bibinfo{volume}{B393}},
  \bibinfo{pages}{23} (\bibinfo{year}{1993}).

\bibitem[{\citenamefont{Buras and Munz}(1995)}]{buras1}
\bibinfo{author}{\bibfnamefont{A.}~\bibnamefont{Buras}} \bibnamefont{and}
  \bibinfo{author}{\bibfnamefont{M.}~\bibnamefont{Munz}},
  \bibinfo{journal}{Phys.\ Rev.} \textbf{\bibinfo{volume}{D52}},
  \bibinfo{pages}{186} (\bibinfo{year}{1995}).

\bibitem[{\citenamefont{Coan and {\it et
  al.}~(CLEO~collaboration)}(1997)}]{cleo4}
\bibinfo{author}{\bibfnamefont{T.}~\bibnamefont{Coan}} \bibnamefont{and}
  \bibinfo{author}{\bibnamefont{{\it et al.}~(CLEO~collaboration)}},
  \bibinfo{journal}{Phys.\ Rev. \ Lett.} \textbf{\bibinfo{volume}{80}},
  \bibinfo{pages}{2289} (\bibinfo{year}{1997}).

\bibitem[{\citenamefont{Melikhov et~al.}(1997)\citenamefont{Melikhov, Nikitin,
  and Simula}}]{qm}
\bibinfo{author}{\bibfnamefont{D.}~\bibnamefont{Melikhov}},
  \bibinfo{author}{\bibfnamefont{N.}~\bibnamefont{Nikitin}}, \bibnamefont{and}
  \bibinfo{author}{\bibfnamefont{S.}~\bibnamefont{Simula}},
  \bibinfo{journal}{Phys.\ Lett.} \textbf{\bibinfo{volume}{B410}},
  \bibinfo{pages}{290} (\bibinfo{year}{1997}).

\bibitem[{\citenamefont{Ali and {\it et al.}}(2000)}]{lcsm}
\bibinfo{author}{\bibfnamefont{A.}~\bibnamefont{Ali}} \bibnamefont{and}
  \bibinfo{author}{\bibnamefont{{\it et al.}}}, \bibinfo{journal}{Phys.\ Rev.}
  \textbf{\bibinfo{volume}{D61}}, \bibinfo{pages}{074024}
  (\bibinfo{year}{2000}).

\bibitem[{\citenamefont{Abe and {\it et al.}~(BELLE~collaboration)}()}]{belle3}
\bibinfo{author}{\bibfnamefont{K.}~\bibnamefont{Abe}} \bibnamefont{and}
  \bibinfo{author}{\bibnamefont{{\it et al.}~(BELLE~collaboration)}},
  \eprint{hep-ex/0109026}.

\bibitem[{\citenamefont{Grossman
  et~al.}(1996{\natexlab{a}})\citenamefont{Grossman, Ligeti, and Nardi}}]{gln}
\bibinfo{author}{\bibfnamefont{Y.}~\bibnamefont{Grossman}},
  \bibinfo{author}{\bibfnamefont{Z.}~\bibnamefont{Ligeti}}, \bibnamefont{and}
  \bibinfo{author}{\bibfnamefont{E.}~\bibnamefont{Nardi}},
  \bibinfo{journal}{Nucl.\ Phys.} \textbf{\bibinfo{volume}{B465}},
  \bibinfo{pages}{369} (\bibinfo{year}{1996}{\natexlab{a}}).

\bibitem[{\citenamefont{Grossman
  et~al.}(1996{\natexlab{b}})\citenamefont{Grossman, Ligeti, and Nardi}}]{gln2}
\bibinfo{author}{\bibfnamefont{Y.}~\bibnamefont{Grossman}},
  \bibinfo{author}{\bibfnamefont{Z.}~\bibnamefont{Ligeti}}, \bibnamefont{and}
  \bibinfo{author}{\bibfnamefont{E.}~\bibnamefont{Nardi}},
  \bibinfo{journal}{Nucl.\ Phys.} \textbf{\bibinfo{volume}{B480}},
  \bibinfo{pages}{753} (\bibinfo{year}{1996}{\natexlab{b}}).

\bibitem[{\citenamefont{Ali et~al.}(1995)\citenamefont{Ali, Giudice, and
  Mannel}}]{ali3}
\bibinfo{author}{\bibfnamefont{A.}~\bibnamefont{Ali}},
  \bibinfo{author}{\bibfnamefont{G.}~\bibnamefont{Giudice}}, \bibnamefont{and}
  \bibinfo{author}{\bibfnamefont{T.}~\bibnamefont{Mannel}},
  \bibinfo{journal}{Z. \ Phys.} \textbf{\bibinfo{volume}{C67}},
  \bibinfo{pages}{417} (\bibinfo{year}{1995}).

\bibitem[{\citenamefont{Buras and Fleischer}(1998)}]{BF}
\bibinfo{author}{\bibfnamefont{A.}~\bibnamefont{Buras}} \bibnamefont{and}
  \bibinfo{author}{\bibfnamefont{R.}~\bibnamefont{Fleischer}},
  \bibinfo{journal}{Adv.\ Ser.\ Direct.\ High Energy Phys.}
  \textbf{\bibinfo{volume}{15}}, \bibinfo{pages}{65} (\bibinfo{year}{1998}).

\bibitem[{\citenamefont{Aubert and {\it et
  al.}~(BABAR~collaboration)}(2001)}]{bbr}
\bibinfo{author}{\bibfnamefont{B.}~\bibnamefont{Aubert}} \bibnamefont{and}
  \bibinfo{author}{\bibnamefont{{\it et al.}~(BABAR~collaboration)}},
  \bibinfo{journal}{Phys.\ Rev.\ Lett.} \textbf{\bibinfo{volume}{87}},
  \bibinfo{pages}{091801} (\bibinfo{year}{2001}).

\bibitem[{\citenamefont{Abe and {\it et
  al.}~(BELLE~collaboration)}(2001)}]{bel}
\bibinfo{author}{\bibfnamefont{K.}~\bibnamefont{Abe}} \bibnamefont{and}
  \bibinfo{author}{\bibnamefont{{\it et al.}~(BELLE~collaboration)}},
  \bibinfo{journal}{Phys.\ Rev.\ Lett.} \textbf{\bibinfo{volume}{87}},
  \bibinfo{pages}{091802} (\bibinfo{year}{2001}).

\bibitem[{\citenamefont{Aubert and {\it et
  al.}~(BABAR~collaboration)}({\natexlab{a}})}]{bbr2}
\bibinfo{author}{\bibfnamefont{B.}~\bibnamefont{Aubert}} \bibnamefont{and}
  \bibinfo{author}{\bibnamefont{{\it et al.}~(BABAR~collaboration)}},
  \eprint{hep-ph/0107074}.

\bibitem[{\citenamefont{Soares and Wolfenstein}(1995)}]{SW}
\bibinfo{author}{\bibfnamefont{J.}~\bibnamefont{Soares}} \bibnamefont{and}
  \bibinfo{author}{\bibfnamefont{L.}~\bibnamefont{Wolfenstein}},
  \bibinfo{journal}{Phys.\ Rev.} \textbf{\bibinfo{volume}{D47}},
  \bibinfo{pages}{1021} (\bibinfo{year}{1995}).

\bibitem[{\citenamefont{Grossman et~al.}(1997)\citenamefont{Grossman, Nir, and
  Worah}}]{nir}
\bibinfo{author}{\bibfnamefont{Y.}~\bibnamefont{Grossman}},
  \bibinfo{author}{\bibfnamefont{Y.}~\bibnamefont{Nir}}, \bibnamefont{and}
  \bibinfo{author}{\bibfnamefont{M.}~\bibnamefont{Worah}},
  \bibinfo{journal}{Phys.\ Lett.} \textbf{\bibinfo{volume}{B407}},
  \bibinfo{pages}{307} (\bibinfo{year}{1997}).

\bibitem[{\citenamefont{Boutigny and {\it et
  al.}~(BABAR~collaboration)}()}]{bbrpb}
\bibinfo{author}{\bibfnamefont{D.}~\bibnamefont{Boutigny}} \bibnamefont{and}
  \bibinfo{author}{\bibnamefont{{\it et al.}~(BABAR~collaboration)}},
  \eprint{BABAR physics book, SLAC-PUB-504 (1998).}

\bibitem[{\citenamefont{Aubert and {\it et
  al.}~(BABAR~collaboration)}({\natexlab{b}})}]{bbr3}
\bibinfo{author}{\bibfnamefont{B.}~\bibnamefont{Aubert}} \bibnamefont{and}
  \bibinfo{author}{\bibnamefont{{\it et al.}~(BABAR~collaboration)}},
  \eprint{hep-ph/0105001}.

\bibitem[{\citenamefont{Briere and {\it et
  al.}~(CLEO~collaboration)}(2001)}]{cleob}
\bibinfo{author}{\bibfnamefont{R.}~\bibnamefont{Briere}} \bibnamefont{and}
  \bibinfo{author}{\bibnamefont{{\it et al.}~(CLEO~collaboration)}},
  \bibinfo{journal}{Phys.\ Rev. \ Lett.} \textbf{\bibinfo{volume}{86}},
  \bibinfo{pages}{3718} (\bibinfo{year}{2001}).

\bibitem[{\citenamefont{Gronau and London}(1990)}]{gronau}
\bibinfo{author}{\bibfnamefont{M.}~\bibnamefont{Gronau}} \bibnamefont{and}
  \bibinfo{author}{\bibfnamefont{D.}~\bibnamefont{London}},
  \bibinfo{journal}{Phys.\ Rev.\ Lett.} \textbf{\bibinfo{volume}{65}},
  \bibinfo{pages}{3381} (\bibinfo{year}{1990}).

\bibitem[{\citenamefont{Atwood et~al.}(1997)\citenamefont{Atwood, Dunietz, and
  Soni}}]{soni}
\bibinfo{author}{\bibfnamefont{D.}~\bibnamefont{Atwood}},
  \bibinfo{author}{\bibfnamefont{I.}~\bibnamefont{Dunietz}}, \bibnamefont{and}
  \bibinfo{author}{\bibfnamefont{A.}~\bibnamefont{Soni}},
  \bibinfo{journal}{Phys.\ Rev.\ Lett.} \textbf{\bibinfo{volume}{78}},
  \bibinfo{pages}{3257} (\bibinfo{year}{1997}).

\bibitem[{\citenamefont{Dunietz and Sachs}(1988)}]{sachs}
\bibinfo{author}{\bibfnamefont{I.}~\bibnamefont{Dunietz}} \bibnamefont{and}
  \bibinfo{author}{\bibfnamefont{R.~G.} \bibnamefont{Sachs}},
  \bibinfo{journal}{Phys.\ Rev.} \textbf{\bibinfo{volume}{D37}},
  \bibinfo{pages}{3186} (\bibinfo{year}{1988}).

\bibitem[{\citenamefont{Aleksan et~al.}(1992)\citenamefont{Aleksan, I.Dunietz,
  and Kayser}}]{aleksan}
\bibinfo{author}{\bibfnamefont{R.}~\bibnamefont{Aleksan}},
  \bibinfo{author}{\bibnamefont{I.Dunietz}}, \bibnamefont{and}
  \bibinfo{author}{\bibfnamefont{B.}~\bibnamefont{Kayser}},
  \bibinfo{journal}{Z.\ Phys.} \textbf{\bibinfo{volume}{C54}},
  \bibinfo{pages}{653} (\bibinfo{year}{1992}).

\bibitem[{\citenamefont{I.Dunietz}(1998)}]{dunietz}
\bibinfo{author}{\bibnamefont{I.Dunietz}}, \bibinfo{journal}{Phys.\ Lett.}
  \textbf{\bibinfo{volume}{B427}}, \bibinfo{pages}{179} (\bibinfo{year}{1998}).

\bibitem[{\citenamefont{Soares}(1991)}]{soares}
\bibinfo{author}{\bibfnamefont{J.}~\bibnamefont{Soares}},
  \bibinfo{journal}{Nucl. \ Phys.} \textbf{\bibinfo{volume}{B367}},
  \bibinfo{pages}{575} (\bibinfo{year}{1991}).

\bibitem[{\citenamefont{Kagan and Neubert}(1998)}]{kagan}
\bibinfo{author}{\bibfnamefont{A.}~\bibnamefont{Kagan}} \bibnamefont{and}
  \bibinfo{author}{\bibfnamefont{M.}~\bibnamefont{Neubert}},
  \bibinfo{journal}{Phys. \ Rev.} \textbf{\bibinfo{volume}{D58}},
  \bibinfo{pages}{094012} (\bibinfo{year}{1998}).

\bibitem[{\citenamefont{Coan and {\it et
  al.}~(CLEO~collaboration)}(2001)}]{cleo3}
\bibinfo{author}{\bibfnamefont{T.}~\bibnamefont{Coan}} \bibnamefont{and}
  \bibinfo{author}{\bibnamefont{{\it et al.}~(CLEO~collaboration)}},
  \bibinfo{journal}{Phys.\ Rev. \ Lett.} \textbf{\bibinfo{volume}{86}},
  \bibinfo{pages}{5661} (\bibinfo{year}{2001}).

\bibitem[{\citenamefont{Abe et~al.}(1998)}]{CDFbmumu}
\bibinfo{author}{\bibfnamefont{F.}~\bibnamefont{Abe}} \bibnamefont{et~al.}
  (\bibinfo{collaboration}{CDF}), \bibinfo{journal}{Phys. Rev.}
  \textbf{\bibinfo{volume}{D57}}, \bibinfo{pages}{3811} (\bibinfo{year}{1998}).

\bibitem[{\citenamefont{Bartl et~al.}(2001)}]{Bartl:2001wc}
\bibinfo{author}{\bibfnamefont{A.}~\bibnamefont{Bartl}} \bibnamefont{et~al.},
  \bibinfo{journal}{Phys. Rev.} \textbf{\bibinfo{volume}{D64}},
  \bibinfo{pages}{076009} (\bibinfo{year}{2001}),
  \eprint[http://arXiv.org/abs]{hep-ph/0103324}.

\bibitem[{\citenamefont{Babu and Kolda}(2000)}]{Babu:1999hn}
\bibinfo{author}{\bibfnamefont{K.~S.} \bibnamefont{Babu}} \bibnamefont{and}
  \bibinfo{author}{\bibfnamefont{C.}~\bibnamefont{Kolda}},
  \bibinfo{journal}{Phys. Rev. Lett.} \textbf{\bibinfo{volume}{84}},
  \bibinfo{pages}{228} (\bibinfo{year}{2000}),
  \eprint[http://arXiv.org/abs]{hep-ph/9909476}.

\bibitem[{\citenamefont{Huang et~al.}(1999)\citenamefont{Huang, Liao, and
  Yan}}]{Huang:1998vb}
\bibinfo{author}{\bibfnamefont{C.-S.} \bibnamefont{Huang}},
  \bibinfo{author}{\bibfnamefont{W.}~\bibnamefont{Liao}}, \bibnamefont{and}
  \bibinfo{author}{\bibfnamefont{Q.-S.} \bibnamefont{Yan}},
  \bibinfo{journal}{Phys. Rev.} \textbf{\bibinfo{volume}{D59}},
  \bibinfo{pages}{011701} (\bibinfo{year}{1999}),
  \eprint[http://arXiv.org/abs]{hep-ph/9803460}.

\bibitem[{\citenamefont{Chankowski and Slawianowska}(2001)}]{Chankowski}
\bibinfo{author}{\bibfnamefont{P.~H.} \bibnamefont{Chankowski}}
  \bibnamefont{and}
  \bibinfo{author}{\bibfnamefont{L.}~\bibnamefont{Slawianowska}},
  \bibinfo{journal}{Phys. Rev.} \textbf{\bibinfo{volume}{D63}},
  \bibinfo{pages}{054012} (\bibinfo{year}{2001}),
  \eprint[http://arXiv.org/abs]{hep-ph/0008046}.

\bibitem[{\citenamefont{Bobeth et~al.}(2001)\citenamefont{Bobeth, Ewerth,
  Kruger, and Urban}}]{Urban}
\bibinfo{author}{\bibfnamefont{C.}~\bibnamefont{Bobeth}},
  \bibinfo{author}{\bibfnamefont{T.}~\bibnamefont{Ewerth}},
  \bibinfo{author}{\bibfnamefont{F.}~\bibnamefont{Kruger}}, \bibnamefont{and}
  \bibinfo{author}{\bibfnamefont{J.}~\bibnamefont{Urban}},
  \bibinfo{journal}{Phys. Rev.} \textbf{\bibinfo{volume}{D64}},
  \bibinfo{pages}{074014} (\bibinfo{year}{2001}),
  \eprint[http://arXiv.org/abs]{hep-ph/0104284}.

\bibitem[{\citenamefont{Isidori and Retico}(2001)}]{Isidori}
\bibinfo{author}{\bibfnamefont{G.}~\bibnamefont{Isidori}} \bibnamefont{and}
  \bibinfo{author}{\bibfnamefont{A.}~\bibnamefont{Retico}},
  \bibinfo{journal}{JHEP} \textbf{\bibinfo{volume}{11}}, \bibinfo{pages}{001}
  (\bibinfo{year}{2001}), \eprint[http://arXiv.org/abs]{hep-ph/0110121}.

\bibitem[{\citenamefont{Dedes et~al.}(2001)\citenamefont{Dedes, Dreiner, and
  Nierste}}]{Dedes:2001fv}
\bibinfo{author}{\bibfnamefont{A.}~\bibnamefont{Dedes}},
  \bibinfo{author}{\bibfnamefont{H.~K.} \bibnamefont{Dreiner}},
  \bibnamefont{and} \bibinfo{author}{\bibfnamefont{U.}~\bibnamefont{Nierste}}
  (\bibinfo{year}{2001}), \eprint[http://arXiv.org/abs]{hep-ph/0108037}.

\bibitem[{\citenamefont{Atwood and {\it et al.}}(preliminary)}]{fnalbrep}
\bibinfo{author}{\bibfnamefont{D.}~\bibnamefont{Atwood}} \bibnamefont{and}
  \bibinfo{author}{\bibnamefont{{\it et al.}}}, \emph{\bibinfo{title}{B physics
  at the tevatron: Run-ii and beyond}} (\bibinfo{year}{preliminary}),
  \bibinfo{note}{fERMILAB-Pub-01/197}.

\bibitem[{\citenamefont{Ambrosanio et~al.}(2001)\citenamefont{Ambrosanio,
  Dedes, Heinemeyer, Su, and Weiglein}}]{ASBS}
\bibinfo{author}{\bibfnamefont{S.}~\bibnamefont{Ambrosanio}},
  \bibinfo{author}{\bibfnamefont{A.}~\bibnamefont{Dedes}},
  \bibinfo{author}{\bibfnamefont{S.}~\bibnamefont{Heinemeyer}},
  \bibinfo{author}{\bibfnamefont{S.}~\bibnamefont{Su}}, \bibnamefont{and}
  \bibinfo{author}{\bibfnamefont{G.}~\bibnamefont{Weiglein}}
  (\bibinfo{year}{2001}), \eprint[http://arXiv.org/abs]{hep-ph/0106255}.

\bibitem[{\citenamefont{Feng et~al.}(2000{\natexlab{a}})\citenamefont{Feng,
  Matchev, and Moroi}}]{Feng:1999mn}
\bibinfo{author}{\bibfnamefont{J.~L.} \bibnamefont{Feng}},
  \bibinfo{author}{\bibfnamefont{K.~T.} \bibnamefont{Matchev}},
  \bibnamefont{and} \bibinfo{author}{\bibfnamefont{T.}~\bibnamefont{Moroi}},
  \bibinfo{journal}{Phys. Rev. Lett.} \textbf{\bibinfo{volume}{84}},
  \bibinfo{pages}{2322} (\bibinfo{year}{2000}{\natexlab{a}}),
  \eprint[http://arXiv.org/abs]{hep-ph/9908309}.

\bibitem[{\citenamefont{Feng et~al.}(2000{\natexlab{b}})\citenamefont{Feng,
  Matchev, and Moroi}}]{Feng:1999zg}
\bibinfo{author}{\bibfnamefont{J.~L.} \bibnamefont{Feng}},
  \bibinfo{author}{\bibfnamefont{K.~T.} \bibnamefont{Matchev}},
  \bibnamefont{and} \bibinfo{author}{\bibfnamefont{T.}~\bibnamefont{Moroi}},
  \bibinfo{journal}{Phys. Rev.} \textbf{\bibinfo{volume}{D61}},
  \bibinfo{pages}{075005} (\bibinfo{year}{2000}{\natexlab{b}}),
  \eprint[http://arXiv.org/abs]{hep-ph/9909334}.

\bibitem[{\citenamefont{Feng et~al.}(2000{\natexlab{c}})\citenamefont{Feng,
  Matchev, and Wilczek}}]{Feng:2000gh}
\bibinfo{author}{\bibfnamefont{J.~L.} \bibnamefont{Feng}},
  \bibinfo{author}{\bibfnamefont{K.~T.} \bibnamefont{Matchev}},
  \bibnamefont{and} \bibinfo{author}{\bibfnamefont{F.}~\bibnamefont{Wilczek}},
  \bibinfo{journal}{Phys. Lett.} \textbf{\bibinfo{volume}{B482}},
  \bibinfo{pages}{388} (\bibinfo{year}{2000}{\natexlab{c}}),
  \eprint[http://arXiv.org/abs]{hep-ph/0004043}.

\bibitem[{\citenamefont{Ellis et~al.}(1998)\citenamefont{Ellis, Falk, and
  Olive}}]{Ellis:1998kh}
\bibinfo{author}{\bibfnamefont{J.~R.} \bibnamefont{Ellis}},
  \bibinfo{author}{\bibfnamefont{T.}~\bibnamefont{Falk}}, \bibnamefont{and}
  \bibinfo{author}{\bibfnamefont{K.~A.} \bibnamefont{Olive}},
  \bibinfo{journal}{Phys. Lett.} \textbf{\bibinfo{volume}{B444}},
  \bibinfo{pages}{367} (\bibinfo{year}{1998}),
  \eprint[http://arXiv.org/abs]{hep-ph/9810360}.

\bibitem[{\citenamefont{Ellis et~al.}(2000{\natexlab{a}})\citenamefont{Ellis,
  Falk, Olive, and Srednicki}}]{Ellis:1999mm}
\bibinfo{author}{\bibfnamefont{J.~R.} \bibnamefont{Ellis}},
  \bibinfo{author}{\bibfnamefont{T.}~\bibnamefont{Falk}},
  \bibinfo{author}{\bibfnamefont{K.~A.} \bibnamefont{Olive}}, \bibnamefont{and}
  \bibinfo{author}{\bibfnamefont{M.}~\bibnamefont{Srednicki}},
  \bibinfo{journal}{Astropart. Phys.} \textbf{\bibinfo{volume}{13}},
  \bibinfo{pages}{181} (\bibinfo{year}{2000}{\natexlab{a}}),
  \eprint[http://arXiv.org/abs]{hep-ph/9905481}.

\bibitem[{\citenamefont{Gomez et~al.}(2000)\citenamefont{Gomez, Lazarides, and
  Pallis}}]{Gomez:1999dk}
\bibinfo{author}{\bibfnamefont{M.~E.} \bibnamefont{Gomez}},
  \bibinfo{author}{\bibfnamefont{G.}~\bibnamefont{Lazarides}},
  \bibnamefont{and} \bibinfo{author}{\bibfnamefont{C.}~\bibnamefont{Pallis}},
  \bibinfo{journal}{Phys. Rev.} \textbf{\bibinfo{volume}{D61}},
  \bibinfo{pages}{123512} (\bibinfo{year}{2000}),
  \eprint[http://arXiv.org/abs]{hep-ph/9907261}.

\bibitem[{\citenamefont{Ellis et~al.}(2001{\natexlab{c}})\citenamefont{Ellis,
  Falk, Ganis, Olive, and Srednicki}}]{Ellis:2001ms}
\bibinfo{author}{\bibfnamefont{J.~R.} \bibnamefont{Ellis}},
  \bibinfo{author}{\bibfnamefont{T.}~\bibnamefont{Falk}},
  \bibinfo{author}{\bibfnamefont{G.}~\bibnamefont{Ganis}},
  \bibinfo{author}{\bibfnamefont{K.~A.} \bibnamefont{Olive}}, \bibnamefont{and}
  \bibinfo{author}{\bibfnamefont{M.}~\bibnamefont{Srednicki}},
  \bibinfo{journal}{Phys. Lett.} \textbf{\bibinfo{volume}{B510}},
  \bibinfo{pages}{236} (\bibinfo{year}{2001}{\natexlab{c}}),
  \eprint[http://arXiv.org/abs]{hep-ph/0102098}.

\bibitem[{\citenamefont{Lahanas and Spanos}(2001)}]{Lahanas:2001yr}
\bibinfo{author}{\bibfnamefont{A.~B.} \bibnamefont{Lahanas}} \bibnamefont{and}
  \bibinfo{author}{\bibfnamefont{V.~C.} \bibnamefont{Spanos}}
  (\bibinfo{year}{2001}), \eprint[http://arXiv.org/abs]{hep-ph/0106345}.

\bibitem[{\citenamefont{Battaglia et~al.}(2001)}]{Battaglia:2001zp}
\bibinfo{author}{\bibfnamefont{M.}~\bibnamefont{Battaglia}}
  \bibnamefont{et~al.} (\bibinfo{year}{2001}),
  \eprint[http://arXiv.org/abs]{hep-ph/0106204}.

\bibitem[{\citenamefont{Ellis et~al.}(2001{\natexlab{d}})\citenamefont{Ellis,
  Feng, Ferstl, Matchev, and Olive}}]{Ellis:2001hv}
\bibinfo{author}{\bibfnamefont{J.~R.} \bibnamefont{Ellis}},
  \bibinfo{author}{\bibfnamefont{J.~L.} \bibnamefont{Feng}},
  \bibinfo{author}{\bibfnamefont{A.}~\bibnamefont{Ferstl}},
  \bibinfo{author}{\bibfnamefont{K.~T.} \bibnamefont{Matchev}},
  \bibnamefont{and} \bibinfo{author}{\bibfnamefont{K.~A.} \bibnamefont{Olive}}
  (\bibinfo{year}{2001}{\natexlab{d}}),
  \eprint[http://arXiv.org/abs]{astro-ph/0110225}.

\bibitem[{\citenamefont{Lewin and Smith}(1996)}]{Lewin:1996}
\bibinfo{author}{\bibfnamefont{J.~D.} \bibnamefont{Lewin}} \bibnamefont{and}
  \bibinfo{author}{\bibfnamefont{P.~F.} \bibnamefont{Smith}},
  \bibinfo{journal}{Astropart. Phys.} \textbf{\bibinfo{volume}{6}},
  \bibinfo{pages}{87} (\bibinfo{year}{1996}).

\bibitem[{\citenamefont{Jungman et~al.}(1996)\citenamefont{Jungman,
  Kamionkowski, and Griest}}]{Jungman:1996df}
\bibinfo{author}{\bibfnamefont{G.}~\bibnamefont{Jungman}},
  \bibinfo{author}{\bibfnamefont{M.}~\bibnamefont{Kamionkowski}},
  \bibnamefont{and} \bibinfo{author}{\bibfnamefont{K.}~\bibnamefont{Griest}},
  \bibinfo{journal}{Phys. Rept.} \textbf{\bibinfo{volume}{267}},
  \bibinfo{pages}{195} (\bibinfo{year}{1996}),
  \eprint[http://arXiv.org/abs]{hep-ph/9506380}.

\bibitem[{\citenamefont{Bernabei et~al.}(1998)}]{Bernabei:1998}
\bibinfo{author}{\bibfnamefont{R.}~\bibnamefont{Bernabei}}
  \bibnamefont{et~al.}, \bibinfo{journal}{Phys. Lett.}
  \textbf{\bibinfo{volume}{B424}}, \bibinfo{pages}{195} (\bibinfo{year}{1998}).

\bibitem[{\citenamefont{Bernabei et~al.}(2001)}]{Bernabei:2001}
\bibinfo{author}{\bibfnamefont{R.}~\bibnamefont{Bernabei}}
  \bibnamefont{et~al.}, \bibinfo{journal}{Nucl. Phys. Proc. Suppl.}
  \textbf{\bibinfo{volume}{91}}, \bibinfo{pages}{361} (\bibinfo{year}{2001}).

\bibitem[{\citenamefont{Belli et~al.}(2001)}]{Belli:2001}
\bibinfo{author}{\bibfnamefont{P.}~\bibnamefont{Belli}} \bibnamefont{et~al.}
  (\bibinfo{year}{2001}), \eprint[http://arXiv.org/abs]{hep-ph/0112018}.

\bibitem[{\citenamefont{Abusaidi et~al.}(2000)}]{Abusaidi:2000}
\bibinfo{author}{\bibfnamefont{R.}~\bibnamefont{Abusaidi}} \bibnamefont{et~al.}
  (\bibinfo{collaboration}{CDMS}), \bibinfo{journal}{Nucl. Instrum. Meth.}
  \textbf{\bibinfo{volume}{A444}}, \bibinfo{pages}{345} (\bibinfo{year}{2000}),
  \eprint[http://arXiv.org/abs]{astro-ph/0002471}.

\bibitem[{\citenamefont{Benoit et~al.}(2001)}]{Benoit:2001}
\bibinfo{author}{\bibfnamefont{A.}~\bibnamefont{Benoit}} \bibnamefont{et~al.}
  (\bibinfo{collaboration}{EDELWEISS}), \bibinfo{journal}{Phys. Lett.}
  \textbf{\bibinfo{volume}{B513}}, \bibinfo{pages}{15} (\bibinfo{year}{2001}),
  \eprint[http://arXiv.org/abs]{astro-ph/0106094}.

\bibitem[{\citenamefont{Smith and Weiner}(2001)}]{Smith:2001hy}
\bibinfo{author}{\bibfnamefont{D.~R.} \bibnamefont{Smith}} \bibnamefont{and}
  \bibinfo{author}{\bibfnamefont{N.}~\bibnamefont{Weiner}},
  \bibinfo{journal}{Phys. Rev.} \textbf{\bibinfo{volume}{D64}},
  \bibinfo{pages}{043502} (\bibinfo{year}{2001}),
  \eprint[http://arXiv.org/abs]{hep-ph/0101138}.

\bibitem[{\citenamefont{Suzuki}(2000)}]{Suzuki:2000}
\bibinfo{author}{\bibfnamefont{Y.}~\bibnamefont{Suzuki}}
  (\bibinfo{collaboration}{Xenon}) (\bibinfo{year}{2000}),
  \eprint[http://arXiv.org/abs]{hep-ph/0008296}.

\bibitem[{\citenamefont{Spooner et~al.}(2000)}]{Spooner:2000}
\bibinfo{author}{\bibfnamefont{N.~J.~C.} \bibnamefont{Spooner}}
  \bibnamefont{et~al.}, p. \bibinfo{pages}{365} (\bibinfo{year}{2000}),
  \bibinfo{note}{prepared for 4th International Symposium on Sources and
  Detection of Dark Matter in the Universe (DM 2000), Marina del Rey,
  California, 23-25 Feb 2000}.

\bibitem[{\citenamefont{Martoff et~al.}(2000)\citenamefont{Martoff,
  Snowden-Ifft, Ohnuki, Spooner, and Lehner}}]{Martoff:2000}
\bibinfo{author}{\bibfnamefont{C.~J.} \bibnamefont{Martoff}},
  \bibinfo{author}{\bibfnamefont{D.~P.} \bibnamefont{Snowden-Ifft}},
  \bibinfo{author}{\bibfnamefont{T.}~\bibnamefont{Ohnuki}},
  \bibinfo{author}{\bibfnamefont{N.}~\bibnamefont{Spooner}}, \bibnamefont{and}
  \bibinfo{author}{\bibfnamefont{M.}~\bibnamefont{Lehner}},
  \bibinfo{journal}{Nucl. Instrum. Meth.} \textbf{\bibinfo{volume}{A440}},
  \bibinfo{pages}{355} (\bibinfo{year}{2000}).

\bibitem[{\citenamefont{Klapdor-Kleingrothaus and
  Majorovits}(2000)}]{Klapdor:2000}
\bibinfo{author}{\bibfnamefont{H.~V.} \bibnamefont{Klapdor-Kleingrothaus}}
  \bibnamefont{and}
  \bibinfo{author}{\bibfnamefont{B.}~\bibnamefont{Majorovits}}
  (\bibinfo{year}{2000}), \eprint[http://arXiv.org/abs]{hep-ph/0103079}.

\bibitem[{\citenamefont{Baudis et~al.}(2000)}]{Baudis:2000}
\bibinfo{author}{\bibfnamefont{L.}~\bibnamefont{Baudis}} \bibnamefont{et~al.}
  (\bibinfo{year}{2000}), \eprint[http://arXiv.org/abs]{hep-ex/0012022}.

\bibitem[{\citenamefont{Gaitskell}(2001)}]{Gaitskell:2001}
\bibinfo{author}{\bibfnamefont{R.~J.} \bibnamefont{Gaitskell}}
  (\bibinfo{year}{2001}), \eprint[http://arXiv.org/abs]{astro-ph/0106200}.

\bibitem[{\citenamefont{Spooner and Kudryavtsev}(2001)}]{Spooner:2001}
\bibinfo{author}{\bibfnamefont{N.~J.~C.} \bibnamefont{Spooner}}
  \bibnamefont{and} \bibinfo{author}{\bibfnamefont{V.~A.}
  \bibnamefont{Kudryavtsev}} (\bibinfo{year}{2001}),
  \eprint[http://arXiv.org/abs]{astro-ph/0111053}.

\bibitem[{\citenamefont{Falk et~al.}()\citenamefont{Falk, Ganis, McDonald,
  Olive, and Srednicki}}]{ssard}
\bibinfo{author}{\bibfnamefont{T.}~\bibnamefont{Falk}},
  \bibinfo{author}{\bibfnamefont{G.}~\bibnamefont{Ganis}},
  \bibinfo{author}{\bibfnamefont{J.}~\bibnamefont{McDonald}},
  \bibinfo{author}{\bibfnamefont{K.~A.} \bibnamefont{Olive}}, \bibnamefont{and}
  \bibinfo{author}{\bibfnamefont{M.}~\bibnamefont{Srednicki}},
  \bibinfo{note}{unpublished}.

\bibitem[{\citenamefont{Ellis et~al.}(2000{\natexlab{b}})\citenamefont{Ellis,
  Ferstl, and Olive}}]{Ellis:2000ds}
\bibinfo{author}{\bibfnamefont{J.~R.} \bibnamefont{Ellis}},
  \bibinfo{author}{\bibfnamefont{A.}~\bibnamefont{Ferstl}}, \bibnamefont{and}
  \bibinfo{author}{\bibfnamefont{K.~A.} \bibnamefont{Olive}},
  \bibinfo{journal}{Phys. Lett.} \textbf{\bibinfo{volume}{B481}},
  \bibinfo{pages}{304} (\bibinfo{year}{2000}{\natexlab{b}}),
  \eprint[http://arXiv.org/abs]{hep-ph/0001005}.

\bibitem[{\citenamefont{Ellis et~al.}(2001{\natexlab{e}})\citenamefont{Ellis,
  Ferstl, and Olive}}]{Ellis:2001qm}
\bibinfo{author}{\bibfnamefont{J.~R.} \bibnamefont{Ellis}},
  \bibinfo{author}{\bibfnamefont{A.}~\bibnamefont{Ferstl}}, \bibnamefont{and}
  \bibinfo{author}{\bibfnamefont{K.~A.} \bibnamefont{Olive}}
  (\bibinfo{year}{2001}{\natexlab{e}}),
  \eprint[http://arXiv.org/abs]{hep-ph/0111064}.

\bibitem[{\citenamefont{Schnee et~al.}(1998)}]{Schnee:1998gf}
\bibinfo{author}{\bibfnamefont{R.~W.} \bibnamefont{Schnee}}
  \bibnamefont{et~al.}, \bibinfo{journal}{Phys. Rept.}
  \textbf{\bibinfo{volume}{307}}, \bibinfo{pages}{283} (\bibinfo{year}{1998}).

\bibitem[{\citenamefont{Bravin et~al.}(1999)}]{Bravin:1999fc}
\bibinfo{author}{\bibfnamefont{M.}~\bibnamefont{Bravin}} \bibnamefont{et~al.}
  (\bibinfo{collaboration}{CRESST-Collaboration}), \bibinfo{journal}{Astropart.
  Phys.} \textbf{\bibinfo{volume}{12}}, \bibinfo{pages}{107}
  (\bibinfo{year}{1999}), \eprint[http://arXiv.org/abs]{hep-ex/9904005}.

\bibitem[{\citenamefont{Klapdor-Kleingrothaus}(2000)}]{GENIUS}
\bibinfo{author}{\bibfnamefont{H.~V.} \bibnamefont{Klapdor-Kleingrothaus}}
  (\bibinfo{year}{2000}), \eprint[http://arXiv.org/abs]{hep-ph/0104028}.

\bibitem[{\citenamefont{Gaitskell and Mandic}()}]{GM:2000web}
\bibinfo{author}{\bibfnamefont{R.}~\bibnamefont{Gaitskell}} \bibnamefont{and}
  \bibinfo{author}{\bibfnamefont{V.}~\bibnamefont{Mandic}},
  \bibinfo{note}{interactive data plotter maintained at
  http://dmtools.berkeley.edu}.

\bibitem[{\citenamefont{Silk et~al.}(1985)\citenamefont{Silk, Olive, and
  Srednicki}}]{Silk:1985ax}
\bibinfo{author}{\bibfnamefont{J.}~\bibnamefont{Silk}},
  \bibinfo{author}{\bibfnamefont{K.~A.} \bibnamefont{Olive}}, \bibnamefont{and}
  \bibinfo{author}{\bibfnamefont{M.}~\bibnamefont{Srednicki}},
  \bibinfo{journal}{Phys. Rev. Lett.} \textbf{\bibinfo{volume}{55}},
  \bibinfo{pages}{257} (\bibinfo{year}{1985}).

\bibitem[{\citenamefont{Freese}(1986)}]{Freese:1985qw}
\bibinfo{author}{\bibfnamefont{K.}~\bibnamefont{Freese}},
  \bibinfo{journal}{Phys. Lett.} \textbf{\bibinfo{volume}{B167}},
  \bibinfo{pages}{295} (\bibinfo{year}{1986}).

\bibitem[{\citenamefont{Krauss et~al.}(1986)\citenamefont{Krauss, Srednicki,
  and Wilczek}}]{Krauss:1985aa}
\bibinfo{author}{\bibfnamefont{L.~M.} \bibnamefont{Krauss}},
  \bibinfo{author}{\bibfnamefont{M.}~\bibnamefont{Srednicki}},
  \bibnamefont{and} \bibinfo{author}{\bibfnamefont{F.}~\bibnamefont{Wilczek}},
  \bibinfo{journal}{Phys. Rev.} \textbf{\bibinfo{volume}{D33}},
  \bibinfo{pages}{2079} (\bibinfo{year}{1986}).

\bibitem[{\citenamefont{Bergstrom
  et~al.}(1998{\natexlab{a}})\citenamefont{Bergstrom, Edsjo, and
  Gondolo}}]{Bergstrom:1998xh}
\bibinfo{author}{\bibfnamefont{L.}~\bibnamefont{Bergstrom}},
  \bibinfo{author}{\bibfnamefont{J.}~\bibnamefont{Edsjo}}, \bibnamefont{and}
  \bibinfo{author}{\bibfnamefont{P.}~\bibnamefont{Gondolo}},
  \bibinfo{journal}{Phys. Rev.} \textbf{\bibinfo{volume}{D58}},
  \bibinfo{pages}{103519} (\bibinfo{year}{1998}{\natexlab{a}}),
  \eprint[http://arXiv.org/abs]{hep-ph/9806293}.

\bibitem[{\citenamefont{Bottino et~al.}(1999)\citenamefont{Bottino, Donato,
  Fornengo, and Scopel}}]{Bottino:1998vw}
\bibinfo{author}{\bibfnamefont{A.}~\bibnamefont{Bottino}},
  \bibinfo{author}{\bibfnamefont{F.}~\bibnamefont{Donato}},
  \bibinfo{author}{\bibfnamefont{N.}~\bibnamefont{Fornengo}}, \bibnamefont{and}
  \bibinfo{author}{\bibfnamefont{S.}~\bibnamefont{Scopel}},
  \bibinfo{journal}{Astropart. Phys.} \textbf{\bibinfo{volume}{10}},
  \bibinfo{pages}{203} (\bibinfo{year}{1999}),
  \eprint[http://arXiv.org/abs]{hep-ph/9809239}.

\bibitem[{\citenamefont{Corsetti and Nath}(2000)}]{Corsetti:1999ma}
\bibinfo{author}{\bibfnamefont{A.}~\bibnamefont{Corsetti}} \bibnamefont{and}
  \bibinfo{author}{\bibfnamefont{P.}~\bibnamefont{Nath}},
  \bibinfo{journal}{Int. J. Mod. Phys.} \textbf{\bibinfo{volume}{A15}},
  \bibinfo{pages}{905} (\bibinfo{year}{2000}),
  \eprint[http://arXiv.org/abs]{hep-ph/9904497}.

\bibitem[{\citenamefont{Feng et~al.}(2001{\natexlab{d}})\citenamefont{Feng,
  Matchev, and Wilczek}}]{Feng:2001zu}
\bibinfo{author}{\bibfnamefont{J.~L.} \bibnamefont{Feng}},
  \bibinfo{author}{\bibfnamefont{K.~T.} \bibnamefont{Matchev}},
  \bibnamefont{and} \bibinfo{author}{\bibfnamefont{F.}~\bibnamefont{Wilczek}},
  \bibinfo{journal}{Phys. Rev.} \textbf{\bibinfo{volume}{D63}},
  \bibinfo{pages}{045024} (\bibinfo{year}{2001}{\natexlab{d}}),
  \eprint[http://arXiv.org/abs]{astro-ph/0008115}.

\bibitem[{\citenamefont{Barger et~al.}(2001{\natexlab{b}})\citenamefont{Barger,
  Halzen, Hooper, and Kao}}]{Barger:2001ur}
\bibinfo{author}{\bibfnamefont{V.~D.} \bibnamefont{Barger}},
  \bibinfo{author}{\bibfnamefont{F.}~\bibnamefont{Halzen}},
  \bibinfo{author}{\bibfnamefont{D.}~\bibnamefont{Hooper}}, \bibnamefont{and}
  \bibinfo{author}{\bibfnamefont{C.}~\bibnamefont{Kao}}
  (\bibinfo{year}{2001}{\natexlab{b}}),
  \eprint[http://arXiv.org/abs]{hep-ph/0105182}.

\bibitem[{\citenamefont{Andres et~al.}(1999)}]{AMANDA}
\bibinfo{author}{\bibfnamefont{E.}~\bibnamefont{Andres}} \bibnamefont{et~al.}
  (\bibinfo{collaboration}{AMANDA}) (\bibinfo{year}{1999}),
  \eprint[http://arXiv.org/abs]{astro-ph/9906205}.

\bibitem[{\citenamefont{Anassontzis et~al.}(2000)}]{NESTOR}
\bibinfo{author}{\bibfnamefont{E.~G.} \bibnamefont{Anassontzis}}
  \bibnamefont{et~al.} (\bibinfo{collaboration}{NESTOR}),
  \bibinfo{journal}{Nucl. Phys. Proc. Suppl.} \textbf{\bibinfo{volume}{85}},
  \bibinfo{pages}{153} (\bibinfo{year}{2000}).

\bibitem[{\citenamefont{Carmona}(2001)}]{ANTARES}
\bibinfo{author}{\bibfnamefont{E.}~\bibnamefont{Carmona}}
  (\bibinfo{collaboration}{ANTARES}), \bibinfo{journal}{Nucl. Phys. Proc.
  Suppl.} \textbf{\bibinfo{volume}{95}}, \bibinfo{pages}{161}
  (\bibinfo{year}{2001}).

\bibitem[{\citenamefont{Leuthold}(1998)}]{IceCube}
\bibinfo{author}{\bibfnamefont{M.}~\bibnamefont{Leuthold}}
  (\bibinfo{year}{1998}), \bibinfo{note}{prepared for International Workshop on
  Simulations and Analysis Methods for Large Neutrino Telescopes, Zeuthen,
  Germany, 6-9 Jul 1998}.

\bibitem[{\citenamefont{Urban et~al.}(1992)}]{Urban:1992ej}
\bibinfo{author}{\bibfnamefont{M.}~\bibnamefont{Urban}} \bibnamefont{et~al.},
  \bibinfo{journal}{Phys. Lett.} \textbf{\bibinfo{volume}{B293}},
  \bibinfo{pages}{149} (\bibinfo{year}{1992}),
  \eprint[http://arXiv.org/abs]{hep-ph/9208255}.

\bibitem[{\citenamefont{Berezinsky et~al.}(1992)\citenamefont{Berezinsky,
  Gurevich, and Zybin}}]{Berezinsky:1992mx}
\bibinfo{author}{\bibfnamefont{V.~S.} \bibnamefont{Berezinsky}},
  \bibinfo{author}{\bibfnamefont{A.~V.} \bibnamefont{Gurevich}},
  \bibnamefont{and} \bibinfo{author}{\bibfnamefont{K.~P.} \bibnamefont{Zybin}},
  \bibinfo{journal}{Phys. Lett.} \textbf{\bibinfo{volume}{B294}},
  \bibinfo{pages}{221} (\bibinfo{year}{1992}).

\bibitem[{\citenamefont{Berezinsky et~al.}(1994)\citenamefont{Berezinsky,
  Bottino, and Mignola}}]{Berezinsky:1994wv}
\bibinfo{author}{\bibfnamefont{V.}~\bibnamefont{Berezinsky}},
  \bibinfo{author}{\bibfnamefont{A.}~\bibnamefont{Bottino}}, \bibnamefont{and}
  \bibinfo{author}{\bibfnamefont{G.}~\bibnamefont{Mignola}},
  \bibinfo{journal}{Phys. Lett.} \textbf{\bibinfo{volume}{B325}},
  \bibinfo{pages}{136} (\bibinfo{year}{1994}),
  \eprint[http://arXiv.org/abs]{hep-ph/9402215}.

\bibitem[{\citenamefont{Bergstrom
  et~al.}(1998{\natexlab{b}})\citenamefont{Bergstrom, Ullio, and
  Buckley}}]{Bergstrom:1998fj}
\bibinfo{author}{\bibfnamefont{L.}~\bibnamefont{Bergstrom}},
  \bibinfo{author}{\bibfnamefont{P.}~\bibnamefont{Ullio}}, \bibnamefont{and}
  \bibinfo{author}{\bibfnamefont{J.~H.} \bibnamefont{Buckley}},
  \bibinfo{journal}{Astropart. Phys.} \textbf{\bibinfo{volume}{9}},
  \bibinfo{pages}{137} (\bibinfo{year}{1998}{\natexlab{b}}),
  \eprint[http://arXiv.org/abs]{astro-ph/9712318}.

\bibitem[{\citenamefont{Tylka}(1989)}]{Tylka:1989xj}
\bibinfo{author}{\bibfnamefont{A.~J.} \bibnamefont{Tylka}},
  \bibinfo{journal}{Phys. Rev. Lett.} \textbf{\bibinfo{volume}{63}},
  \bibinfo{pages}{840} (\bibinfo{year}{1989}).

\bibitem[{\citenamefont{Turner and Wilczek}(1990)}]{Turner:1990kg}
\bibinfo{author}{\bibfnamefont{M.~S.} \bibnamefont{Turner}} \bibnamefont{and}
  \bibinfo{author}{\bibfnamefont{F.}~\bibnamefont{Wilczek}},
  \bibinfo{journal}{Phys. Rev.} \textbf{\bibinfo{volume}{D42}},
  \bibinfo{pages}{1001} (\bibinfo{year}{1990}).

\bibitem[{\citenamefont{Kamionkowski and Turner}(1991)}]{Kamionkowski:1991ty}
\bibinfo{author}{\bibfnamefont{M.}~\bibnamefont{Kamionkowski}}
  \bibnamefont{and} \bibinfo{author}{\bibfnamefont{M.~S.}
  \bibnamefont{Turner}}, \bibinfo{journal}{Phys. Rev.}
  \textbf{\bibinfo{volume}{D43}}, \bibinfo{pages}{1774} (\bibinfo{year}{1991}).

\bibitem[{\citenamefont{Moskalenko and Strong}(1999)}]{Moskalenko:1999sb}
\bibinfo{author}{\bibfnamefont{I.~V.} \bibnamefont{Moskalenko}}
  \bibnamefont{and} \bibinfo{author}{\bibfnamefont{A.~W.}
  \bibnamefont{Strong}}, \bibinfo{journal}{Phys. Rev.}
  \textbf{\bibinfo{volume}{D60}}, \bibinfo{pages}{063003}
  (\bibinfo{year}{1999}), \eprint[http://arXiv.org/abs]{astro-ph/9905283}.

\bibitem[{\citenamefont{Barrau}(2001)}]{AMS}
\bibinfo{author}{\bibfnamefont{A.}~\bibnamefont{Barrau}}
  (\bibinfo{collaboration}{AMS}) (\bibinfo{year}{2001}),
  \eprint[http://arXiv.org/abs]{astro-ph/0103493}.

\bibitem[{\citenamefont{Kane et~al.}(2001)\citenamefont{Kane, Wang, and
  Wells}}]{Kane:2001fz}
\bibinfo{author}{\bibfnamefont{G.~L.} \bibnamefont{Kane}},
  \bibinfo{author}{\bibfnamefont{L.-T.} \bibnamefont{Wang}}, \bibnamefont{and}
  \bibinfo{author}{\bibfnamefont{J.~D.} \bibnamefont{Wells}}
  (\bibinfo{year}{2001}), \eprint[http://arXiv.org/abs]{hep-ph/0108138}.

\bibitem[{\citenamefont{Baltz et~al.}(2001)\citenamefont{Baltz, Edsjo, Freese,
  and Gondolo}}]{Baltz:2001ir}
\bibinfo{author}{\bibfnamefont{E.~A.} \bibnamefont{Baltz}},
  \bibinfo{author}{\bibfnamefont{J.}~\bibnamefont{Edsjo}},
  \bibinfo{author}{\bibfnamefont{K.}~\bibnamefont{Freese}}, \bibnamefont{and}
  \bibinfo{author}{\bibfnamefont{P.}~\bibnamefont{Gondolo}}
  (\bibinfo{year}{2001}), \eprint[http://arXiv.org/abs]{astro-ph/0109318}.

\bibitem[{\citenamefont{Silk and Stebbins}(1992)}]{Silk:1992bh}
\bibinfo{author}{\bibfnamefont{J.}~\bibnamefont{Silk}} \bibnamefont{and}
  \bibinfo{author}{\bibfnamefont{A.}~\bibnamefont{Stebbins}}
  (\bibinfo{year}{1992}), \bibinfo{note}{cFPA-TH-92-09}.

\bibitem[{\citenamefont{Chardonnet et~al.}(1996)\citenamefont{Chardonnet,
  Mignola, Salati, and Taillet}}]{Chardonnet:1996ca}
\bibinfo{author}{\bibfnamefont{P.}~\bibnamefont{Chardonnet}},
  \bibinfo{author}{\bibfnamefont{G.}~\bibnamefont{Mignola}},
  \bibinfo{author}{\bibfnamefont{P.}~\bibnamefont{Salati}}, \bibnamefont{and}
  \bibinfo{author}{\bibfnamefont{R.}~\bibnamefont{Taillet}},
  \bibinfo{journal}{Phys. Lett.} \textbf{\bibinfo{volume}{B384}},
  \bibinfo{pages}{161} (\bibinfo{year}{1996}),
  \eprint[http://arXiv.org/abs]{astro-ph/9606174}.

\bibitem[{\citenamefont{Bottino et~al.}(1998)\citenamefont{Bottino, Donato,
  Fornengo, and Salati}}]{Bottino:1998tw}
\bibinfo{author}{\bibfnamefont{A.}~\bibnamefont{Bottino}},
  \bibinfo{author}{\bibfnamefont{F.}~\bibnamefont{Donato}},
  \bibinfo{author}{\bibfnamefont{N.}~\bibnamefont{Fornengo}}, \bibnamefont{and}
  \bibinfo{author}{\bibfnamefont{P.}~\bibnamefont{Salati}},
  \bibinfo{journal}{Phys. Rev.} \textbf{\bibinfo{volume}{D58}},
  \bibinfo{pages}{123503} (\bibinfo{year}{1998}),
  \eprint[http://arXiv.org/abs]{astro-ph/9804137}.

\bibitem[{\citenamefont{Donato et~al.}(2000)\citenamefont{Donato, Fornengo, and
  Salati}}]{Donato:1999gy}
\bibinfo{author}{\bibfnamefont{F.}~\bibnamefont{Donato}},
  \bibinfo{author}{\bibfnamefont{N.}~\bibnamefont{Fornengo}}, \bibnamefont{and}
  \bibinfo{author}{\bibfnamefont{P.}~\bibnamefont{Salati}},
  \bibinfo{journal}{Phys. Rev.} \textbf{\bibinfo{volume}{D62}},
  \bibinfo{pages}{043003} (\bibinfo{year}{2000}),
  \eprint[http://arXiv.org/abs]{hep-ph/9904481}.

\bibitem[{SUS()}]{SUSYWG}
\bibinfo{note}{Joint LEP~2 Supersymmetry Working Group, \\ {\tt
  http://lepsusy.web.cern.ch/lepsusy/Welcome.html}}.

\bibitem[{\citenamefont{Matchev and
  Pierce}(1999{\natexlab{a}})}]{Matchev:1999nb}
\bibinfo{author}{\bibfnamefont{K.~T.} \bibnamefont{Matchev}} \bibnamefont{and}
  \bibinfo{author}{\bibfnamefont{D.~M.} \bibnamefont{Pierce}},
  \bibinfo{journal}{Phys. Rev.} \textbf{\bibinfo{volume}{D60}},
  \bibinfo{pages}{075004} (\bibinfo{year}{1999}{\natexlab{a}}),
  \eprint[http://arXiv.org/abs]{hep-ph/9904282}.

\bibitem[{\citenamefont{Baer et~al.}(2000)\citenamefont{Baer, Drees, Paige,
  Quintana, and Tata}}]{Baer:1999bq}
\bibinfo{author}{\bibfnamefont{H.}~\bibnamefont{Baer}},
  \bibinfo{author}{\bibfnamefont{M.}~\bibnamefont{Drees}},
  \bibinfo{author}{\bibfnamefont{F.}~\bibnamefont{Paige}},
  \bibinfo{author}{\bibfnamefont{P.}~\bibnamefont{Quintana}}, \bibnamefont{and}
  \bibinfo{author}{\bibfnamefont{X.}~\bibnamefont{Tata}},
  \bibinfo{journal}{Phys. Rev.} \textbf{\bibinfo{volume}{D61}},
  \bibinfo{pages}{095007} (\bibinfo{year}{2000}),
  \eprint[http://arXiv.org/abs]{hep-ph/9906233}.

\bibitem[{\citenamefont{Barger and Kao}(1999)}]{Barger:1998hp}
\bibinfo{author}{\bibfnamefont{V.~D.} \bibnamefont{Barger}} \bibnamefont{and}
  \bibinfo{author}{\bibfnamefont{C.}~\bibnamefont{Kao}},
  \bibinfo{journal}{Phys. Rev.} \textbf{\bibinfo{volume}{D60}},
  \bibinfo{pages}{115015} (\bibinfo{year}{1999}),
  \eprint[http://arXiv.org/abs]{hep-ph/9811489}.

\bibitem[{\citenamefont{Matchev and
  Pierce}(1999{\natexlab{b}})}]{Matchev:1999yn}
\bibinfo{author}{\bibfnamefont{K.~T.} \bibnamefont{Matchev}} \bibnamefont{and}
  \bibinfo{author}{\bibfnamefont{D.~M.} \bibnamefont{Pierce}},
  \bibinfo{journal}{Phys. Lett.} \textbf{\bibinfo{volume}{B467}},
  \bibinfo{pages}{225} (\bibinfo{year}{1999}{\natexlab{b}}),
  \eprint[http://arXiv.org/abs]{hep-ph/9907505}.

\bibitem[{\citenamefont{Lykken and Matchev}(2000)}]{Lykken:1999kp}
\bibinfo{author}{\bibfnamefont{J.~D.} \bibnamefont{Lykken}} \bibnamefont{and}
  \bibinfo{author}{\bibfnamefont{K.~T.} \bibnamefont{Matchev}},
  \bibinfo{journal}{Phys. Rev.} \textbf{\bibinfo{volume}{D61}},
  \bibinfo{pages}{015001} (\bibinfo{year}{2000}),
  \eprint[http://arXiv.org/abs]{hep-ph/9903238}.

\bibitem[{\citenamefont{Demina et~al.}(2000)\citenamefont{Demina, Lykken,
  Matchev, and Nomerotski}}]{Demina:1999ty}
\bibinfo{author}{\bibfnamefont{R.}~\bibnamefont{Demina}},
  \bibinfo{author}{\bibfnamefont{J.~D.} \bibnamefont{Lykken}},
  \bibinfo{author}{\bibfnamefont{K.~T.} \bibnamefont{Matchev}},
  \bibnamefont{and}
  \bibinfo{author}{\bibfnamefont{A.}~\bibnamefont{Nomerotski}},
  \bibinfo{journal}{Phys. Rev.} \textbf{\bibinfo{volume}{D62}},
  \bibinfo{pages}{035011} (\bibinfo{year}{2000}),
  \eprint[http://arXiv.org/abs]{hep-ph/9910275}.

\end{thebibliography}

\end{document}